\documentclass[twocolumn,trackchanges]{aastex631}

\usepackage{amsmath}
\usepackage{epsf}
\usepackage{color}
\usepackage{enumitem}
\usepackage{mathtools}
\usepackage{float}
\usepackage[mathcal]{eucal}
 \let\mathscr\relax
\usepackage[scr]{rsfso}

\usepackage{pgffor}

\newcommand{\sol}{$_{\odot}$}
\newcommand{\muv}{$M_\mathrm{UV}$}
\newcommand{\bsed}{$\beta_\mathrm{SED}$}
\newcommand{\bpl}{$\beta_\mathrm{PL}$}

\usepackage{makecell, tabularx}

\graphicspath{{./}{figures/}}

\received{}
\revised{}
\accepted{}

\submitjournal{ApJL}

\shorttitle{Galaxy UV Colors at $z \sim 9-16$}
\shortauthors{Morales et al.}

\begin{document}

\title{Rest-Frame UV Colors for Faint Galaxies at $z \sim 9-16$ with the \textit{JWST} NGDEEP Survey}

\correspondingauthor{Alexa Morales}
\email{alexa.morales@utexas.edu}

\author[0000-0003-4965-0402]{Alexa M.\ Morales}\altaffiliation{NSF Graduate Research Fellow}
\affiliation{Department of Astronomy, The University of Texas at Austin, 2515 Speedway, Austin, TX, 78712, USA}

\author[0000-0001-8519-1130]{Steven L. Finkelstein}
\affiliation{Department of Astronomy, The University of Texas at Austin, 2515 Speedway, Austin, TX, 78712, USA}

\author[0000-0002-9393-6507]{Gene C. K. Leung}
\affiliation{Department of Astronomy, The University of Texas at Austin, 2515 Speedway, Austin, TX, 78712, USA}

\author[0000-0002-9921-9218]{Micaela B. Bagley}
\affiliation{Department of Astronomy, The University of Texas at Austin, 2515 Speedway, Austin, TX, 78712, USA}

\author[0000-0001-7151-009X]{Nikko J. Cleri}
\affiliation{Department of Physics and Astronomy, Texas A\&M University, College Station, TX, 77843-4242 USA}
\affiliation{George P.\ and Cynthia Woods Mitchell Institute for Fundamental Physics and Astronomy, Texas A\&M University, College Station, TX, 77843-4242 USA}

\author[0000-0003-2842-9434]{Romeel Dav\'e}
\affiliation{Institute for Astronomy, University of Edinburgh, Blackford Hill, Edinburgh, EH9 3HJ UK}
\affiliation{Department of Physics and Astronomy, University of the Western Cape, Robert Sobukwe Rd, Bellville, Cape Town 7535, South Africa}

\author[0000-0001-5414-5131]{Mark Dickinson}
\affiliation{NSF's National Optical-Infrared Astronomy Research Laboratory, 950 N. Cherry Ave., Tucson, AZ 85719, USA}

\author[0000-0001-7113-2738]{Henry C. Ferguson}
\affiliation{Space Telescope Science Institute, 3700 San Martin Drive, Baltimore, MD 21218, USA}

\author[0000-0001-6145-5090]{Nimish P. Hathi}
\affiliation{Space Telescope Science Institute, 3700 San Martin Drive, Baltimore, MD 21218, USA}

\author{Ewan Jones}
\affiliation{Institute for Astronomy, University of Edinburgh, Blackford Hill, Edinburgh, EH9 3HJ UK}

\author[0000-0002-6610-2048]{Anton M. Koekemoer}
\affiliation{Space Telescope Science Institute, 3700 San Martin Drive, Baltimore, MD 21218, USA}

\author[0000-0001-7503-8482]{Casey Papovich}
\affiliation{Department of Physics and Astronomy, Texas A\&M University, College Station, TX, 77843-4242 USA}
\affiliation{George P.\ and Cynthia Woods Mitchell Institute for Fundamental Physics and Astronomy, Texas A\&M University, College Station, TX, 77843-4242 USA}

\author[0000-0003-4528-5639]{Pablo G. P\'erez-Gonz\'alez}
\affiliation{Centro de Astrobiolog\'{\i}a (CAB), CSIC-INTA, Ctra. de Ajalvir km 4, Torrej\'on de Ardoz, E-28850, Madrid, Spain}

\author[0000-0003-3382-5941]{Nor Pirzkal}
\affiliation{ESA/AURA Space Telescope Science Institute} 

\author{Britton Smith}
\affiliation{Institute for Astronomy, University of Edinburgh, Blackford Hill, Edinburgh, EH9 3HJ UK}

\author[0000-0003-3903-6935]{Stephen M.~Wilkins} %
\affiliation{Astronomy Centre, University of Sussex, Falmer, Brighton BN1 9QH, UK}
\affiliation{Institute of Space Sciences and Astronomy, University of Malta, Msida MSD 2080, Malta}

\author[0000-0003-3466-035X]{{L. Y. Aaron} {Yung}}
\affiliation{Astrophysics Science Division, NASA Goddard Space Flight Center, 8800 Greenbelt Rd, Greenbelt, MD 20771, USA}
\affiliation{Space Telescope Science Institute, 3700 San Martin Drive, Baltimore, MD 21218, USA}

% \author{The NGDEEP Collaboration}

\begin{abstract}
We present measurements of the rest-frame UV spectral slope, $\beta$, for a sample of 36 faint star-forming galaxies at $z \sim 9-16$ discovered in one of the deepest \textit{JWST} NIRCam surveys to date, the Next Generation Deep Extragalactic Exploratory Public (NGDEEP) Survey. %While our sample provides notable insights, we discuss the impact of potential incompleteness, an important consideration for samples at these redshifts. 
We use robust photometric measurements for UV-faint galaxies (down to $M_\mathrm{UV} \sim -16$), originally published in \citet{leung2023}, and measure values of the UV spectral slope via photometric power-law fitting to both the observed photometry and to stellar population models obtained through spectral energy distribution (SED) fitting with \textsc{Bagpipes}. We obtain a median and 68\% confidence interval for $\beta$ from photometric power-law fitting of $\beta_\mathrm{PL} = -2.7^{+0.5}_{-0.5}$ and from SED-fitting, $\beta_\mathrm{SED} = -2.3^{+0.2}_{-0.1}$ for the full sample. We show that when only 2-3 photometric detections are available, SED-fitting has a lower scatter and reduced biases than photometric power-law fitting.  We quantify this bias and find that after correction, the median $\beta_\mathrm{SED,corr} = -2.5^{+0.2}_{-0.2}$.  We measure physical properties for our galaxies with \textsc{Bagpipes} and find that our faint ($M_\mathrm{UV} = -18.1^{+0.7}_{-0.9}$) sample is low mass ($\mathrm{log}[M_{\ast}/M_\odot] = 7.7^{+0.5}_{-0.5}$), fairly dust-poor ($A_\mathrm{v} = 0.1^{+0.2}_{-0.1}$ mag), and modestly young ($\mathrm{log[age]} = 7.8^{+0.2}_{-0.8}$ yr) with a median star formation rate of $\mathrm{log(SFR)} = -0.3^{+0.4}_{-0.4} M_\odot\mathrm{/yr}$. We find no strong evidence for ultra-blue UV spectral slopes ($\beta \sim -3$) within our sample, as would be expected for exotically metal-poor ($Z/Z$\sol $<$ 10$^{-3}$) stellar populations with very high Lyman-continuum escape fractions. Our observations are consistent with model predictions that galaxies of these stellar masses at $z\sim9-16$ should have only modestly low metallicities ($Z/Z$\sol $\sim$ 0.1--0.2).

\end{abstract}

% =========================================================================%

\section{Introduction} \label{sec:intro}

Analyzing galaxies in the rest-frame ultraviolet (UV) regime provides pivotal insights into the early universe. Massive stars within active star-forming galaxies emit UV radiation and, when observed, can offer information on various astrophysical properties. One key parameter of interest is the rest-frame UV spectral slope, $\beta$, a measure of the steepness of the UV continuum \citep[where $f_\lambda\propto \lambda^\beta$;][]{Calzetti1994,Meurer1999}. This serves as a potent tool in investigating underlying physical processes within galaxies, including dust attenuation, stellar mass, star formation rate, and metallicity \citep[e.g.]{Finkelstein2012,Wilkins2013,Bouwens2014,Rogers2014}. 

Historically, studies investigating the rest-frame UV spectral slope have relied on data obtained from ground and space-based observatories, predominantly the \textit{Hubble Space Telescope} (\textit{HST}).  Such observations have significantly advanced our understanding of galaxy properties from $z\sim 2-10$, revealing the intricate connection between the rest-frame UV spectral slope and other galaxy properties \citep{Hathi2008, Wilkins2011,Finkelstein2012, Bouwens2012,Dunlop2012,Dunlop2013,Hathi2013,Bouwens2014,Bhatawdekar2021,Tacchella2022}. However, recent advancements in observational capabilities, particularly with the advent of  \textit{JWST}, have revolutionized our ability to explore the rest-frame UV spectral slope with unprecedented precision and sensitivity to higher redshifts.

\textit{JWST} \citep{Gardner2006_JWST,Gardner2023_JWST}, equipped with its suite of cutting-edge imaging and spectroscopic instruments, such as the NIRCam \citep{Rieke2023NIRCam} instrument used for this work, is providing a remarkable leap forward in our understanding of the evolution of galaxy properties. With its improved sensitivity in the near-infrared and mid-infrared wavelength range, \textit{JWST} enables the precise determination of the rest-frame UV spectral slope for a larger sample of galaxies further back in time than observed before (at $z\gtrsim 10$). 

One enticing question for this early epoch is whether the first generations of stars can be observed.
%Despite the substantial progress made in our understanding of galaxy formation and evolution, the detection and characterization of exotically metal-poor stellar populations via observations still have yet to be done. 
Such stellar populations are expected to be extremely metal-poor \citep[$Z \leq 10^{-3} Z_\odot$;][]{Schaerer2003}, if not completely metal-free (e.g., Population III), resulting in very blue values of $\beta < -3$.  Prior to \textit{JWST}, there were few convincing ultra-blue galaxies (with $\beta$ measured via photometric power-law fitting) that had been found \citep{Schaerer2002,Bouwens2010,Labbe2010,Ono2010,Jiang2020}. The direct detection of these exotic populations has been challenging due to several factors, including their expected brief lifetimes, and that when they explode and pollute the surrounding interstellar medium future generations of stars contain metals, leaving a narrow window in time to detect metal-free stars.  % and are expected to span to lower masses, living longer, making them easier to detect. 
Lastly, prior to {\it JWST}, we were limited technologically, unable to probe to $z >$ 10 where such objects are expected to exist.  %Our current observational limitations further complicate efforts to detect these Population III stars. 
With the advent of {\it JWST}, it is therefore imperative to explore what constraints can be placed on stellar populations at the highest redshifts. 
Here using the first \textit{JWST} public deep field, the Next Generation Deep Extragalactic Exploratory Public (NGDEEP) Survey, we search for such objects, emphasizing exploring methods of accurately measuring the UV spectral slope.

This paper aims to investigate the range of  UV spectral slopes observed in faint galaxies at $z >$ 10, as well as explore potential correlations between  $\beta$ and various galaxy properties. UV-faint galaxies have already been characterized from $z\sim6-7$ with \textit{HST}. These galaxies have an average UV slope $\beta \sim -2.4$, indicative of low-metallicity, low-dust populations \citep{Wilkins2011,Finkelstein2012,Dunlop2013,Bouwens2014,Wilkins2016}. Early studies with {\it JWST} \citep{Topping2022,Cullen2023}  have measured $\beta \lesssim -3.0$ for a few galaxies, though the uncertainties are large thus, no convincing discovery of exotic stellar populations has been made. Here we aim to push these observations to lower luminosities at higher redshift while also presenting a methodology to reduce the scatter on our measurements of $\beta$. %By comparing the findings from current studies with \textit{JWST} to the previous work conducted with the \textit{HST}, we seek to assess the advancements made and gain deeper insights into the physical mechanisms governing galaxy evolution for these faint, low-mass sources. 

This paper is structured as follows. In Section~\ref{sec:data}, we describe the NGDEEP survey and the data reduction process. In Section~\ref{sec:methods}, we discuss our galaxy sample and our methods for photometric power-law and SED-fitting to derive $\beta$. In Section~\ref{sec:results}, we describe our findings from fitting observations and simulations to models and explore correlations between $\beta$ and galaxy parameters. In Section~\ref{sec:disc}, we discuss our results and present our conclusions in Section~\ref{sec:conclusions}. 
We use the \cite{planck2020} cosmology of $H_0=67.4 ~ \mathrm{km~s}^{-1} ~\mathrm{Mpc}^{-1}$, $\Omega_\mathrm{m}=0.315$ and $\Omega_\mathrm{\Lambda} = 0.685$ and all magnitudes are given in the AB system \citep{oke1983}.

% =========================================================================%

\section{Data}\label{sec:data}

\subsection{NGDEEP Imaging and Photometry}\label{sec:ngdeep}
NGDEEP is a deep imaging and slitless spectroscopic \textit{JWST} Cycle 1 treasury program (PID 2079, PIs: S. Finkelstein, C. Papovich, N. Pirzkal) targeting constraints in feedback processes in galaxies across cosmic time \citep{bagley2023_ngdeep}. NGDEEP's parallel ultra-deep NIRCam imaging in the Hubble Ultra-Deep Field ``Parallel-2" (HUDF-Par2) field should ultimately detect $\gtrsim 100$ galaxies at redshifts $z = 9\sim16$ over $5 \ \mathrm{arcmin}^2$ at $1 - 5 \mu \mathrm{m}$ on the deepest \textit{HST} F814W imaging in the sky (m$_\mathrm{lim,F814W} \sim$ 30) from the HUDF12 campaign \citep{Ellis2013,Koekemoer2013}. NGDEEP should ultimately reach $m\sim30.8$, up to $\sim 0.5 \ \mathrm{mag}$ deeper than Guaranteed Time Observation (GTO) NIRCam surveys in some filters, and will provide robust measurements for $\beta$ and galaxy morphologies for UV-faint, low-mass galaxies. While the full program was planned for early 2023, only half of the program was obtained due to a temporary suspension of operations for NIRISS, causing the NGDEEP observations to be pushed to the edge of the visibility window. The next visibility window satisfying the PA requirement of the parallel observations will occur in early 2024 when the remaining observations are expected to be taken. In this study, we report results using NIRCam data from the first half of the NGDEEP program, which is among the deepest data yet obtained by {\it JWST}. 

We use the NIRCam and \textit{HST}/ACS F814W imaging mosaics and photometric catalogs presented in \citet{leung2023}.  We refer the reader there for full details but summarize the reduction and cataloging process here. The NIRCam imaging data are reduced using the {\it JWST} pipeline\footnote{\url{https://github.com/spacetelescope/jwst}} \citep{Bushouse2022_JWST_Pipeline} with custom modifications. We process the exposures through Stage 1 of the pipeline, applying custom procedures to remove snowballs, wisps, and $1/f$ noise. The images are then processed through Stage 2. Since only pre-flight flats are available for the short wavelength filters through the CRDS, we apply custom sky flats to these filters, and we use the recently updated CRDS flats for the long wavelength filters. Before combining the exposures into mosaics through Stage 3, a custom version of the \texttt{TweakReg} routine is used to align the NIRCam images with the ACS mosaics \citep{Koekemoer2011,Koekemoer2013}, whose astrometry is tied to {\it Gaia} DR3\footnote{\url{https://www.cosmos.esa.int/web/gaia/dr3}}. Finally, we perform a custom background subtraction procedure on all the mosaics.

Aperture photometry was performed using \textsc{Source Extractor} \citep{bertin1996_sextractor}.  Colors were measured using small elliptical Kron apertures, accounting for the varying PSF across 0.8--5 $\mu$m via matching the point-spread-function (PSF) of bands bluer than F277W to the F277W PSF.  For F356W and F444W, source-by-source correction factors were derived to account for missing flux in the F277W-defined aperture.  Total fluxes were derived first by scaling to the flux ratio in F277W between the flux in the small elliptical aperture to the default \textsc{Source Extractor} ``MAG\_AUTO" Kron parameters, with a residual aperture correction (typically 5--10\%) measured and applied following source injection simulations.  Flux uncertainties were measured empirically from the data and assigned to each object for each filter based on the size of its elliptical aperture.

% =========================================================================%

\section{Methodology}\label{sec:methods}

%Here, we describe the process of obtaining our final sample. 
In Section~\ref{sec:sc}, we describe our sample of high redshift galaxies. In Section~\ref{sec:betamethodsed}, we describe the process of deriving the UV spectral slope SED-fitting (and other galaxy properties). In Section~\ref{sec:betamethodpl}, we describe the process of deriving the UV spectral slope from photometric power-law fitting to the observed photometry. In Section~\ref{sec:datasets}, we compare our results with other observations.

% ----------------------------------------------------------

\subsection{Galaxy Sample}\label{sec:sc}

We make use of the galaxy sample from \cite{leung2023}, summarizing their sample selection and completeness here.  They incorporate a combination of flux detection signal-to-noise ratios (S/N) with quantities derived from photometric redshift fitting with \textsc{EAZY} \citep{brammer_eazy}, using the full probability density functions of the photometric redshift, $P(z)$, and the best-fit redshift, $\mathrm{z\_a}$. Their final sample was then visually inspected to reject any spurious sources and to ensure a real dropout in all filter bands blueward of the Lyman break corresponding to the estimated redshift.  We here make use of 36 of the full sample of 38 galaxies from \cite{leung2023}, which span photometric redshifts $z\sim8.5-15.5$ and UV magnitudes $-21 \leq M_\mathrm{{UV}} \leq -16$.  The two galaxies we excluded are red sources whose fits with all photometric data points with \textsc{Bagpipes} \citep{Carnall2021} yield unphysical results. As discussed in \cite{leung2023}, this may indicate that these sources have an AGN component contributing to their emission, similar to other sources identified at brighter magnitudes and lower redshifts \citep[e.g.][]{Kocevski2023,Labbe2023b,Matthee2023,Barro2023}.

To explore any potential sample bias against UV-red objects, we make use of the same source injection simulations described in \citet{leung2023}. They measured the completeness of their photometric selection by injecting and attempting to recover 50,000 mock objects with varied F277W magnitudes, colors, and surface brightness profiles across images, measuring completeness via the proportion recovered in both photometric and sample selection criteria within specified magnitude bins. Galaxy size, a potential influence on sample completeness, is incorporated as an estimation parameter. Any bias in size measurements, identified by comparing input and recovered half-light radii, is corrected during the completeness calculation to ensure accurate data evaluation and selection robustness.  We explore the completeness as a function of $\beta$ and find that our sample selection remains sensitive to significantly redder UV slopes than our reddest object, and we observe that the completeness is only $\sim15\%$ lower at $\beta = -1.5$ than $\beta = -2.5$.  We conclude that our photometric selection criteria do not bias the observed $\beta$ distribution.

% ----------------------------------------------------------%

\subsection{Deriving the UV spectral slope from SED fitting}
\label{sec:betamethodsed}
For each source in the sample, we run \textsc{Bagpipes} \citep{Carnall2021}, a Bayesian SED-fitting code where photometric data is provided as input along with a suite of user-defined priors, and the posterior distributions of galaxy properties and corresponding model spectra are returned. Our priors span a wide parameter space to ensure we are not omitting valuable information (see Table~\ref{tab:priors}).  We aim to measure $\beta$ from these model spectra directly, following \citet{Finkelstein2012}.

We made two modifications to the \textsc{Bagpipes} source code: (1) we ensure that the \textsc{Bagpipes} is consistent with the  \textsc{EAZY} results used to select the galaxy sample by implementing an additional redshift prior that is the cumulative distribution function (CDF) of our \textsc{EAZY} P(z) while allowing the redshift range tested to be free from $z=0-20$. (2) We also incorporate a new free parameter added to \textsc{Bagpipes}, the Lyman continuum escape fraction, $f_\mathrm{esc}$, to allow for the flexibility of measuring bluer colors within our fitting ($f_\mathrm{{esc}}$ was otherwise by default set to zero, which can lead to nebular continuum dominating at young ages, and thus $\beta > -3$ even for dust-free, metal-poor populations). By incorporating this $f_\mathrm{{esc}}$ component, alongside other tuneable priors, we are able to allow \textsc{Bagpipes} to generate a model reaching a minimum blue `floor' of $\beta_\mathrm{SED} = -3.25$ when $f_\mathrm{{esc}} = 1$ (compared to $\beta_\mathrm{SED} = -2.44$ when $f_\mathrm{{esc}}=0$). Ideally, this $\beta$ floor would be bluer than the bluest observed value (which, as we have shown, is not possible if $f_\mathrm{{esc}}$ is fixed to zero). This should result in the recovery of the correct median $\beta$ without any systematic bias. For this work, we make use of the default BC03 stellar models \citep{BC03_models}, and \cite{Calzetti1994} dust models \textsc{Bagpipes} provides, though we note that future work using Population III star models may be able to reach bluer values.

% ------------- Figure 1 ----------- %
\begin{figure}[ht!]
    \centering
    \includegraphics[width=0.48\textwidth]{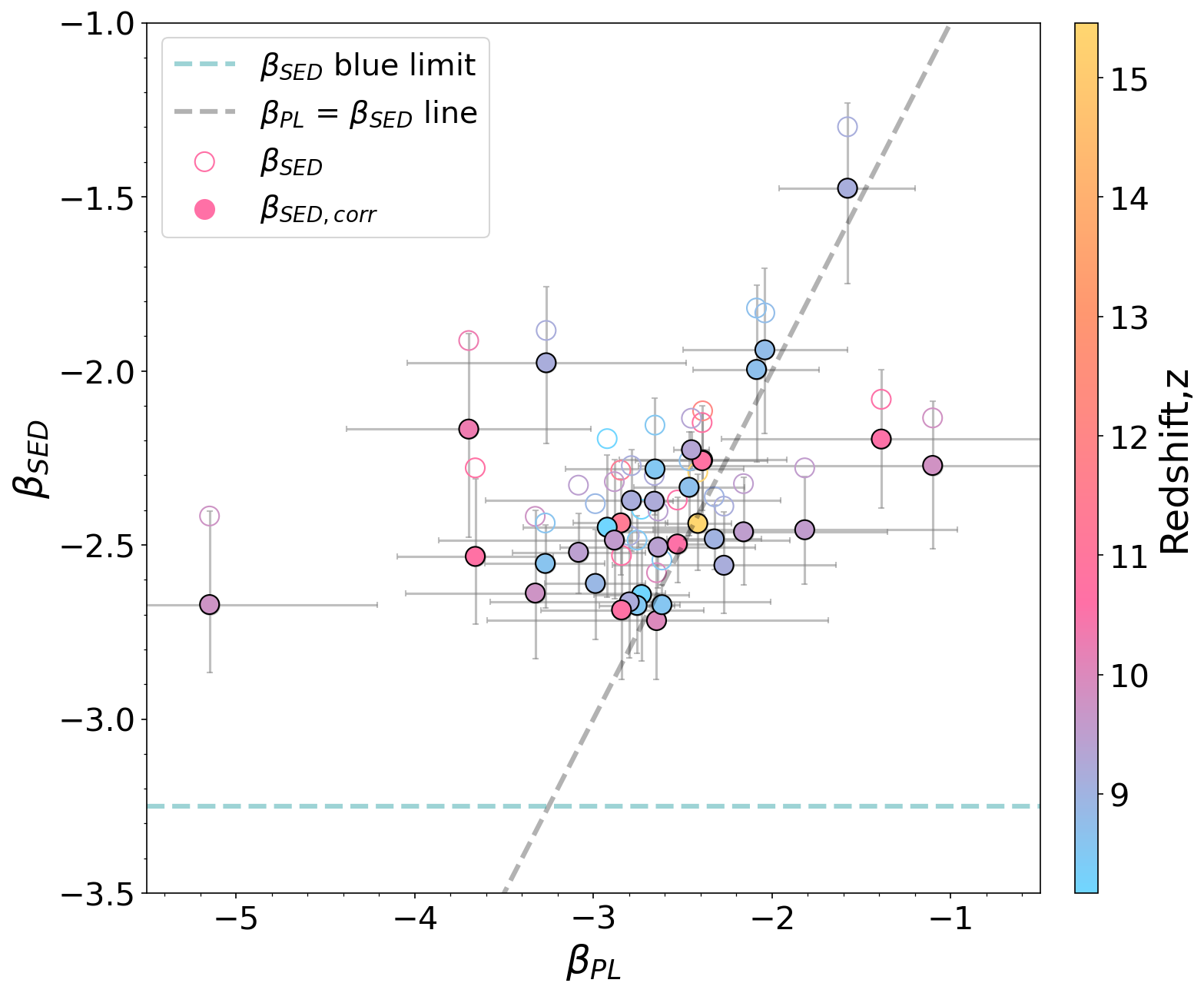}
    \caption{UV $\beta$ slope derived from photometric power-law fitting versus from SED-fitting color-coded by redshift. Original $\beta$ values are shown as open circles, while the filled circles have their SED-fitting-based $\beta$ corrected for measurement bias (discussed in Section~\ref{res_sims}). The blue dashed horizontal line is the bluest \bsed \ can go with \textsc{Bagpipes}, and the grey dashed line is the one-to-one line when equating $\beta_\mathrm{SED}$ to $\beta_\mathrm{PL}$. This relation shows that photometric power-law fitting, on average, yields bluer UV slopes and larger errors than SED-fitting.}
    \label{fig:bvb}
\end{figure}
% ------------- Figure 1 ----------- %

% ------------- Table 1 ----------- %

\begin{table*}[ht!]
\caption{\textsc{Bagpipes} priors defined for this work.}
\label{tab:priors}
\begin{tabular}{lll}
\hline
\hline

\textbf{Parameter}                                             & \textbf{Range}         & \textbf{Description}         \\                                             \hline
                      % \\
\textbf{SFH Component}              &               &                                                                                        \\
sfh{[}`age'{]}                                        & (0.002, 13.0) & Age of the galaxy in Gyr                                                               \\
sfh{[}`tau'{]}                                        & (0.1, 14.0)   & Delayed decay time in Gyr                                                                    \\
sfh{[}`metallicity'{]}                                & (0.0, 2.5)    & Metallicity in units of $Z_\odot$                                                      \\
sfh{[}`massformed'{]}                                 & (4.0, 13.0)   & $\mathrm{log}_{10}$ of total mass formed in units of $M_\odot$                         \\
\textbf{Dust Component}              &               &                                                                                        \\
dust{[}`type'{]}                                      & Calzetti      & Shape of the attenuation curve                                                         \\
dust{[}`Av'{]}                                        & (0.0, 4.0)    & Dust attenuation in units of magnitude                                                 \\
dust{[}`eta'{]}                                       & 1.0           & Multiplicative factor on $A_\mathrm{v}$ for stars in birth clouds                               \\
\textbf{Nebular Component}           &               &                                                                                        \\
nebular{[}`logU'{]}                                   & (-4.0, -1.0)  & $\mathrm{log}_{10}$ of the ionization parameter                                        \\
nebular{[}`fesc'{]}                                   & (0.0, 1.0)     & Lyman continuum escape fraction                                                                        \\
\textbf{Additional Fit Instructions} &               &                                                                                        \\
fit\_instructions{[}`redshift'{]}                     & (0.0, 20.0)   & Redshift range tested                                                                  \\
fit\_instructions{[}`redshift\_prior'{]}              & CDF(z) function & CDF of the \textsc{EAZY} P(z) as prior on redshift \\
\hline
\hline
\end{tabular}
\vspace{5mm}
\end{table*}
% ------------- Table 1 ----------- %

We utilize the median (and the difference between the median and the 68\% confidence bounds) for redshift, stellar mass, UV magnitude, dust attenuation, star formation rate, and mass-weighted age posteriors as our final values (and error bars) in this work. For each source, we set \textsc{Bagpipes} to fit model SEDs to the photometric data points and their errors, return the best-fit SED model, and return 1000 draws from the resulting posteriors for various galaxy properties as a function of these SEDs.

We measured the UV slope, \bsed, from each of the 1000 posterior model spectra via a power-law fit to all model flux density data points from $\lambda_\mathrm{rest} = 1500-3000 \AA$, while excluding a 10\AA\ region around wavelengths where we expect high-redshift emission lines to lie ($\lambda_\mathrm{rest} = 1549, 1909, 2796, 2803 \AA$), resulting in an average of 131 data points per model SED for the UV slope calculation. We adopt the median and the difference from the 1$\sigma$ bounds from these measurements as our final \bsed\ value and uncertainties, where $\mathrm{log}(f_\lambda) = \beta \mathrm{log}(\lambda)+ \mathrm{log}(y_\mathrm{int})$. This approach, while deviating from the wavelength windows of \cite{Calzetti1994}, similarly avoids strong emission lines. For our SED figures (Figures~\ref{fig:seds} and \ref{fig:app_seds}), we display the lowest $\chi^2$ model SED, the central 68\%, and the full range covered by the full model posterior.

% ----------------------------------------------------------%

\subsection{Deriving the UV spectral slope, $\beta$, from photometric power-law fitting}
\label{sec:betamethodpl}

Here we describe our process of measuring the UV spectral slope via photometric power-law fitting to the observed photometry. This is a more commonly used method to measure the UV spectral slope that is not reliant on stellar population models, though is very sensitive to the number of photometric data points available. 
When measuring the UV spectral slope with photometric power-law fitting to the observed photometry, \bpl, we redshift the rest-frame $\lambda = 1500 - 3000 \AA$ regime to the corresponding median redshift estimated from \textsc{EAZY}, $\mathrm{z\_a}$, for each galaxy in the sample and keep the filters whose central wavelength within each filter curve falls fully within the redshifted wavelength range. The redshift range of our sample, $z\sim9-16$, results in typically 2-3 photometric data points being available to fit a line (namely F200W, F277W, F356W, and at the highest redshifts, F444W). Once the photometric data points within the wavelength range are determined, we fit the data points to a line (as defined at the end of Section~\ref{sec:betamethodsed}) and run this process through \textsc{Emcee} \citep{Foreman-Mackey2013_emcee}. This procedure maximizes the likelihood that the model described by three free parameters matches the observed photometry for a given source: $\beta$, $y_\mathrm{int}$, and fractional error factor, log(f) where it is assumed that the likelihood function and its uncertainties are simply a Gaussian where the variance is underestimated by some fractional amount. 
Results are derived from the median and 68th percentile of the posterior distribution on these three parameters from a chain consisting of $5000$ steps and a burn-in of 100 steps (i.e., the MCMC process is iterated over 5000 times to estimate these parameters).

UV spectral slopes for our sample from both measurement methods are shown in Figure~\ref{fig:bvb}. 
This plot shows both the original and corrected $\beta$ values (corrections applied are discussed in Section~\ref{res_sims}). We show that the SED-fitting method yields smaller uncertainties on average (see Figure~\ref{fig:seds} for examples of this) -- this is simply due to the use of more data points along the specified wavelength range mentioned in Section~\ref{sec:betamethodsed} given that the SED is a good fit to the data (future works with varying stellar models may help alleviate any limitations our current model may have, see \S 5). After the correction is applied, we find that \bsed \ is in fairly good agreement with \bpl \ for some objects, while the \bpl\ measurements show significantly larger scatter, extending both to much bluer and much redder values than \bsed.  Specifically, while no objects reach $\beta_\mathrm{SED} < -3$, there are many whose $\beta_\mathrm{PL} < -3$.

% ----------------------------------------------------------%

\subsection{Literature Comparison Samples}\label{sec:datasets}

In comparing our work to other observations, we wanted to ensure we used datasets with selection strategies similar to the ones we implemented, and that the redshift range overlapped with this work. The observational datasets we used to compare values of $\beta$ versus redshift, stellar mass, and UV magnitude are as follows: UV slope values and corresponding galaxy parameters at $z=9-11$ by \cite{Tacchella2022} whose sample consists of 11 bright (H-band magnitude $<$ 26.6) galaxy candidates selected in the \textit{HST} CANDELS fields from \cite{Finkelstein2022_CANDELS} with photometry spanning the UV/optical with \textit{HST} and near-infrared with \textit{Spitzer}/IRAC. The work of \cite{Topping2022} utilizes the initial NIRCam data release at $z=7-11$ for 123 galaxies from the \textit{JWST} Cosmic Evolution Early Release Science (CEERS, \cite{Finkelstein2017_CEERSProp}) survey, this survey overlaps with that of the \textit{HST} EGS field. At $z=8-16$, we use the 61 galaxy observations in \cite{Cullen2023} from three \textit{JWST} fields: SMACS J0723, GLASS, and CEERS \citep{Pontoppidan2022} and the ground-based COSMOS/UltraVISTA survey \citep{McCracken2012}. Finally, we include a sample of 18 galaxies in the NGDEEP field at $z = 8-16$ from the works of \cite{Austin2023}.

% =========================================================================%

\section{Results}\label{sec:results}

In this Section, we describe the evolution of the UV spectral slope and its ties to galaxy properties.
In Section~\ref{res_bvb}, we showcase six galaxies in our sample that span various extremes in properties to showcase our sample and $\beta$ measurements.  In Section~\ref{res_sims}, we discuss the potential for measurement biases, using a sample of simulated galaxies which we fit with \textsc{Bagpipes}, deriving a correction factor to our observed sample. In Section~\ref{res_params}, we present our results for our corrected $\beta$ values as a function of various galaxy parameters.

% ------------- Figure 2 ----------- %

\begin{figure*}[ht!]
\centering
  \includegraphics[width=\textwidth]{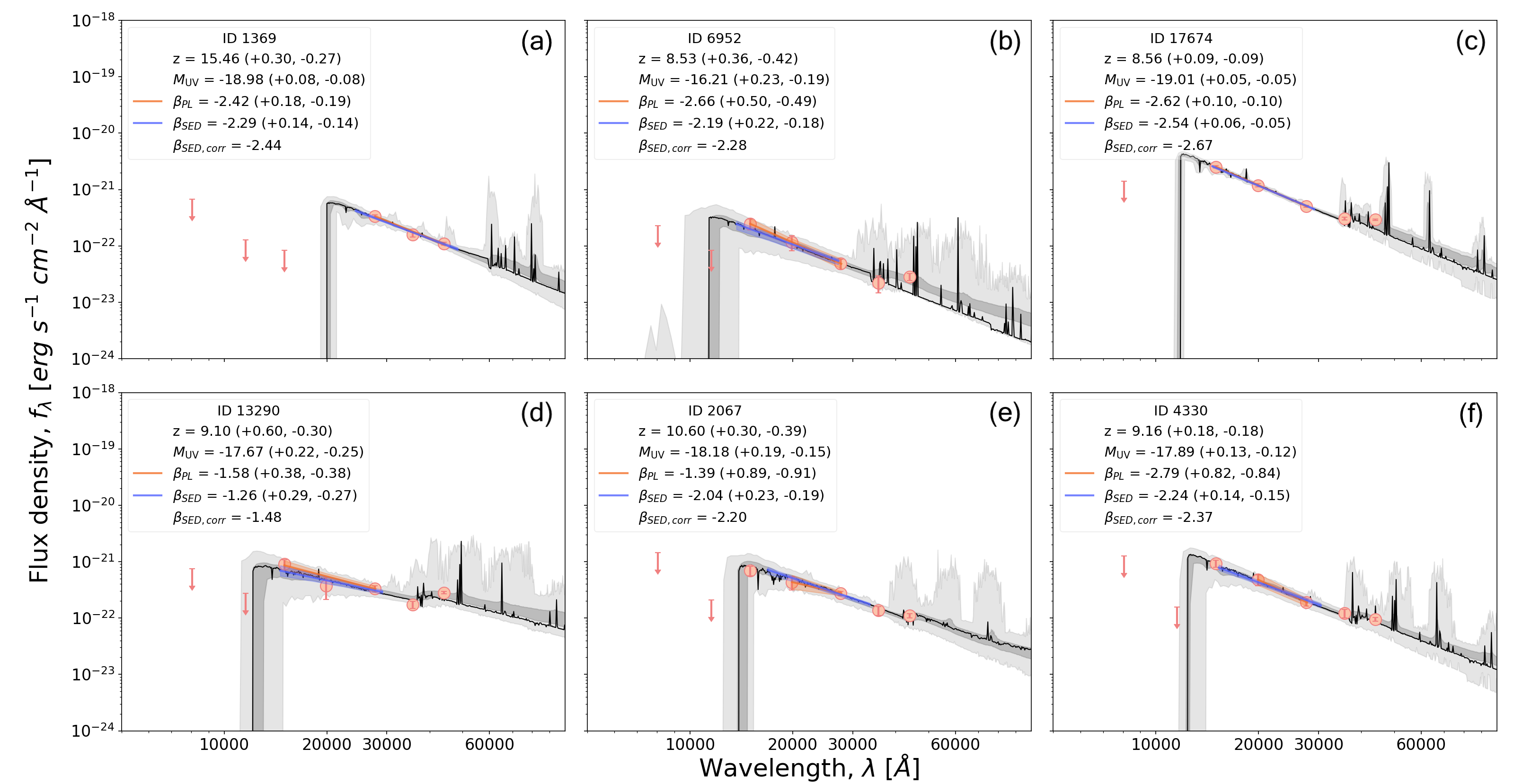}
\caption{Examples of \textsc{Bagpipes} SED fits to the observed photometry (orange circles) for six cases: (a) The highest redshift source in our sample at $z_\mathrm{SED} = 15.5^{+0.3}_{-0.3}$, (b) one of the faintest sources in our sample at $M_\mathrm{UV} = -16.2^{+0.3}_{-0.2}$, (c) one of the bluest sources in our sample with $\beta_\mathrm{SED,corr} = -2.7^{+0.1}_{-0.1}$, (d) one of the `reddest' sources in our sample with $\beta_\mathrm{SED,corr} = -1.5^{+0.3}_{-0.2}$, (e) a source where \bpl \ is redder than \bsed, and (f) a source where \bpl \ is bluer than \bsed. The black line is the SED model with the lowest $\chi^2$ from \textsc{Bagpipes}, the light grey shaded region represents the full range of SEDs being fit, and the darker grey shaded region is the $1\sigma$ bounds of the posterior sample of SED models. The blue (orange) shaded regions and lines represent the best-fit line to the photometry, where the UV spectral slopes and errors are obtained from when fit via the SED (photometric power-law) method.  It is clear, especially in the last two cases, that scatter in a single photometric band can affect \bpl\ significantly more than \bsed.}
\label{fig:seds}
\vspace{5mm}
\end{figure*}

% ----------------------------------------------------------%

\subsection{Highlighting Six Example Galaxies}\label{res_bvb}

In Figure~\ref{fig:seds}, we show six galaxies from our sample of 36 galaxies in the NGDEEP field with their corresponding photometry and model SEDs. The figure shows SEDs for the highest redshift galaxy in our sample at $z_\mathrm{SED} = 15.5^{+0.3}_{-0.3}$, whose photometric power-law and SED-fitted UV spectral slope are both very blue, $\beta_\mathrm{SED} = -2.4^{+0.1}_{-0.1}$ ($-2.3$ before correction) and $\beta_\mathrm{PL} = -2.4^{+0.2}_{-0.2}$. This is consistent with a low-metallicity, young stellar population, though spectroscopic follow-up is needed both to confirm the redshift as well as the metallicity. We also obtained photometric data for a $z = 8.5^{+0.3}_{-0.4}$ galaxy whose UV magnitude is quite faint, $M_\mathrm{UV} = -16.2^{+0.2}_{-0.2}$ and still exhibits very blue colors both with photometric power-law and SED-fitting, $\beta_\mathrm{SED} = -2.3^{+0.2}_{-0.2}$ ($-2.2$ before correction) and $\beta_\mathrm{PL} = -2.7^{+0.5}_{-0.5}$, highlighting the large scatter in \bpl\ at faint measured fluxes.

We also show SEDs for one of the bluest sources in our sample at $z= 8.6^{+0.1}_{-0.1}$ with a $\beta_\mathrm{SED} = -2.7^{+0.1}_{-0.1}$ ($-2.5$ before correction) and one of the `reddest' sources in our sample with a $\beta_\mathrm{SED} = -1.5^{+0.3}_{-0.2}$ ($-1.3$ before correction). Although when modeled with \textsc{Bagpipes}, the bluest galaxy is considered to be a fairly young galaxy, $\sim 10^7 $ years old (similar to the bulk of our sample), the lack of dust attenuation and relatively high specific star formation rate both play a role in the extremely blue slope. The redder galaxy, on the other hand, is slightly fainter, with a bit more dust attenuation contributing to our measurement. Lastly, we show two examples of galaxies whose \bsed \ fits are either bluer or redder than the measured photometric power-law fits, \bpl. However, in both instances, we show that our SED-fitting method yields smaller errors than photometric power-law fitting overall. As discussed, depending on the filter wavelength coverage and the redshift range, the power-law method can be limited to just two data points, yielding not only larger uncertainties but also larger scatter in the measured value. 

% ------------- Figure 2 ----------- %

% ------------- Table 2 ----------- %

\begin{table*}[ht!]
\caption{Redshift and UV spectral slopes for this sample.}
\label{tab:beta_slope}
\centering
\begin{tabularx}{0.85\textwidth}{ccccccccc}
\hline
\hline
ID     & RA       & Dec       & $\mathrm{z\_a}^a$  & ${\mathrm{z}_\mathrm{SED}}^b$      & ${\beta_\mathrm{PL}}^c$     &  \# Filters    & $\beta_\mathrm{SED}$         & Adjusted \\

     &        &        &  &     &     &  for \bpl     &        &  ${\beta_\mathrm{SED}}^d$ \\
\hline
250   & 53.249451 & -27.883313 & $11.62^{+0.21}_{-0.33}$ & $11.38^{+0.18}_{-0.21}$   & $-2.85^{+0.27}_{-0.26}$ & 3 & $-2.29^{+0.15}_{-0.13}$ & -2.44                             \\
1191  & 53.266583 & -27.876581 & $12.31^{+1.23}_{-0.30}$ & $12.10^{+0.21}_{-0.21}$   & $-2.39^{+0.47}_{-0.47}$ & 2 & $-2.11^{+0.16}_{-0.16}$ & -2.26                             \\
1369  & 53.249467 & -27.875710 & $15.82^{+0.12}_{-0.81}$ & $15.46^{+0.30}_{-0.27}$   & $-2.42^{+0.18}_{-0.19}$ & 3 & $-2.29^{+0.14}_{-0.14}$ & -2.44                             \\
1716  & 53.251057 & -27.796992 & $11.29^{+0.30}_{-1.47}$ & $10.27^{+0.69}_{-0.75}$   & $-3.70^{+0.69}_{-0.68}$ & 3 & $-1.91^{+0.31}_{-0.27}$ & -2.17                             \\
2067  & 53.239797 & -27.800244 & $10.75^{+0.42}_{-0.66}$ & $10.57^{+0.36}_{-0.36}$   & $-1.39^{+0.89}_{-0.91}$ & 2 & $-2.08^{+0.20}_{-0.20}$ & -2.20                             \\
2470  & 53.248546 & -27.802743 & $8.56^{+0.15}_{-0.63}$  & $8.17^{+0.24}_{-0.27}$    & $-2.73^{+0.27}_{-0.27}$ &  3 & $-2.40^{+0.19}_{-0.14}$ & -2.64                             \\
2497  & 53.248775 & -27.803090 & $9.19^{+0.60}_{-0.21}$  & $9.16^{+0.18}_{-0.18}$    & $-2.27^{+0.62}_{-0.63}$ & 2 & $-2.39^{+0.14}_{-0.15}$ & -2.56                             \\
3514  & 53.256875 & -27.807957 & $11.23^{+0.27}_{-0.90}$ & $10.72^{+0.45}_{-0.60}$   & $-3.66^{+0.44}_{-0.44}$ & 3 & $-2.28^{+0.19}_{-0.22}$ & -2.53                             \\
4134  & 53.245546 & -27.814372 & $10.69^{+0.18}_{-0.30}$ & $10.57^{+0.15}_{-0.15}$   & $-2.53^{+0.22}_{-0.22}$ & 2 & $-2.37^{+0.11}_{-0.13}$ & -2.50                             \\
4330  & 53.264834 & -27.816024 & $9.16^{+0.96}_{-0.18}$  & $9.13^{+0.18}_{-0.15}$    & $-2.79^{+0.82}_{-0.84}$ & 2 & $-2.27^{+0.15}_{-0.15}$ & -2.37                             \\
4674  & 53.245601 & -27.817588 & $8.62^{+0.18}_{-0.21}$  & $8.50^{+0.15}_{-0.12}$    & $-2.76^{+0.21}_{-0.21}$ & 3& $-2.49^{+0.14}_{-0.26}$ & -2.67                             \\
4740  & 53.257473 & -27.817828 & $9.19^{+0.72}_{-0.21}$  & $9.19^{+0.24}_{-0.18}$    & $-2.80^{+0.78}_{-0.79}$ & 2& $-2.47^{+0.16}_{-0.21}$ & -2.66                             \\
4919  & 53.262137 & -27.818774 & $9.34^{+1.02}_{-0.24}$  & $9.58^{+0.45}_{-0.39}$    & $-1.82^{+0.85}_{-0.86}$ & 2& $-2.28^{+0.16}_{-0.15}$ & -2.46                             \\
5118  & 53.261009 & -27.819895 & $9.25^{+0.09}_{-0.06}$  & $9.22^{+0.03}_{-0.03}$    & $-2.66^{+0.11}_{-0.11}$ & 2& $-2.30^{+0.04}_{-0.04}$ & -2.37                             \\
5947  & 53.248954 & -27.822988 & $8.89^{+1.50}_{-0.57}$  & $8.80^{+0.33}_{-0.36}$    & $-2.04^{+0.46}_{-0.46}$ & 3& $-1.83^{+0.24}_{-0.23}$ & -1.94                             \\
6134  & 53.248902 & -27.823695 & $9.07^{+0.09}_{-0.12}$  & $9.01^{+0.06}_{-0.09}$    & $-2.32^{+0.26}_{-0.26}$ & 2& $-2.36^{+0.09}_{-0.10}$ & -2.48                             \\
6477  & 53.250469 & -27.825099 & $8.74^{+0.33}_{-0.30}$  & $8.71^{+0.24}_{-0.21}$    & $-2.46^{+0.31}_{-0.31}$ & 3& $-2.26^{+0.14}_{-0.16}$ & -2.33                             \\
6952  & 53.239303 & -27.827168 & $8.68^{+0.45}_{-0.75}$  & $8.53^{+0.36}_{-0.42}$    & $-2.66^{+0.50}_{-0.49}$ & 3& $-2.16^{+0.24}_{-0.20}$ & -2.28                             \\
6980  & 53.256828 & -27.827244 & $10.06^{+0.45}_{-0.60}$ & $9.76^{+0.33}_{-0.30}$    & $-3.33^{+0.73}_{-0.73}$ & 2& $-2.42^{+0.19}_{-0.24}$ & -2.64                             \\
7530  & 53.237017 & -27.829892 & $8.71^{+0.57}_{-0.45}$  & $8.71^{+0.30}_{-0.30}$    & $-2.09^{+0.36}_{-0.35}$ & 3& $-1.82^{+0.27}_{-0.24}$ & -2.00                             \\
7722  & 53.242580 & -27.830637 & $10.93^{+0.30}_{-0.39}$ & $10.66^{+0.30}_{-0.36}$   & $-2.84^{+0.45}_{-0.46}$ & 3& $-2.53^{+0.20}_{-0.25}$ & -2.69                             \\
8042  & 53.237915 & -27.832320 & $8.53^{+0.18}_{-0.99}$  & $8.17^{+0.36}_{-0.51}$    & $-2.92^{+0.47}_{-0.47}$ & 3& $-2.19^{+0.20}_{-0.21}$ & -2.45                             \\
8165  & 53.234410 & -27.833172 & $10.48^{+0.36}_{-0.96}$ & $9.82^{+0.48}_{-0.48}$    & $-5.15^{+0.95}_{-0.94}$ & 2& $-2.42^{+0.20}_{-0.27}$ & -2.67                             \\
8427  & 53.235623 & -27.834282 & $8.98^{+0.18}_{-0.30}$  & $8.89^{+0.12}_{-0.18}$    & $-2.99^{+0.28}_{-0.28}$ & 3& $-2.38^{+0.16}_{-0.15}$ & -2.61                             \\
8461  & 53.235797 & -27.834499 & $10.36^{+0.27}_{-1.26}$ & $9.53^{+0.64}_{-0.43}$    & $-2.88^{+0.98}_{-0.98}$ & 2& $-2.32^{+0.17}_{-0.23}$ & -2.49                             \\
8894  & 53.241008 & -27.828822 & $9.34^{+0.87}_{-0.06}$  & $9.43^{+0.39}_{-0.18}$    & $-2.64^{+0.55}_{-0.54}$ & 2& $-2.40^{+0.12}_{-0.14}$ & -2.51                             \\
9261  & 53.270902 & -27.841204 & $8.74^{+0.15}_{-0.24}$  & $8.62^{+0.15}_{-0.15}$    & $-3.27^{+0.34}_{-0.33}$ & 3& $-2.44^{+0.13}_{-0.11}$ & -2.55                             \\
9555  & 53.258685 & -27.847324 & $10.45^{+0.39}_{-0.84}$ & $10.00^{+0.45}_{-0.51}$   & $-2.65^{+0.95}_{-0.96}$ & 2& $-2.58^{+0.17}_{-0.19}$ & -2.72                             \\
10296 & 53.276910 & -27.850568 & $10.48^{+0.33}_{-0.30}$ & $9.46^{+0.09}_{-0.03}$    & $-3.08^{+0.37}_{-0.37}$ & 2& $-2.33^{+0.12}_{-0.11}$ & -2.52                             \\
11522 & 53.242062 & -27.855079 & $10.84^{+0.36}_{-0.30}$ & $10.75^{+0.18}_{-0.21}$   & $-2.39^{+0.37}_{-0.37}$ & 2& $-2.15^{+0.14}_{-0.14}$ & -2.26                             \\
12453 & 53.280001 & -27.858265 & $9.88^{+0.78}_{-0.54}$  & $9.82^{+0.42}_{-0.42}$    & $-1.10^{+1.65}_{-1.81}$ & 2& $-2.13^{+0.24}_{-0.18}$ & -2.27                             \\
13290 & 53.258417 & -27.861651 & $8.95^{+1.59}_{-0.18}$  & $9.10^{+0.63}_{-0.30}$    & $-1.58^{+0.38}_{-0.38}$ & 3& $-1.30^{+0.27}_{-0.24}$ & -1.48                             \\
13406 & 53.240619 & -27.862122 & $9.31^{+1.05}_{-0.15}$  & $9.52^{+0.51}_{-0.33}$    & $-2.16^{+0.80}_{-0.81}$ & 2& $-2.32^{+0.15}_{-0.16}$ & -2.46                             \\
13782 & 53.244589 & -27.863587 & $9.22^{+1.53}_{-0.18}$  & $9.13^{+0.51}_{-0.30}$    & $-3.26^{+0.78}_{-0.78}$ & 2& $-1.88^{+0.23}_{-0.22}$ & -1.98                             \\
17672 & 53.249058 & -27.815575 & $9.40^{+0.03}_{-0.09}$  & $9.32^{+0.04}_{-0.04}$    & $-2.45^{+0.10}_{-0.10}$ & 2& $-2.14^{+0.06}_{-0.05}$ & -2.23                             \\
17674 & 53.233517 & -27.816677 & $8.56^{+0.15}_{-0.15}$  & $8.56^{+0.09}_{-0.09}$    & $-2.62^{+0.10}_{-0.10}$ & 3& $-2.54^{+0.06}_{-0.06}$ & -2.67                             \\

\hline
\hline

\end{tabularx}
\tablecomments{
\textit{a} = Redshift determined from \textsc{EAZY} P(z)
\\
\textit{b} = Redshift determined from \textsc{Bagpipes} with the \textsc{EAZY} redshift prior applied
\\
\textit{c} = Photometric power-law fit of the UV spectral slope as determined with \textsc{Emcee}
\\
\textit{d} = Adjusted \bsed \ values with correction applied as discussed in Section~\ref{res_sims}
\\}

\vspace{5mm}
\end{table*}

% ------------- Table 2 ----------- %

% ----------------------------------------------------------%

\subsection{Verifying our methodology with simulations}\label{res_sims}
\begin{figure*}
    \centering
    \includegraphics[width=\textwidth]{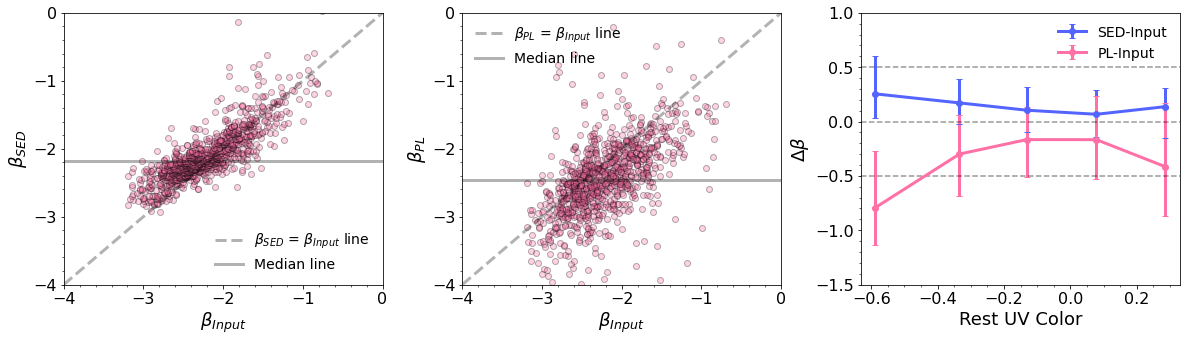}

    \caption{The results of our source-injection simulations.  The first two panels show the input value of $\beta$ versus the recovered SED-fitted (left) and photometric power-law (middle) $\beta$.  The SED method has a tighter scatter, though exhibits a clear bias to redder measured values at true values of $\beta \lesssim -$2. The photometric power-law method shows a significantly larger scatter, as well as a larger bias (in the blue direction) at all input values of $\beta$.  We quantify these biases in the right panel, showing the difference in $\beta$ as a function of the rest-UV color for our simulated sources. These colors are chosen as they can be measured for real galaxies (where the intrinsic $\beta$ is not known). We show that the SED-fitting method results in a more accurate recovery of $\beta$ than photometric power-law fitting. However, as there is still a significant bias, we use this measured $\Delta \beta$ value as a correction factor applied to  \bsed \ for the observed galaxies in our sample (see Figure~\ref{fig:bvb}). As the scatter and bias are both worse for \bpl, and we use \bsed\ for our analysis below, we do not apply these corrections to \bpl. 
    }
    \vspace{5mm}
    \label{fig:sims}
\end{figure*}

To test the accuracy and assess any bias in our SED and photometric power-law measurements of the UV spectral slope, we utilize source injection simulations.  Full details of this process are described in \citet{leung2023}, but we summarize briefly here. The simulated sources are created with a range of F277W magnitudes, colors, and surface brightness profiles. Once added to a given image, photometry and resulting photometric redshifts are defined in the same manner as what is done on real images. Utilizing the same selection criteria as the observed galaxies, we randomly draw a sample of 1000 simulated galaxies that are defined to have `true' input UV spectral slopes, stellar mass, and UV magnitudes. We ensure these galaxies span a wide enough parameter space for each property to fully align with the dynamic range of the observations. With this sample, we run \textsc{Bagpipes} and \textsc{Emcee} in the same fashion as the observed sample in order to obtain $\beta$, UV magnitude, and stellar mass.  We then explore whether there exist any measurement biases by taking the difference between \bsed, \bpl, and the input $\beta$ as functions of input $\beta$.  As shown in the left two panels of Figure~\ref{fig:sims}, we can see that SED-fitting yields a closer fit to the `true' UV slope than with photometric power-law fitting, where anything past $\Delta\beta \pm 0.5$ is to be considered a strong outlier (for SED-fitting, this fraction is 75 out of 1000 galaxies and for photometric power-law fitting, this fraction is 274 out of 1000 galaxies). In particular, the recovered photometric power-law $\beta$ values preferentially scatter towards bluer slopes.

Although our SED-fitting method yields tighter agreement between the input and recovered UV spectral slope, we still hit a ``minimum $\beta$" plateau in what is returned from \textsc{Bagpipes}, as evidenced by the good agreement at $\beta > -2$, which begins to flatten at bluer input colors. 
Photometric power-law fitting, with its larger spread, essentially shows that more filters are needed past $2-3$ at these high redshifts in order to tighten measurements. This, however, would require a series of medium-band filters in the rest-frame UV regime that could probe just as deep as the current filters being used, which would be very costly observationally.  

% ------------- Figure 3 ----------- %

% ------------- Figure 3 ----------- %

Knowing that SED-fitting and photometric power-law fitting have limitations in their ability to precisely measure the UV spectral slope, we use these simulations to derive a bias-correction factor. As we cannot know the true value of $\beta$ for real sources, we derive this correction as a function of observed rest-UV color.  For each source, both simulated and observed, we measure a rest UV color by taking the difference, in magnitude, of the first two filters redward of the filter where the Lyman-break is expected to lie.

To adjust for discrepancies in measured versus true UV slopes, we calculate a correction factor, $\Delta \beta$, which is the difference between measured UV slopes (from SED or photometric fitting) and known input values. We calculate this quantity in UV color bins of $0.25$, using only those with over 20 data points. To apply this correction to a given real object, we interpolate the $\Delta \beta$ curves to a given observed UV color value, assigning edge values to outliers. As illustrated in Figure~\ref{fig:bvb}, applying this correction even just to \bsed\ aligns original and adjusted \bsed\ values closer to \bpl\ across various redshifts (detailed in Table~\ref{tab:beta_slope}). The Pearson R correlation coefficient analysis shows an improvement from $\rho = 0.38$ to $\rho = 0.48$ post-correction, indicating alleviation of bias in the SED-derived UV slopes, evidenced by a shift towards bluer \bsed\ within the error margins of the sources. These corrected \bsed\ values will be used as the baseline for all subsequent analyses within this study.

We note that this process was similarly done with photometric power-law fitting. However, we obtain an even stronger bias, given the small number of data points being fit. As such, we do not apply these corrections to the photometric power-law fitted values, though we do show the derived correction values for both methods in the right-hand panel of Figure~\ref{fig:sims}.

% ----------------------------------------------------------

\subsection{Analysis of \bsed \ and other Bagpipes galaxy parameters}
\label{res_params}

\begin{figure*}[t!]
    \centering
    \includegraphics[width=0.9\textwidth]{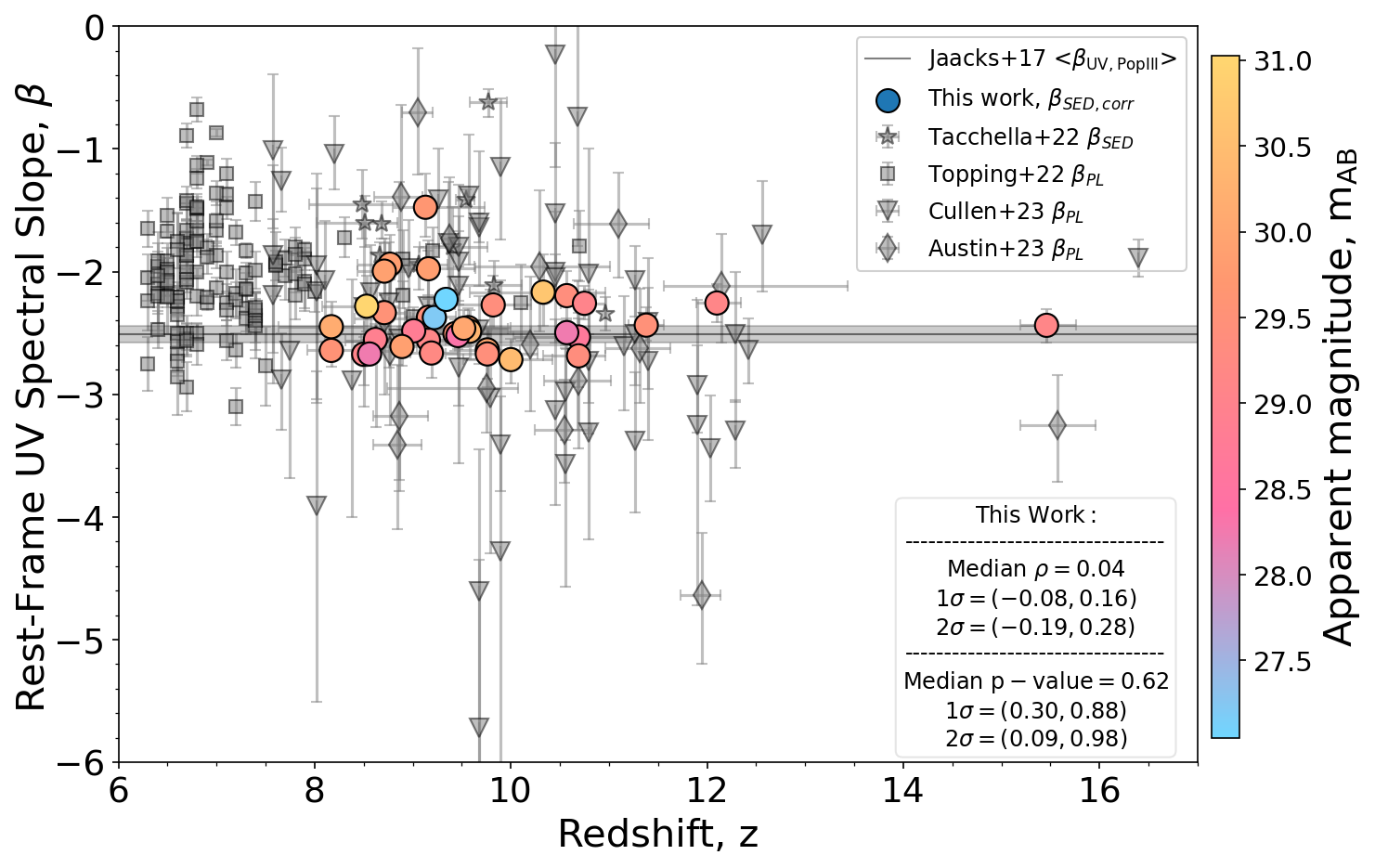}

    \caption{UV spectral slope, $\beta$, as a function of redshift (color-coded by rest-UV apparent magnitude, $\mathrm{m_{AB}}$, defined at $\lambda_\mathrm{rest} = 1500 \AA$) from this work are shown as circles.  Squares are from \cite{Topping2022}, stars are from \cite{Tacchella2022}, and diamonds are from \cite{Austin2023}. The black line and shaded region at $\beta_\mathrm{SED} = -2.51 \pm 0.07$ is the baseline spectral slope for halos enriched by only Population III stars from \cite{Jaacks2018}. This work is consistent with the works of others, however, no clear trends are shown with UV spectral slope as a function of redshift when all data is combined. The corresponding Spearman R correlation coefficient is listed for this work only in the lower right. Within our sample, we see no evidence for ultra-blue UV spectral slopes.}

    \label{fig:bvz}
\end{figure*}

\begin{figure*}[ht!]
    \centering
    \includegraphics[width=\textwidth]{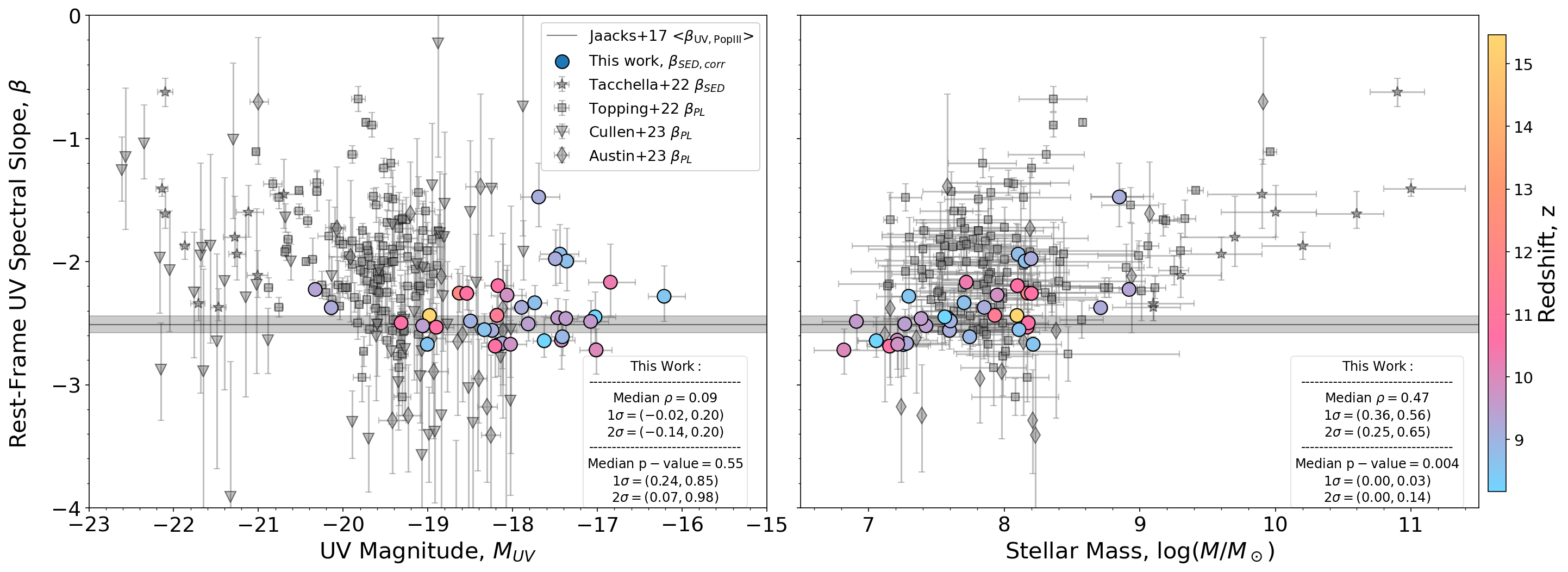}

    \caption{UV $\beta$ slope as a function of UV magnitude and stellar mass (color-coded by redshift) from this work are shown as circles; all other symbols and the shaded line are the same as Figure~\ref{fig:bvz}.  Where our sample overlaps with previous works, our results are consistent, albeit with lower scatter. With our sample alone, we see a moderately strong correlation with stellar mass, but a negligible correlation with UV magnitude. When combining our sample with previous work, we see a weak trend between the UV spectral slope and UV magnitude, and a somewhat stronger trend with stellar mass, where brighter, more massive galaxies exhibit redder UV colors than faint, low-mass counterparts. The corresponding Spearman R correlation coefficients are listed for this work only on the lower right of each panel.}
    \vspace{5mm}
    \label{fig:params1}
\end{figure*}

\begin{figure*}[ht!]
    \centering
    \includegraphics[width=\textwidth]{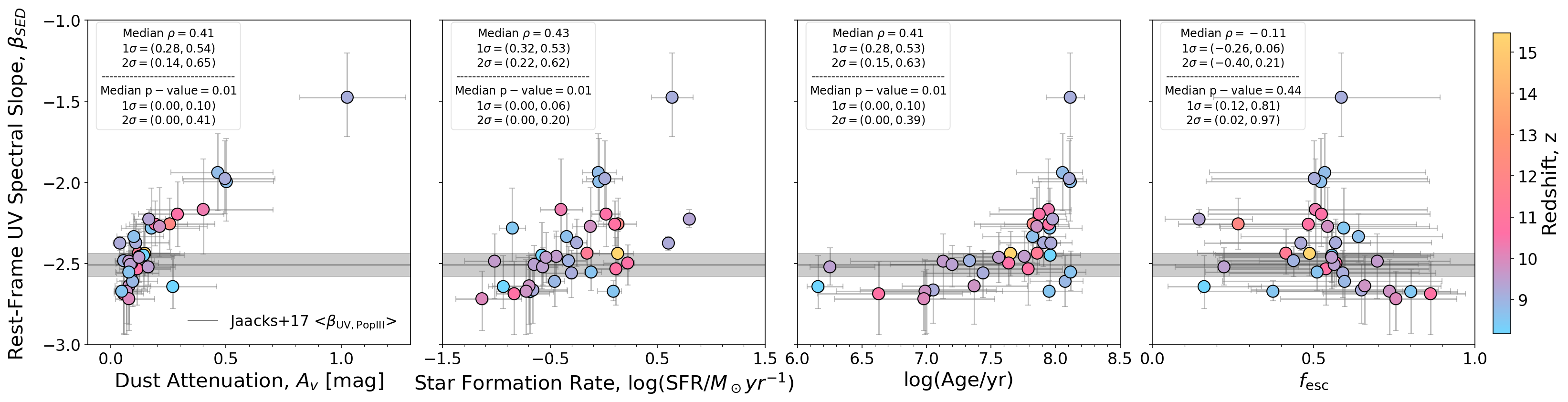}

    \caption{UV $\beta$ slope as a function of dust attenuation, star formation rate, age, and escape fraction (color-coded by redshift). The black line and shaded region at $\beta_\mathrm{SED} = -2.51 \pm 0.07$ is the baseline spectral slope for halos enriched by only Population III stars from \cite{Jaacks2018}. SFR, dust attenuation, and age show a moderately positive correlation with $\beta$. While, finally, $f_\mathrm{esc}$  exhibit a negligible correlation with $\beta$. The corresponding Spearman R correlation coefficients are listed for each parameter in the upper left of each panel.}
    \vspace{5mm}
    \label{fig:params}
\end{figure*}

We compare the results of the \textsc{Bagpipes} posteriors for galaxy parameters versus the bias-corrected values of \bsed, and discuss any correlations. Previous works have explored correlations mainly between $\beta$ and stellar mass \citep{Finkelstein2010,Finkelstein2012}, UV absolute magnitude \citep{Bouwens2010,Bouwens2012}, and dust attenuation \citep{Calzetti1994,Meurer1999}. Here, we explore monotonic trends with these parameters as well as with star formation rate (SFR), mass-weighted age, and $f_\mathrm{esc}$ using Spearman R correlation coefficients\footnote{We define correlation strengths as the following: (1) Negligible $= 0.00 < |\rho| < 0.20 $, (2) Weak $= 0.21 < |\rho| < 0.40 $, (3) Moderate $= 0.41 < |\rho| < 0.60 $, (4) Strong $= 0.61 < |\rho| < 0.80 $, and (5) Very strong $= 0.81 < |\rho| < 1.00 $. We also note statistical significance in the correlations as defined with p-value, where: (1) Significant = p-value $\leq 0.05$ and (2) Non-significant = p-value $> 0.05$}, $\rho$ \citep{Spearman}. We perform a Monte Carlo resampling to estimate the Spearman correlation coefficient and the corresponding p-value between the corrected \bsed \  and other galaxy parameters, accounting for their uncertainties. We repeatedly add normally distributed random noise to the data and recalculate the $\rho$ and p-value. The median $\rho$ and p-value, along with their 68\% and 95\% confidence intervals, are noted in the legends of Figures~\ref{fig:bvz}, \ref{fig:params1}, and \ref{fig:params}.  We offer the median and difference from the 68th confidence bounds for each parameter in Table~\ref{tab:BP_params}. For Figures~\ref{fig:bvz} and \ref{fig:params1}, we plot our corrected \bsed \ values versus redshift, UV magnitude, and stellar mass and offer comparisons with other works that have been done at high redshift \citep{Tacchella2022,Topping2022, Cullen2023, Austin2023}.

% ------------- Figure 4 ----------- %

% ------------- Figure 4 ----------- %

Here, we examine the strength of the monotonic correlations between the UV spectral slope, $\beta$, and various galactic parameters as presented in Figures~\ref{fig:bvz} to \ref{fig:params}, ranked from the strongest to the weakest correlations. The relationship between $\beta$ and stellar mass exhibits the most pronounced positive monotonic correlation, with a median Spearman correlation coefficient, $\rho = 0.47$, and a $1\sigma$ confidence interval ranging from 0.36 to 0.56. These values suggest a moderate-to-weak correlation strength. Similarly, star formation rate and mass-weighted age both demonstrate moderate-to-weak positive correlations with $\beta$, with median $\rho$ values of 0.43 ($1\sigma$ CI $= [0.32-0.53]$) and 0.41 ($1\sigma$ CI $= [0.28-0.53]$), respectively; this suggests that older and more actively star-forming galaxies tend to have redder UV colors. Dust attenuation, $A_\mathrm{v}$, also shows a moderate-to-weak relationship with the UV spectral slope (median $\rho$ of 0.41 ($1\sigma$ CI $= [0.28-0.54]$)) where larger amounts of dust attenuation along the line of sight lead to redder UV slopes \citep[and vice versa,][]{Calzetti1994,Meurer1999}. Finally, redshift, UV magnitude, and the Lyman-continuum escape fraction each show negligible correlations with the UV spectral slope, where each median $|\rho| < 0.12$ ($1\sigma$ CI $= [-0.08 - 0.16], [-0.02 - 0.20], \mathrm{and} [-0.26 - 0.06]$, respectively). While combining diverse datasets can potentially reveal more significant trends in $\beta$, variations in sample selection and analytical methods introduce potential systematic uncertainties that are challenging to quantify. We also abstain from calculating a combined Spearman correlation coefficient across all studies due to the absence of uncertainties in some data, which restricts the application of Monte Carlo resampling techniques defined above.

% 
% ------------- Figure 5 ----------- %

% ------------- Figure 5 ----------- %

% ------------- Figure 6 ----------- %

% ------------- Figure 6 ----------- %

% ------------- Table 3 ----------- %

\begin{table*}[ht!]

\caption{\textsc{Bagpipes} median and $1\sigma$ posteriors for galaxy parameters.}
\label{tab:BP_params}
\centering

\begin{tabularx}{0.88\linewidth}{cccccccc}
\hline\hline
ID    & $\mathrm{z}_\mathrm{SED}$ & $M_\mathrm{UV}$      & $\mathrm{log}(M_*/M_\odot)$ & $A_\mathrm{v}$ [mag] & log(Age/yr)    & log(SFR/$M_\odot \mathrm{yr}^{-1}$) & $f_\mathrm{esc}$    \\
\hline
250   & $11.38^{+0.18}_{-0.21}$   & $-18.18^{+0.1}_{-0.09}$  & $7.93^{+0.16}_{-0.33}$    & $0.12^{+0.14}_{-0.81}$   & $7.86^{+0.20}_{-0.42}$ & $-0.16^{+0.12}_{-0.14}$                & $0.41^{+0.33}_{-0.27}$ \\
1191  & $12.10^{+0.21}_{-0.21}$   & $-18.63^{+0.11}_{-0.09}$ & $8.17^{+0.11}_{-0.33}$    & $0.26^{+0.20}_{-0.14}$   & $7.83^{+0.20}_{-0.46}$ & $0.13^{+0.17}_{-0.18}$                 & $0.27^{+0.31}_{-0.19}$ \\
1369  & $15.46^{+0.30}_{-0.27}$   & $-18.98^{+0.08}_{-0.08}$ & $8.09^{+0.19}_{-0.31}$    & $0.15^{+0.15}_{-0.11}$   & $7.65^{+0.22}_{-0.48}$ & $0.13^{+0.15}_{-0.30}$                 & $0.49^{+0.35}_{-0.30}$ \\
1716  & $10.27^{+0.69}_{-0.75}$   & $-16.85^{+0.29}_{-0.26}$ & $7.72^{+0.21}_{-0.26}$    & $0.40^{+0.30}_{-0.23}$   & $7.94^{+0.18}_{-0.38}$ & $-0.40^{+0.20}_{-0.19}$                & $0.51^{+0.35}_{-0.33}$ \\
2067  & $10.57^{+0.36}_{-0.36}$   & $-18.17^{+0.18}_{-0.16}$ & $8.10^{+0.16}_{-0.23}$    & $0.29^{+0.20}_{-0.17}$   & $7.87^{+0.21}_{-0.38}$ & $0.02^{+0.13}_{-0.13}$                 & $0.52^{+0.34}_{-0.35}$ \\
2470  & $8.17^{+0.24}_{-0.27}$    & $-17.63^{+0.16}_{-0.13}$ & $7.06^{+0.20}_{-0.13}$    & $0.27^{+0.19}_{-0.15}$   & $6.16^{+0.19}_{-0.08}$ & $-0.94^{+0.21}_{-0.14}$                & $0.16^{+0.18}_{-0.11}$ \\
2497  & $9.16^{+0.18}_{-0.18}$    & $-18.24^{+0.11}_{-0.11}$ & $7.60^{+0.24}_{-0.22}$    & $0.09^{+0.12}_{-0.06}$   & $7.44^{+0.34}_{-0.44}$ & $-0.30^{+0.13}_{-0.27}$                & $0.59^{+0.29}_{-0.33}$ \\
3514  & $10.72^{+0.45}_{-0.60}$   & $-18.90^{+0.21}_{-0.18}$ & $8.17^{+0.20}_{-0.33}$    & $0.11^{+0.15}_{-0.08}$   & $7.79^{+0.27}_{-0.55}$ & $0.11^{+0.13}_{-0.23}$                 & $0.54^{+0.32}_{-0.34}$ \\
4134  & $10.57 ^{+0.15}_{-0.15}$  & $-19.32^{+0.08}_{-0.09}$ & $8.18^{+0.15}_{-0.22}$    & $0.09^{+0.12}_{-0.07}$   & $7.64^{+0.24}_{-0.42}$ & $0.22^{+0.06}_{-0.22}$                 & $0.57^{+0.29}_{-0.33}$ \\
4330  & $9.13^{+0.18}_{-0.15}$    & $-17.90^{+0.12}_{-0.12}$ & $7.85^{+0.14}_{-0.22}$    & $0.11^{+0.14}_{-0.08}$   & $7.91^{+0.22}_{-0.34}$ & $-0.25^{+0.09}_{-0.07}$                & $0.57^{+0.30}_{-0.36}$ \\
4674  & $8.50^{+0.15}_{-0.12}$    & $-18.17^{+0.11}_{-0.13}$ & $7.26^{+0.20}_{-0.18}$    & $0.06^{+0.08}_{-0.04}$   & $7.00^{+0.34}_{-0.42}$ & $-0.69^{+0.22}_{-0.21}$                & $0.80^{+0.15}_{-0.28}$ \\
4740  & $9.19^{+0.24}_{-0.18}$    & $-18.12^{+0.15}_{-0.13}$ & $7.28^{+0.29}_{-0.22}$    & $0.07^{+0.09}_{-0.05}$   & $7.05^{+0.49}_{-0.56}$ & $-0.66^{+0.32}_{-0.26}$                & $0.65^{+0.24}_{-0.34}$ \\
4919  & $9.58^{+0.45}_{-0.39}$    & $-17.47^{+0.15}_{-0.14}$ & $7.58^{+0.20}_{-0.29}$    & $0.12^{+0.15}_{-0.09}$   & $7.76^{+0.29}_{-0.47}$ & $-0.44^{+0.11}_{-0.17}$                & $0.56^{+0.31}_{-0.34}$ \\
5118  & $9.22^{+0.03}_{-0.03}$    & $-20.14^{+0.03}_{-0.03}$ & $8.71^{+0.05}_{-0.08}$    & $0.04^{+0.06}_{-0.03}$   & $7.96^{+0.08}_{-0.15}$ & $0.60^{+0.03}_{-0.02}$                 & $0.46^{+0.22}_{-0.30}$ \\
5947  & $8.80^{+0.33}_{-0.36}$    & $-17.44^{+0.22}_{-0.23}$ & $8.10^{+0.16}_{-0.16}$    & $0.46^{+0.24}_{-0.20}$   & $8.05^{+0.16}_{-0.35}$ & $-0.05^{+0.17}_{-0.15}$                & $0.53^{+0.31}_{-0.36}$ \\
6134  & $9.01^{+0.06}_{-0.09}$    & $-18.50^{+0.06}_{-0.06}$ & $7.60^{+0.14}_{-0.20}$    & $0.06^{+0.08}_{-0.04}$   & $7.33^{+0.25}_{-0.37}$ & $-0.33^{+0.17}_{-0.22}$                & $0.44^{+0.22}_{-0.23}$ \\
6477  & $8.71^{+0.24}_{-0.21}$    & $-17.74^{+0.15}_{-0.13}$ & $7.70^{+0.19}_{-0.23}$    & $0.10^{+0.13}_{-0.07}$   & $7.82^{+0.26}_{-0.39}$ & $-0.35^{+0.08}_{-0.11}$                & $0.64^{+0.26}_{-0.36}$ \\
6952  & $8.53^{+0.36}_{-0.42}$    & $-16.21^{+0.25}_{-0.18}$ & $7.30^{+0.16}_{-0.24}$    & $0.18^{+0.21}_{-0.13}$   & $7.95^{+0.25}_{-0.38}$ & $-0.85^{+0.12}_{-0.12}$                & $0.59^{+0.29}_{-0.38}$ \\
6980  & $9.76^{+0.33}_{-0.30}$    & $-17.42^{+0.16}_{-0.14}$ & $7.21^{+0.30}_{-0.29}$    & $0.08^{+0.10}_{-0.06}$   & $7.37^{+0.42}_{-0.56}$ & $-0.70^{+0.21}_{-0.34}$                & $0.66^{+0.24}_{-0.36}$ \\
7530  & $8.71^{+0.30}_{-0.30}$    & $-17.36^{+0.22}_{-0.17}$ & $8.15^{+0.14}_{-0.19}$    & $0.50^{+0.21}_{-0.18}$   & $8.11^{+0.12}_{-0.29}$ & $-0.05^{+0.15}_{-0.15}$                & $0.52^{+0.34}_{-0.36}$ \\
7722  & $10.66^{+0.30}_{-0.36}$   & $-18.21^{+0.15}_{-0.16}$ & $7.15^{+0.29}_{-0.16}$    & $0.06^{+0.09}_{-0.04}$   & $6.63^{+0.65}_{-0.47}$ & $-0.84^{+0.33}_{-0.17}$                & $0.86^{+0.11}_{-0.33}$ \\
8042  & $8.17^{+0.36}_{-0.51}$    & $-17.03^{+0.24}_{-0.20}$ & $7.56^{+0.17}_{-0.25}$    & $0.14^{+0.18}_{-0.10}$   & $7.96^{+0.24}_{-0.43}$ & $-0.58^{+0.12}_{-0.12}$                & $0.56^{+0.32}_{-0.36}$ \\
8165  & $9.82^{+0.48}_{-0.48}$    & $-18.03^{+0.19}_{-0.19}$ & $7.22^{+0.38}_{-0.23}$    & $0.07^{+0.10}_{-0.05}$   & $6.99^{+0.60}_{-0.69}$ & $-0.72^{+0.39}_{-0.27}$                & $0.74^{+0.19}_{-0.41}$ \\
8427  & $8.89^{+0.12}_{-0.18}$    & $-17.41^{+0.12}_{-0.13}$ & $7.75^{+0.12}_{-0.19}$    & $0.10^{+0.12}_{-0.07}$   & $8.07^{+0.14}_{-0.27}$ & $-0.46^{+0.09}_{-0.07}$                & $0.60^{+0.30}_{-0.38}$ \\
8461  & $9.53^{+0.64}_{-0.43}$    & $-17.08^{+0.21}_{-0.20}$ & $6.91^{+0.33}_{-0.25}$    & $0.08^{+0.11}_{-0.06}$   & $7.13^{+0.50}_{-0.72}$ & $-1.02^{+0.33}_{-0.28}$                & $0.70^{+0.23}_{-0.37}$ \\
8894  & $9.43^{+0.39}_{-0.18}$    & $-17.82^{+0.10}_{-0.10}$ & $7.28^{+0.29}_{-0.23}$    & $0.08^{+0.10}_{-0.06}$   & $7.20^{+0.39}_{-0.48}$ & $-0.65^{+0.26}_{-0.28}$                & $0.56^{+0.26}_{-0.34}$ \\
9261  & $8.62^{+0.15}_{-0.15}$    & $-18.33^{+0.12}_{-0.11}$ & $8.11^{+0.12}_{-0.19}$    & $0.08^{+0.10}_{-0.06}$   & $8.12^{+0.12}_{-0.27}$ & $-0.12^{+0.10}_{-0.06}$                & $0.51^{+0.32}_{-0.35}$ \\
9555  & $10.00^{+0.45}_{-0.51}$   & $-17.02^{+0.19}_{-0.19}$ & $6.82^{+0.35}_{-0.22}$    & $0.08^{+0.11}_{-0.06}$   & $6.98^{+0.55}_{-0.69}$ & $-1.14^{+0.37}_{-0.24}$                & $0.76^{+0.19}_{-0.36}$ \\
10296 & $9.46^{+0.09}_{-0.03}$    & $-19.06^{+0.07}_{-0.07}$ & $7.42^{+0.15}_{-0.11}$    & $0.16^{+0.12}_{-0.10}$   & $6.25^{+0.35}_{-0.15}$ & $-0.57^{+0.14}_{-0.10}$                & $0.22^{+0.23}_{-0.15}$ \\
11522 & $10.75^{+0.18}_{-0.21}$   & $-18.54^{+0.11}_{-0.10}$ & $8.20^{+0.13}_{-0.21}$    & $0.19^{+0.14}_{-0.11}$   & $7.95^{+0.15}_{-0.34}$ & $-0.10^{+0.09}_{-0.10}$                & $0.48^{+0.34}_{-0.34}$ \\
12453 & $9.82^{+0.42}_{-0.42}$    & $-18.07^{+0.23}_{-0.19}$ & $7.95^{+0.21}_{-0.29}$    & $0.21^{+0.21}_{-0.15}$   & $7.86^{+0.25}_{-0.44}$ & $-0.13^{+0.15}_{-0.16}$                & $0.54^{+0.32}_{-0.34}$ \\
13290 & $9.10^{+0.63}_{-0.30}$    & $-17.69^{+0.25}_{-0.21}$ & $8.85^{+0.18}_{-0.21}$    & $1.03^{+0.25}_{-0.21}$   & $8.11^{+0.11}_{-0.19}$ & $0.63^{+0.20}_{-0.19}$                 & $0.59^{+0.31}_{-0.40}$ \\
13406 & $9.52^{+0.51}_{-0.33}$    & $-17.37^{+0.14}_{-0.14}$ & $7.37^{+0.25}_{-0.28}$    & $0.12^{+0.14}_{-0.09}$   & $7.56^{+0.39}_{-0.47}$ & $-0.54^{+0.12}_{-0.28}$                & $0.56^{+0.30}_{-0.35}$ \\
13782 & $9.13^{+0.51}_{-0.30}$    & $-17.50^{+0.22}_{-0.19}$ & $8.20^{+0.14}_{-0.16}$    & $0.50^{+0.22}_{-0.19}$   & $8.11^{+0.11}_{-0.24}$ & $0.01^{+0.16}_{-0.17}$                 & $0.50^{+0.35}_{-0.32}$ \\
17672 & $9.32^{+0.04}_{-0.04}$    & $-20.33^{+0.03}_{-0.03}$ & $8.92^{+0.05}_{-0.07}$    & $0.17^{+0.07}_{-0.06}$   & $7.98^{+0.11}_{-0.15}$ & $0.80^{+0.05}_{-0.04}$                 & $0.15^{+0.17}_{-0.11}$ \\
17674 & $8.56^{+0.09}_{-0.09}$    & $-19.01^{+0.05}_{-0.05}$ & $8.21^{+0.21}_{-0.35}$    & $0.05^{+0.07}_{-0.03}$   & $7.95^{+0.26}_{-0.48}$ & $0.09^{+0.08}_{-0.14}$                 & $0.37^{+0.40}_{-0.23}$ \\
\hline\hline
\end{tabularx}

% \end{center}
\vspace{5mm}

\end{table*}

% ------------- Table 3 ----------- %

% =========================================================================%

\section{Discussion}\label{sec:disc}
Here we discuss uncertainties that affect our SED modeling (Section~\ref{disc_caveats}), a comparison with previous work at this redshift range (Section~\ref{disc_comp}), a comparison with model predictions (Section~\ref{disc_preds}), and a discussion on the implications of our results (Section~\ref{disc_implications}).

% ----------------------------------------------------------

\subsection{Modeling caveats}\label{disc_caveats}
When utilizing \textsc{Bagpipes} to build our model SEDs, we set several priors that allow a large range of star-forming galaxy SED models to be fit. While testing to see how blue our models could go with the set list of priors, stellar grids, and dust laws, we were able to reach a blue floor of $\beta_\mathrm{SED} = -3.25$. Both our sample of 36 galaxies and the simulated 1000 galaxies were shown to reach an artificial plateau at $\beta_\mathrm{SED} \sim -2.8$ regardless of this floor (see Figures~\ref{fig:bvb} and \ref{fig:sims}). Though we attempt to correct for this bias, future work can improve upon our methodology by adopting bluer stellar grid models, such as the Yggdrasil stellar models \citep{Yggdrasil} that vary in initial mass function (IMF), metallicity, and star formation history. 

The works of \cite{Bouwens2010}, \cite{Rogers2013}, and \cite{Dunlop2013} discuss the possibility of bias in the UV spectral slope, mainly for fainter galaxies at higher redshifts. This bias is possible because fainter red galaxies may not satisfy sample selection criteria, while fainter blue galaxies would, fostering a blue bias, particularly for UV-faint galaxies.

We also conducted tests to evaluate the impact of introducing a free $f_\mathrm{{esc}}$ parameter, with a flat prior between zero and one, to allow SEDs without a strong nebular continuum. For our sample, we run \textsc{Bagpipes} with and without $f_\mathrm{{esc}}$ varying, and measure $\beta$ in both instances. We find that fixing $f_{\mathrm{esc}} = 0$ versus allowing it to be a free parameter imparts a minimal change in the recovered UV spectral slope.  We quantify this via a nebular reddening estimate, $\Delta\beta_{\mathrm{neb}} = \beta_{\mathrm{SED, corr, f_{esc}=0}} - \beta_{\mathrm{SED,corr}} = -0.01 \pm 0.1$. This indicates that nebular emission does not significantly redden the slope or alter the overall SED shape for this sample. We acknowledge that our reported $f_{\mathrm{esc}}$ values, even when this parameter is incorporated, are not tightly constrained and remain highly uncertain. Furthermore, although this incorporation allows $\beta$ values to become bluer than when not considered, as previously mentioned, we are still limited by how blue the modeling can go with and without the prior added.

% ----------------------------------------------------------

\subsection{Comparison with previous work}\label{disc_comp}

In this work, we directly compare our results to the most recent works within the same redshift range as mentioned in Section~\ref{sec:datasets}.  We first compare the results for $\beta$, \muv, and stellar mass for our sample with that of \cite{Austin2023} given their overlapping scope with respect to the field and redshift range of observations. We find that their median $M_\mathrm{{UV}} \sim -18.9$ is slightly brighter than our sample with $M_\mathrm{{UV}} \sim -18.1$ and our median stellar masses are comparable at $\sim 10^8 \mathrm{M_\odot}$. Any differences in \muv \ could be due to their photometric data reduction (the \citealt{leung2023} reduction is significantly deeper) and/or the subsequent SED fitting to the observational data (for this analysis, we quote their values for their fits with \textsc{LePhare}, see their Table 2). Their average $\beta \sim -2.61$ is bluer than ours and spans a wider range of values than our work, at $\beta_\mathrm{SED} \sim -2.45$, but this may be a result of their use of photometric power-law fitting as their primary approach to measuring the UV spectral slope. 

The work of \cite{Topping2022} studied 123 galaxies in the \textit{JWST} CEERS field at $z=7-11$. These galaxies, utilizing CEERS data, do not reach depths comparable to NGDEEP, and as such, the average luminosities of these galaxies are slightly brighter than those of our sample. Here, the median UV magnitude for their sample is $M_\mathrm{{UV}} \sim -19.6$, corresponding to slightly redder UV slopes averaging $\beta \sim -2.0$ (measured by photometric power-law fitting). \cite{Cullen2023} use data from $z=8-16$ with \textit{JWST} SMACS J0723, GLASS, and CEERS surveys along with the COSMOS/UltraVISTA survey. Their work does not note any stellar mass values for their sample of galaxies. However, the diversified survey inputs indicate an average UV magnitude marginally brighter than our data set, at $M_\mathrm{{UV}} \sim -19.3$ and a resulting median UV slope of $\beta \sim -2.3$ (also measured by photometric power-law fitting; SED-fitting done with \textsc{EAZY} to obtain \muv).

The work of \cite{Tacchella2022} aimed primarily to study high-mass galaxies in the \textit{HST} CANDELS fields at $z>9$, at $\sim 10^9 - 10^{10} \mathrm{M_\odot}$. These galaxies were also primarily brighter and redder than our sample at $M_\mathrm{{UV}} \sim -21.5$ with a median $\beta \sim -1.8$ when measured via SED-fitting, and others we compared to, but offered a bright and high mass-end to any trends in the UV spectral slope with \muv \ and stellar mass. With this addition to the study, we could primarily note that our lower mass galaxies have significantly bluer UV slopes than their work. 

\cite{Bouwens2012} found a luminosity dependence and thus a UV magnitude relationship with the UV spectral slope from $z=4-7$. Specifically, they stated that broad \muv \ ranges ($-21.5 < M_\mathrm{UV} < -16.5$ for their work) would be needed to quantify a true relationship between $\beta$ and UV magnitude. It is possible that these trends could depend on where \muv \ is defined. As shown \cite{Bouwens2012}, if one defines \muv \ at $\lambda_\mathrm{rest} \sim 1500 \AA$, there is, on average, no correlation between \muv \ and $\beta$. However, if one defines this at slightly redder wavelengths, i.e. $\lambda_\mathrm{rest} \sim 2000 \AA$, there is evidence for a strong negative correlation as $\beta$ gets bluer. The works we compare this analysis to have also defined \muv \ at the rest-wavelength $1500 \AA$ regime, and as such, we have yet to see, at high redshift, strong trends in UV magnitude and the UV spectral slope.

Although we expect $\beta$ to decrease as we move earlier in cosmic history, where on average, for the whole population, younger, smaller galaxies with just a few generations of stellar populations existed, the weak trends with redshift we see here could be due to the sample size for each redshift bin we have in our population.  When we only look at our sample, there is no significant correlation between $\beta$ and redshift, while one emerges when combined with other works. Future analyses with larger sample sizes that span both UV-bright and faint galaxies and wider ranges of redshift will be more capable of probing any evolution of the UV spectral slope, as well as minimizing the effects of cosmic variance. We also note that previous works, e.g., \cite{Topping2022,Austin2023,Cullen2023}, estimate the UV spectral slope by applying a power-law fit to the photometry. While this is a reasonable approach, as we have discussed, this method can be less reliable when photometry is limited. While our preferred SED method has reduced scatter, it is not free of its own biases.  Improvement can be made via higher spectral resolution data (e.g., medium-band imaging or prism spectroscopy) with \textit{JWST} in the rest-frame UV-regime of these galaxy SEDs would need to be provided.

% ----------------------------------------------------------
\subsection{Comparison to model predictions }\label{disc_preds}

Here, we compare the results of our observations and the other observational datasets utilized for this work alongside three cosmological simulations: the Santa Cruz semi-analytic model (\textsc{SC-SAM}) \citep{Yung2023}, the First Light And Reionisation Epoch Simulations \citep[\textsc{FLARES},][]{FLARES-I, FLARES-II, FLARES-V}, and the \textsc{SIMBA-EoR} simulations \citep[][Jones et al. in prep.]{Dave2019}.

We match the results from our observations with those found in other studies, using the predictions given by the Santa Cruz semi-analytic model \citep[\textsc{SC-SAM}][]{Yung2023} at $z=9$ (compared to galaxies in redshift bin $z=8.5-9.5$) and $z=11$ (compared to galaxies from $z=9.5-12$, as binned in \citealt{leung2023}). These predictions are made with halo merger trees extracted from the suite of dark matter-only simulations \textsc{gureft}, which provide robust halo merger histories with 170 snapshots stored between $40 \gtrsim z \gtrsim 6$ \citep{Yung2023b}. The \textsc{SC-SAM} tracks the full star formation and chemical enrichment histories of predicted galaxies under the influence of a set of key physical processes (see \citealt{Somerville2015, Yung2019, Yung2022} and references therein) and couples them with the stellar population synthesis (SPS) model Binary Population and Spectral Synthesis \citep[BPASS;][] {Eldridge2017, Byrne2022} to produce high-resolution SEDs for each galaxy. 

We also compare observations to the predictions from the First Light And Reionisation Epoch Simulations \citep[\textsc{FLARES},][]{FLARES-I, FLARES-II, FLARES-V}  at $z=9$. \textsc{FLARES} re-simulates 40 regions selected from a large ($3.2$ Gpc)$^3$ dark matter-only simulation using a variant of \textsc{eagle} \citep{schaye_eagle_2015, crain_eagle_2015} physics model. By simulating a wide range of environments and statistically combining them \textsc{flares} is able to probe a larger dynamic range of galaxies than possible with comparable volume periodic simulations. 

Finally, we compare all observational datasets to the predictions from the \textsc{Simba-EoR} simulations \citep[][Jones et al. in prep.]{Dave2019}. \textsc{Simba-EoR} builds on the \textsc{Simba} model \citep{Dave2019}, adding a sophisticated subgrid model for the interstellar medium that directly tracks $H_2$ and dust co-evolution via a fully coupled chemical network modulated by a local interstellar radiation field, and run at higher resolution.  This results in significantly more early star formation versus \textsc{Simba}, leading to good agreement with the $z>9$ UV luminosity function (UVLF) (\cite{finkelstein2023}).

\begin{figure*}[ht!]
    \centering
    \vspace{5mm}
    \includegraphics[width=\textwidth]{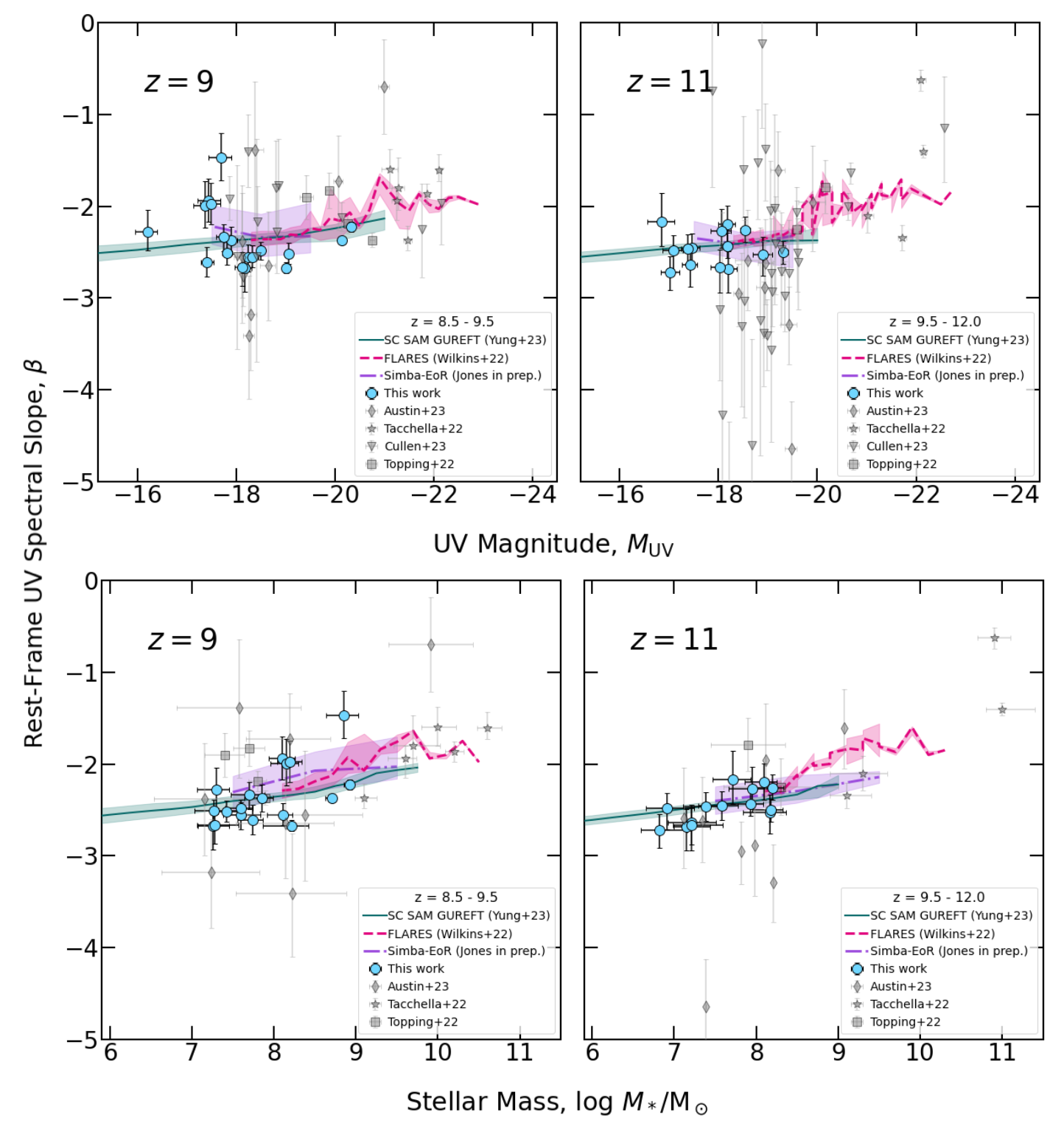}
    \caption{(\textit{Upper panel}) Rest-UV magnitude (\muv) versus rest-frame UV spectral slope ($\beta$) for simulated galaxies at $z=9$ and 11. Our results from SED-fitting (cyan circles) are compared to simulations and observational works from others. The green line marks the median, and the shaded area marks the 16th and 84th percentiles for galaxies in bins of \muv \ galaxies predicted by SC-SAM with merger trees from \textsc{gureft} \citep{Yung2023, Yung2023b}, \textsc{flares}  \citep[pink median track with shaded 16th and 84th percentiles,][]{FLARES-I, FLARES-II, FLARES-V}, \textsc{Simba-EoR} \citep[purple median track with shaded 16th and 84th percentiles,][Jones et al. in prep.]{Dave2019}, \cite{Austin2023} (diamond), \cite{Tacchella2022} (star), \cite{Cullen2023} (triangle), and \cite{Topping2022} (square). (\textit{Lower panel}) Stellar mass, log($M_*/M_\odot$), versus rest-frame UV spectral slope ($\beta$) for observed and simulated galaxies at $z=9, 11$. All data points and tracks are denoted in a similar manner to the upper panel. The green line marks the median, and the shaded area marks the 16th and 84th percentiles for galaxies in bins of $M_*$ from \textsc{gureft}, \textsc{flares}  \citep[pink median track with shaded 16th and 84th percentiles,][]{FLARES-I, FLARES-II, FLARES-V}, \textsc{Simba-EoR} \citep[purple median track with shaded 16th and 84th percentiles,][Jones et al. in prep.]{Dave2019}, this work (cyan circles), \cite{Austin2023} (diamond), \cite{Tacchella2022} (star), and \cite{Topping2022} (square).} 
    \label{fig:gureft}
    \vspace{10mm}
\end{figure*}

For each galaxy in the simulations, we make use of their UV spectral slope, stellar mass, and UV magnitude. In Figure~\ref{fig:gureft}, we compare our work and that of previous works utilized in earlier sections to the \textsc{gureft}, \textsc{flares}, and \textsc{Simba-EoR} models. We find that the measurements of \bsed, \muv, and $M_*$ as derived from \textsc{Bagpipes} for our sample match well with predictions in both redshift bins in comparison to \textsc{gureft} predictions (the brighter, more massive galaxies in our $z=9,11$ samples are also in agreement with the \textsc{flares} and \textsc{Simba-EoR} simulations). Notably, we do not observe any ultra-blue UV spectral slopes ($\beta < -3$), which is in line with these simulations not predicting such extreme values for the galaxy redshifts and masses we examine. Although they are also in agreement with our sample, the \textsc{gureft} and \textsc{flares} simulations do not consider any Population III stellar grids, and our observations do not drive this inclusion for galaxies at the redshift and mass range considered here.  Future works incorporating these Population III stellar grids, such as the Yggdrasil models \citep{Yggdrasil}, will be necessary to explain any potential future observations of $\beta < -3$. 

% ------------- Figure 7 ----------- %

% ------------- Figure 7 ----------- %

% ----------------------------------------------------------

\subsection{Implications}\label{disc_implications}

\textit{JWST}, which, as of now, can detect candidate galaxies out to $z\sim 16$ as shown in this work and others, may provide insights into the early universe and the existence of some of the earliest generations of stars. However, it remains significantly challenging to distinguish between Population III stars and low-metallicity Population II stars due to the absence of distinctive photometric features. While potential signatures might be discovered in future works through spectroscopic analysis, observing at substantial depths will be needed to acquire this detailed information.

Although direct observation of Population III stars remains challenging, the galaxies we observe at high redshift are not without their importance. Many of these galaxies are likely low-metallicity, as inferred from the relatively blue average values we have obtained for $\beta$, and by the evidence for low levels of dust attenuation and low stellar mass, providing a view of the universe when it was younger and less chemically enriched. For example, the local galaxy NGC1705 has $\beta \sim -$2.5 \citep{Calzetti1994}, similar to our median value, with a measured gas-phase abundance of $\sim$20\% Solar.  We thus expect that the typical galaxy in our sample is composed of mainly Population II stars, like galaxies we observe at lower redshifts \citep{Greif2006}. However, their low metallicities mean they can serve as analogs for the conditions under which Population III stars might have formed. This contrasts slightly with findings by \cite{PG_C2023}. They observed a heightened activity in the universe during its early phases, especially around $z\gtrsim11$. Such elevated activity, as they state, could imply a notable presence of Population III stars, as these primordial stars might contribute significantly to the heightened UV photon production. However, our data indicates a more prevalent role for Population II stars. Together, these studies raise intriguing questions about the early universe's stellar populations and their contributions to its early luminosity.

Building on this, the characteristics of galaxies housing different stellar populations are anticipated to vary across several parameters. Here, we discuss the differences in what results are expected from Population III stars and what we are obtaining in this work. 

Dust attenuation is expected to be zero in galaxies with Population III stars \citep{Bromm1999}. Given that these stars are thought to be the first generation and formed in a dust (and metal) free environment, the corresponding galaxies would have significantly less dust than those with Population II stars along the line of sight. Although with \textsc{Bagpipes}, we are obtaining $A_\mathrm{v}$ values on the order of zero, based on our values of $\beta$, our galaxies are consistent with having low but not zero metallicity stellar populations. In such cases, follow-up spectroscopy is vital to verify such situations (e.g., measurements of the rest-frame UV spectral slope directly from the spectra; Balmer decrements are unlikely to be detected at these high redshifts currently).

In terms of age, galaxies with Population III stars are expected to be among the youngest in the universe (existing at high redshifts), as these stars are theorized to have formed within a few hundred million years following the Big Bang \citep{Bromm1999,Greif2006}. Deep imaging and spectroscopic surveys like NGDEEP are perhaps our best chance at catching a glimpse of these stellar populations, or at least a direct descendant of them (i.e., Population II stars enriched by Population III remnants). Our galaxy sample has ages that span on the order of $1-100$ Myr, with the average age being $\mathrm{log(Age)} = 7.8^{+0.2}_{-0.8}$ years old. The work of \cite{Yggdrasil} studies the SED tracks of Population I, II, and III (III.1, III.2, and III (Kroupa IMF); see Section 2 of their work for more information on the different Pop III types). Defining stellar populations by Type A, B, and C (where Type A has an $f_\mathrm{esc}=0$, Type B an $0<f_\mathrm{esc}<1$, and Type C an $f_\mathrm{esc}=1$), we compare the UV slopes of our galaxy sample versus what they predict different stellar populations have when binned by age at 1, 10, and 100 Myr (see their Table 2 for full list of UV slopes for instantaneous-burst models). Our galaxy sample is fully consistent with that of Population II tracks for both Type A and C, where Type A tracks range from $-2.2 < \beta < -2.0$ and Type C tracks range from $-3.1 < \beta < -2.0$ for $1-100$ Myr. We find some evidence that our sample follows that of the less extreme Population III type cases when binned by age due to the uncertainty in $f_\mathrm{esc}$ (these UV slopes span $-3.5 < \beta < -2.3$ depending on Population III type, stellar population type, and age).

We thus conclude that the galaxies we have observed with NGDEEP do not show evidence of harboring ultra-low metallicity and/or Population III stars.  While we lack current observational data for galaxies harboring Population III stars, theoretical predictions and simulations offer valuable insight for distinguishing potential differences when comparing them to galaxies inhabited by Population I and II stars. The forthcoming observations from \textit{JWST} will play a pivotal role in validating these predictions, thus enhancing our comprehension of the early universe. Despite the fact that this study has not confirmed the presence of such exotic stellar populations, there remains a possibility that they exist at higher redshifts and/or perhaps exhibit fainter luminosities than what we can detect at this moment.

% =========================================================================%

\section{Conclusions}\label{sec:conclusions}
Using new data from the ultra-deep \textit{JWST} NGDEEP survey, we analyze the evolution of the rest-frame UV spectral slope as a function of redshift from $z\sim 9-16$. We measure the UV spectral slope both via SED-fitting with \textsc{Bagpipes} and via photometric power-law fitting to the observed photometry. We compare our observed UV slopes to the measured stellar mass, UV magnitude, redshift, age, SFR, dust attenuation, and $\mathrm{f_{esc}}$. With these comparisons, we reach the following conclusions:

\begin{enumerate}
    \item We quantify the bias for both SED-fitting and power-law methods with source-injection simulations. We show that for both observed and simulated galaxies, the SED-fitting technique with \textsc{Bagpipes} yields more accurate (less scatter) UV spectral slope results than photometric power-law fitting. Both methods still exhibit biases -- we quantify this bias for the SED fitting method and use it to correct our UV slope measurements.  After this correction, we find a reasonable agreement between the SED fitting and power-law fitting methodologies (we note we do not apply any corrections to the photometric power-law measurements), though the latter exhibits significant scatter to both much redder and bluer colors. 
    \item When measuring $\beta$ via the SED-fitting method for the entire sample, we obtain an average UV slope, $\beta_\mathrm{SED} = -2.46^{+0.24}_{-0.19}$ ($-2.30^{+0.19}_{-0.13}$ before bias correction), compared to $\beta_\mathrm{PL} = -2.65^{+0.52}_{-0.51}$ via photometric power-law fitting. These average values indicate that our sample of galaxies is predominantly low-metallicity, but we do not find evidence for any galaxies whose stellar populations could be considered exotic (i.e., enriched by Population III stars, $\beta \leq -3$). 
    \item We find moderately positive correlations between the UV spectral slope and dust attenuation, age, stellar mass, and star formation rate as derived from \textsc{Bagpipes}.  We find weak-to-no correlations between the UV spectral slope and the escape fraction, UV magnitude, and redshift. Further observations, with future data reductions and the long-awaited second half of data from this survey, will increase our sample size and perhaps offer a more accurate representation of galaxies during this epoch.

    \item We compare our observations with the results of the \textsc{SC-SAM} \citep{Yung2023}, \textsc{FLARES} \citep[][]{FLARES-I, FLARES-II, FLARES-V}, and the \textsc{SIMBA-EoR} simulations \citep[][Jones et al. in prep.]{Dave2019} and find that both show no evidence of ultra-blue UV slopes ($\beta < -3$) at these redshifts and stellar masses. 

\end{enumerate}

Despite achieving solid measurements of the UV spectral slope, the precise mechanisms driving the observed changes remain a subject of discussion. Further clarity on this matter requires a larger sample of these stellar masses and redshifts. Currently, we hypothesize that variations in the UV spectral slope in the observed galaxies are predominantly influenced by dust attenuation, $A_\mathrm{v}$, and stellar metallicity. Incorporating bluer SED models in future studies with \textsc{Bagpipes} could potentially offer pivotal insights into these variations. A more in-depth analysis involving dust growth or perhaps varying dust attenuation curves \citep{Salim2018}, coupled with assessments of stellar mass, SFR, and age, is essential to outline detailed attributes of the sample more accurately. The second half of data collection for this survey, complemented by future deep-imaging surveys with \textit{JWST}, seems promising in offering a detailed story on the evolutionary dynamics of the UV spectral slope during these early epochs.
% =========================================================================%

\section{Acknowledgements}\label{sec:acknowledgements}

AMM thanks Michael Topping for access to \cite{Topping2022} data for the purpose of this analysis.  We acknowledge that the location where this work took place, the University of Texas at Austin, which sits on indigenous land. The Tonkawa lived in central Texas, and the Comanche and Apache moved through this area. We pay our respects to all the American Indian and Indigenous Peoples and communities who have been or have become a part of these lands and territories in Texas, on this piece of Turtle Island. 

AMM acknowledges support from the National Science Foundation Graduate Research Fellowship Program under Grant Number DGE 2137420. Any opinions, findings, conclusions, or recommendations expressed in this material are those of the author(s) and do not necessarily reflect the views of the National Science Foundation.  AMM and SLF acknowledge support from NASA via STScI JWST-GO-2079.

PGP-G acknowledges support from grant PGC2018-093499-B-I00 and PID2022-139567NB-I00 funded by Spanish Ministerio de Ciencia e Innovación MCIN/AEI/10.13039/501100011033, FEDER, UE.

\noindent
\textit{Software}: \verb|IPython| \citep{Perez2007a}, \verb|matplotlib| \citep{Hunter2007a}, \verb|NumPy| \citep{VanderWalt2011a}, \verb|SciPy| \citep{Oliphant2007a}, \verb|Astropy| \citep{Robitaille2013}, \verb|Bagpipes| \citep{Carnall2021}, \verb|Emcee| \citep{Foreman-Mackey2013_emcee}, \verb|EAZY| \citep{brammer_eazy} .
% =========================================================================%
\bibliographystyle{aasjournal}

\bibliography{library}

% =========================================================================%
\appendix 
\label{appendix}

\vspace{-2mm}

\begin{figure}[htb]
  \includegraphics[width=.33\linewidth]{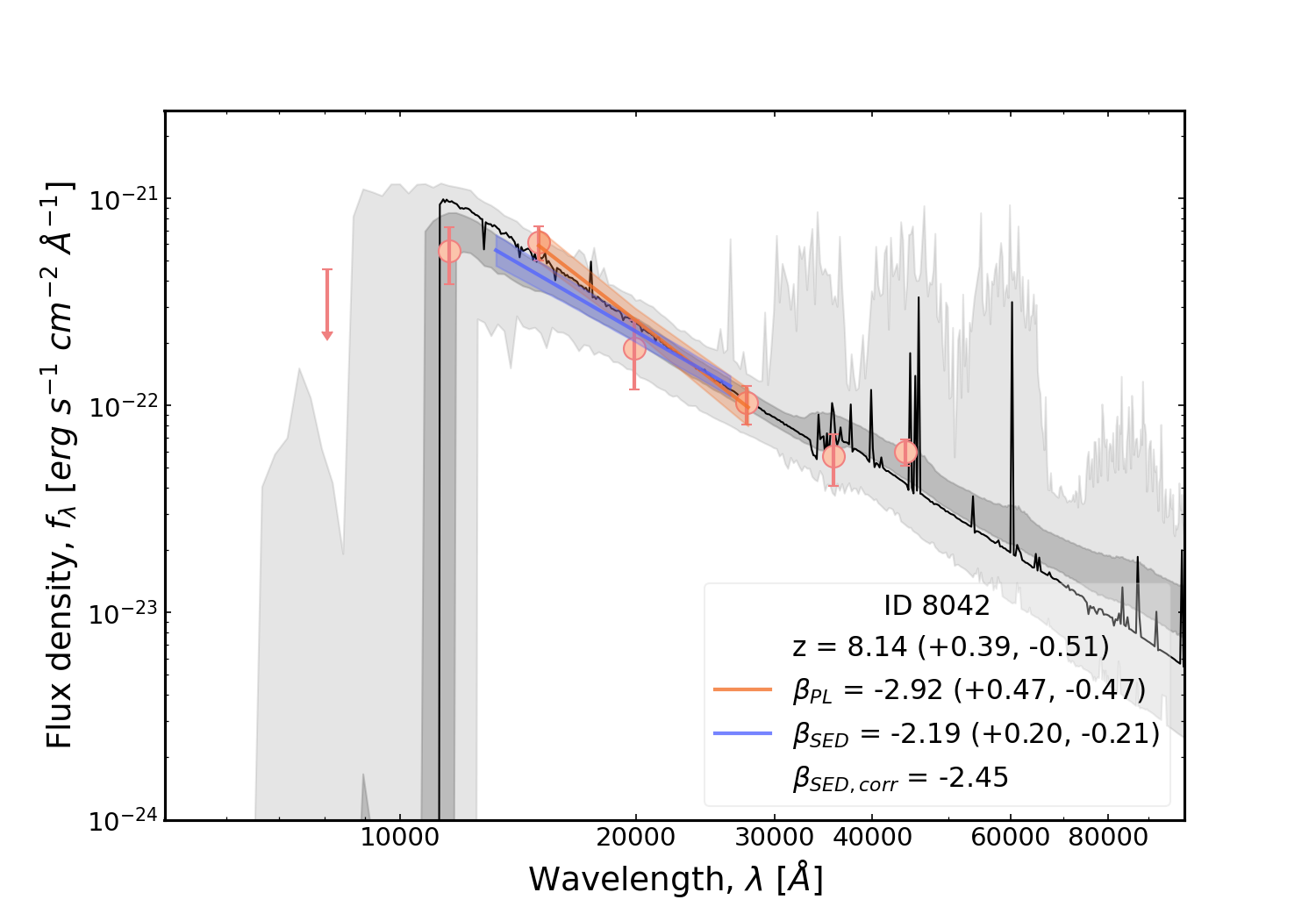}%
\hfill 
  \includegraphics[width=.33\linewidth]{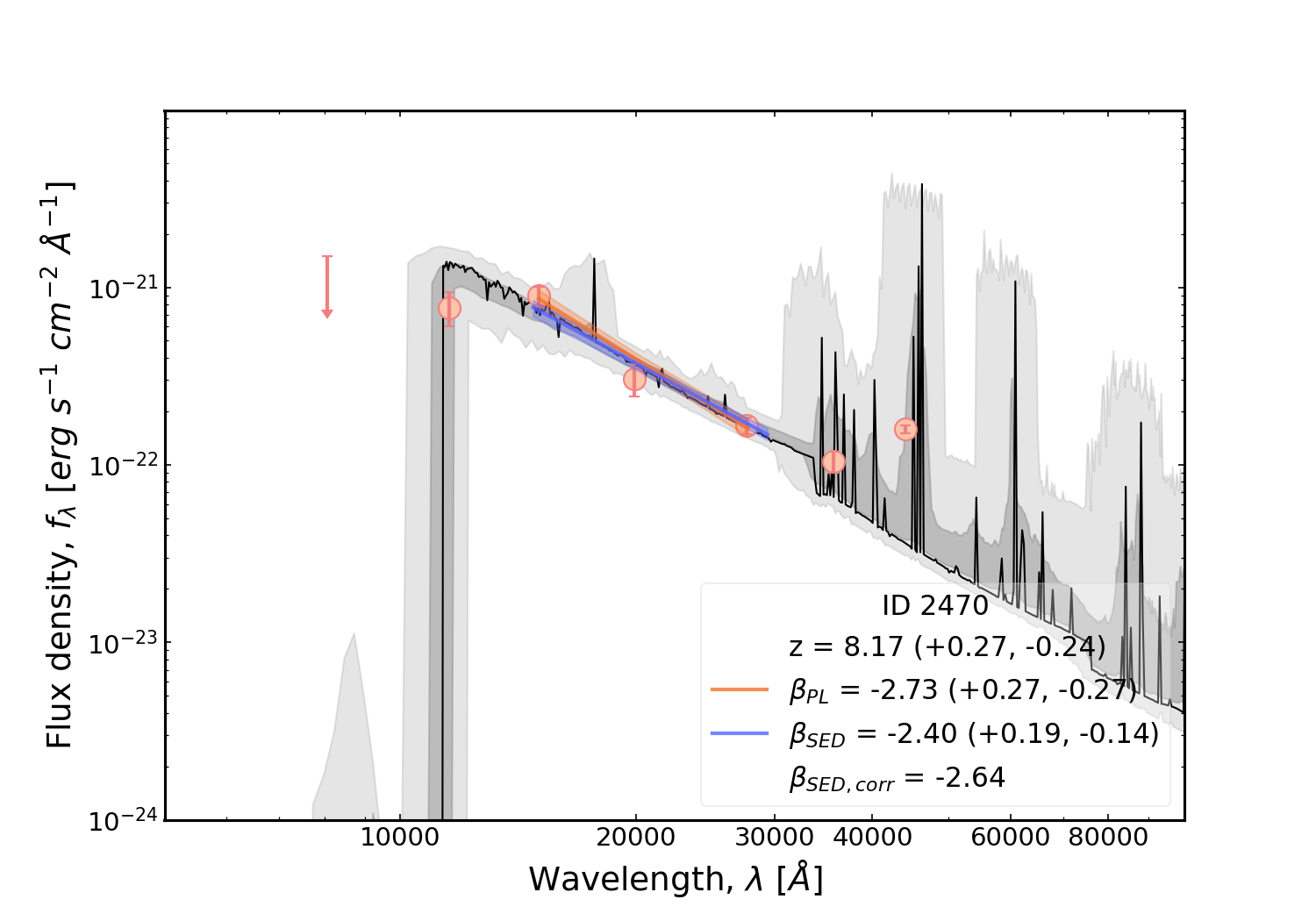}
  \includegraphics[width=.33\linewidth]{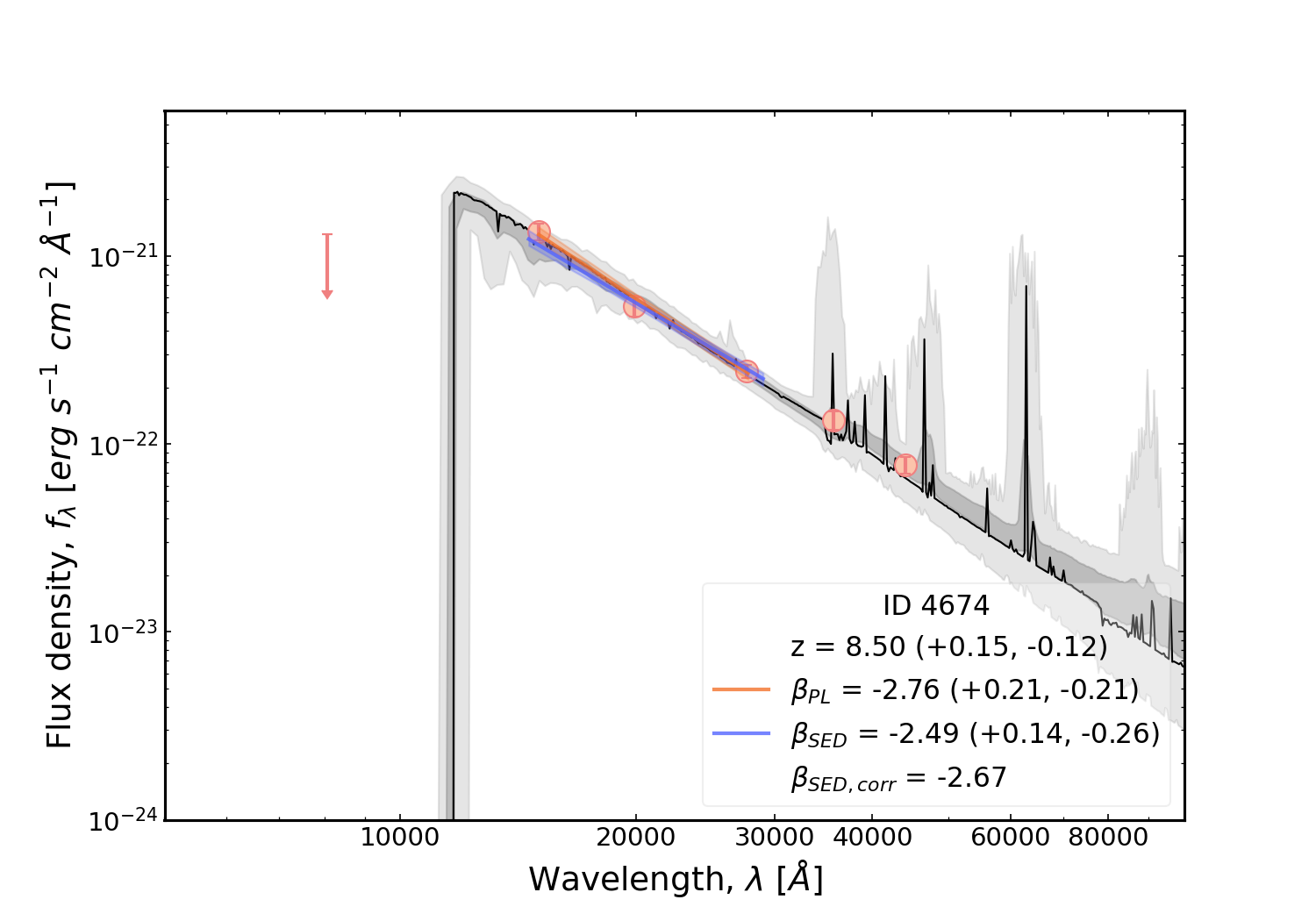} %
\hfill
  \includegraphics[width=.33\linewidth]{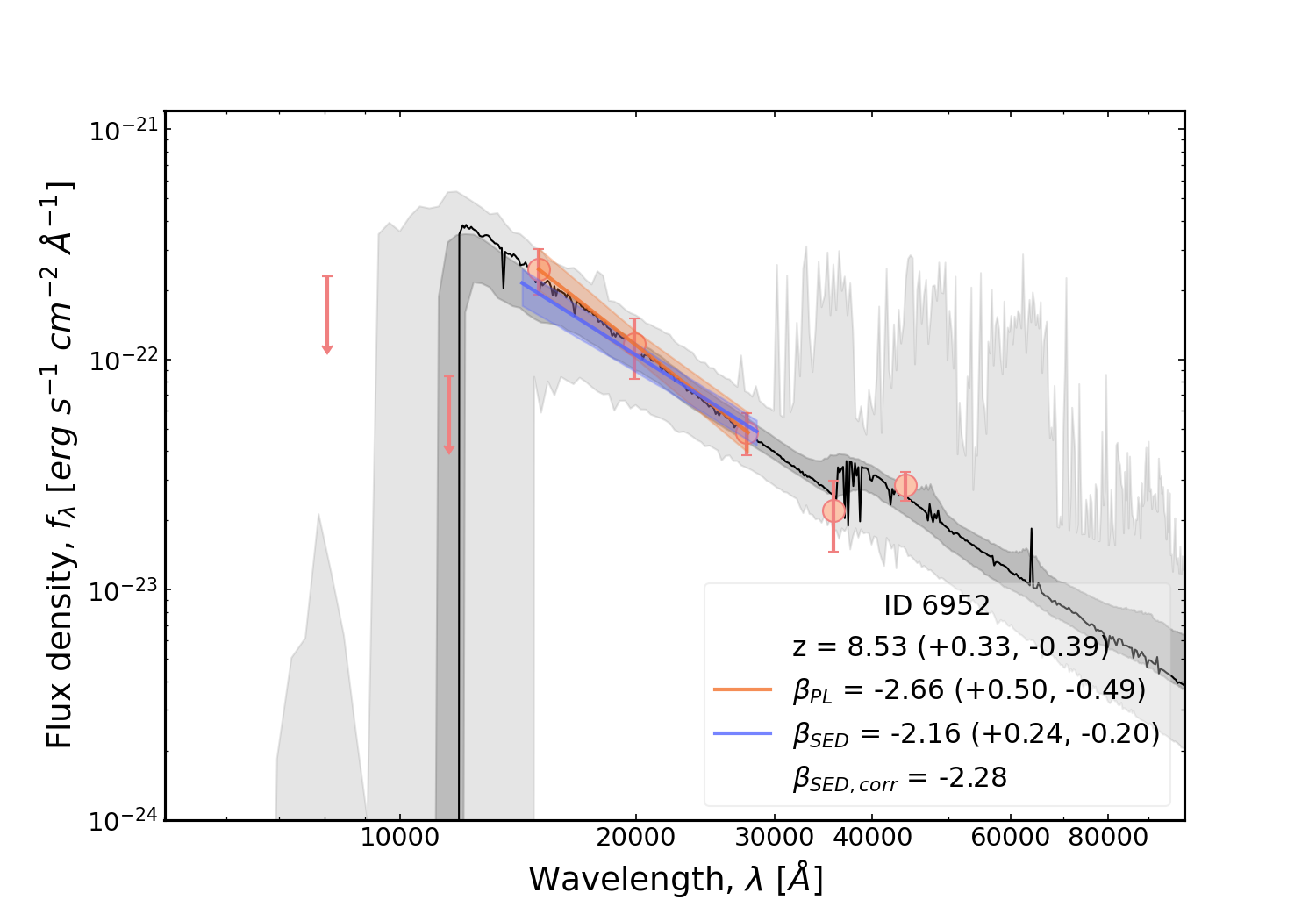}
  \includegraphics[width=.33\linewidth]{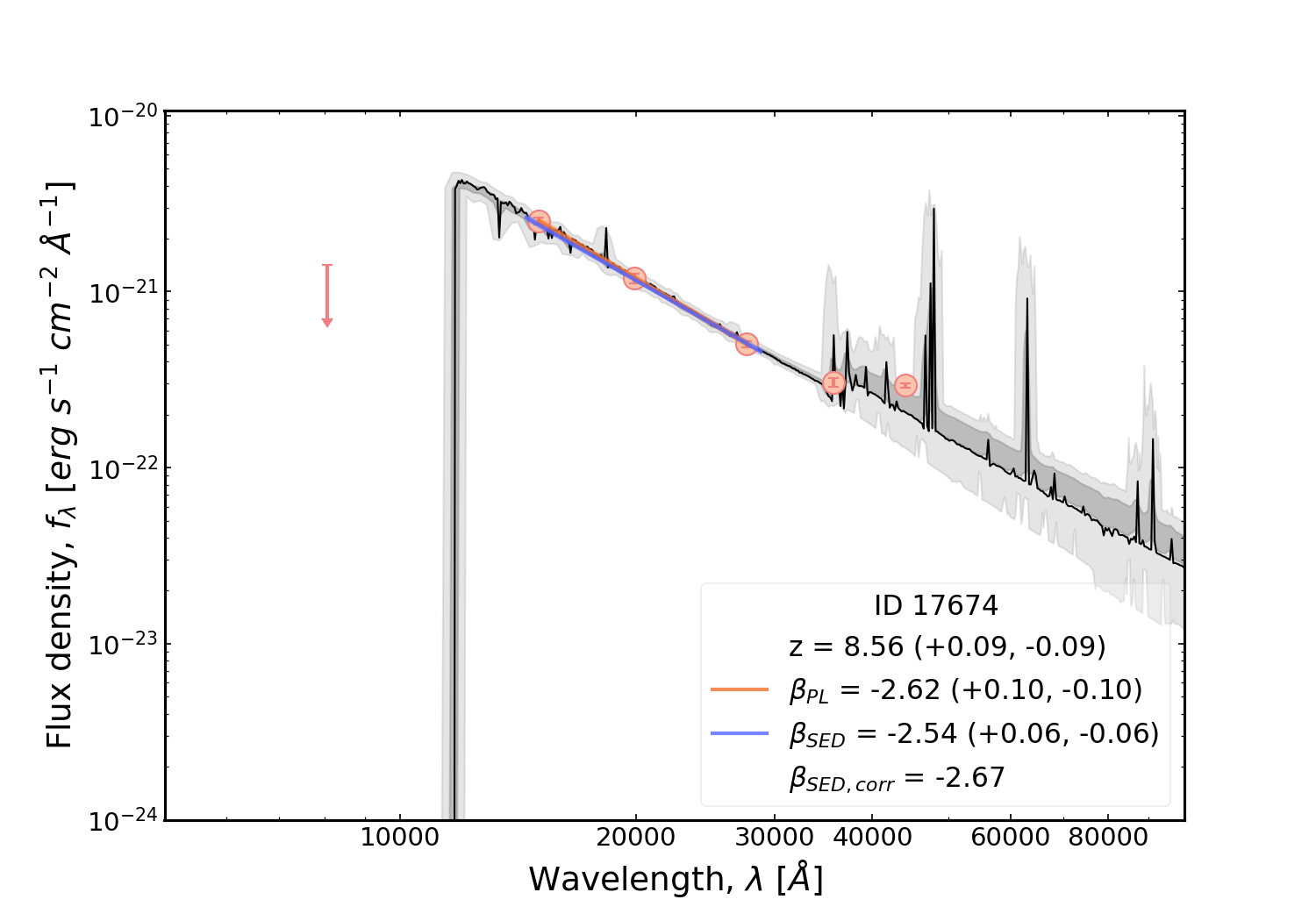}%
\hfill
  \includegraphics[width=.33\linewidth]{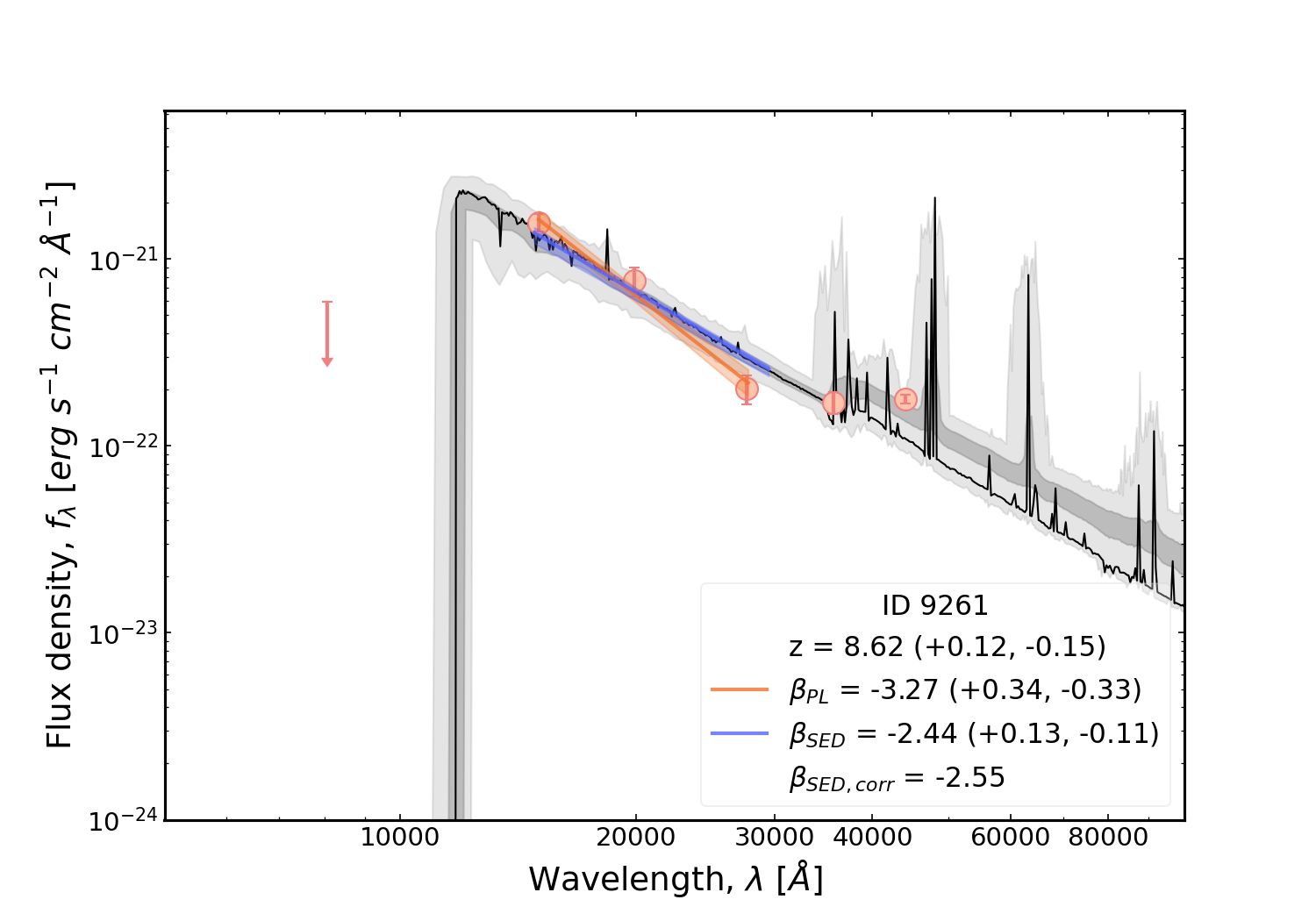}
  \includegraphics[width=.33\linewidth]{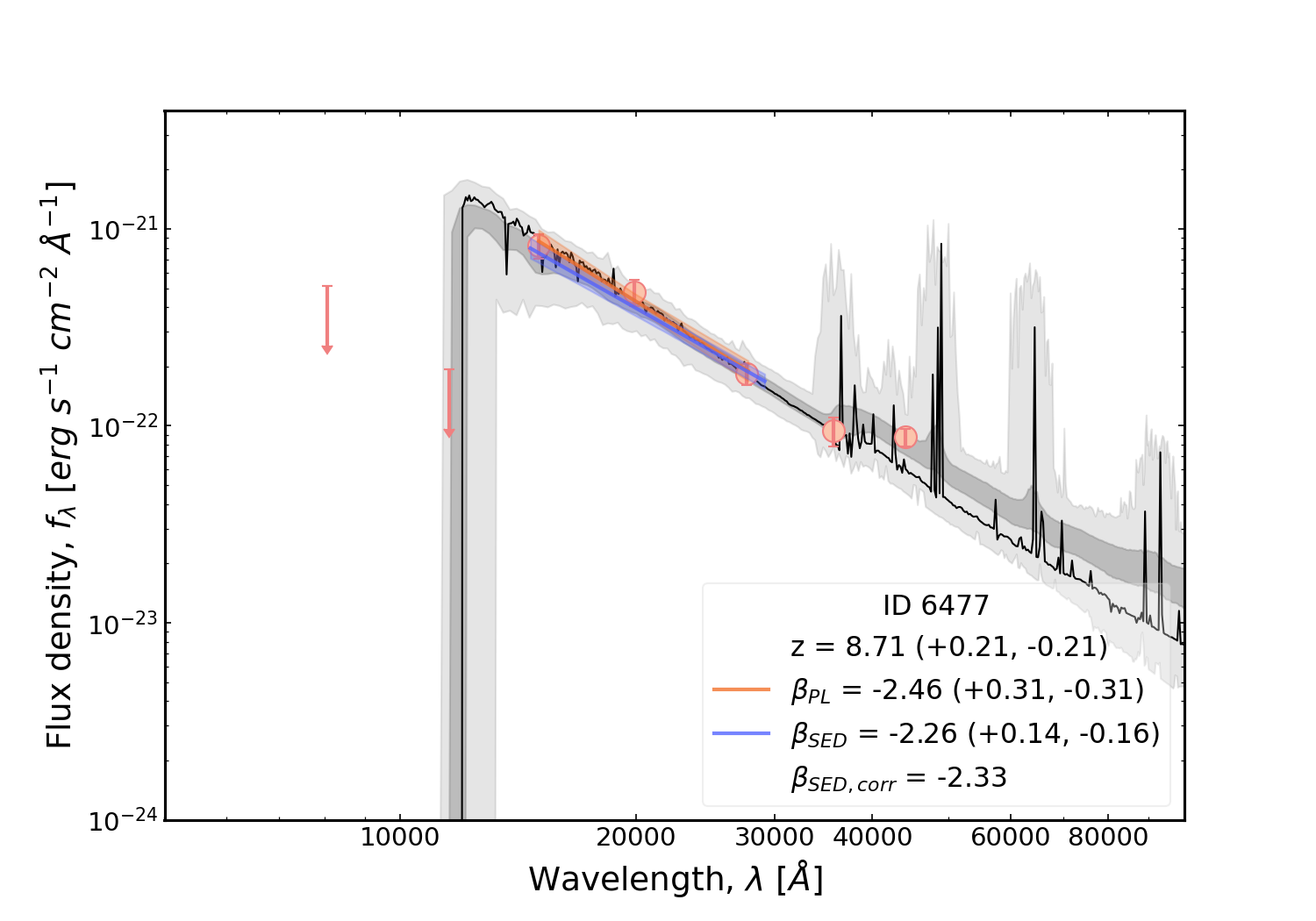}%
\hfill
  \includegraphics[width=.33\linewidth]{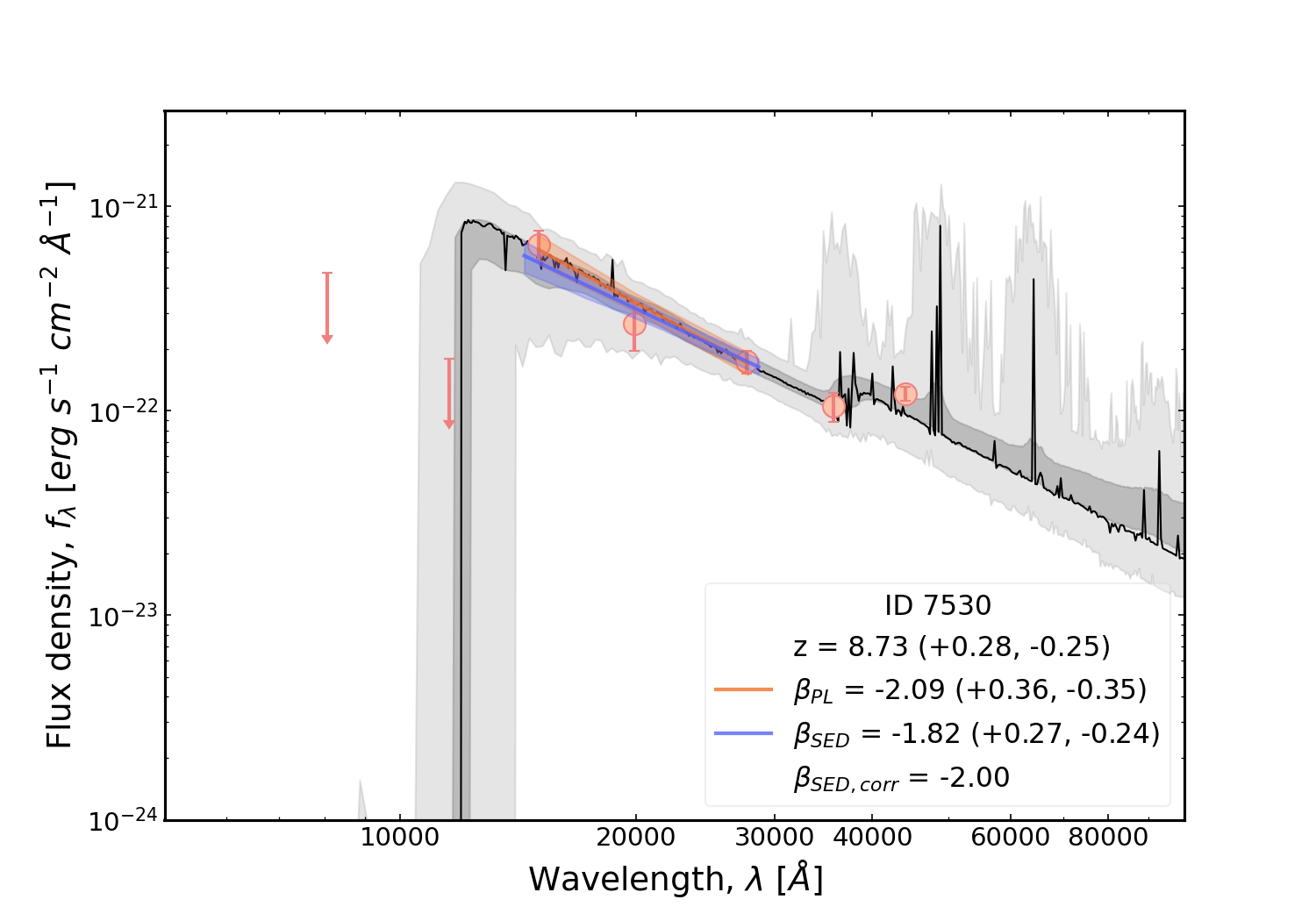}
  \includegraphics[width=.33\linewidth]{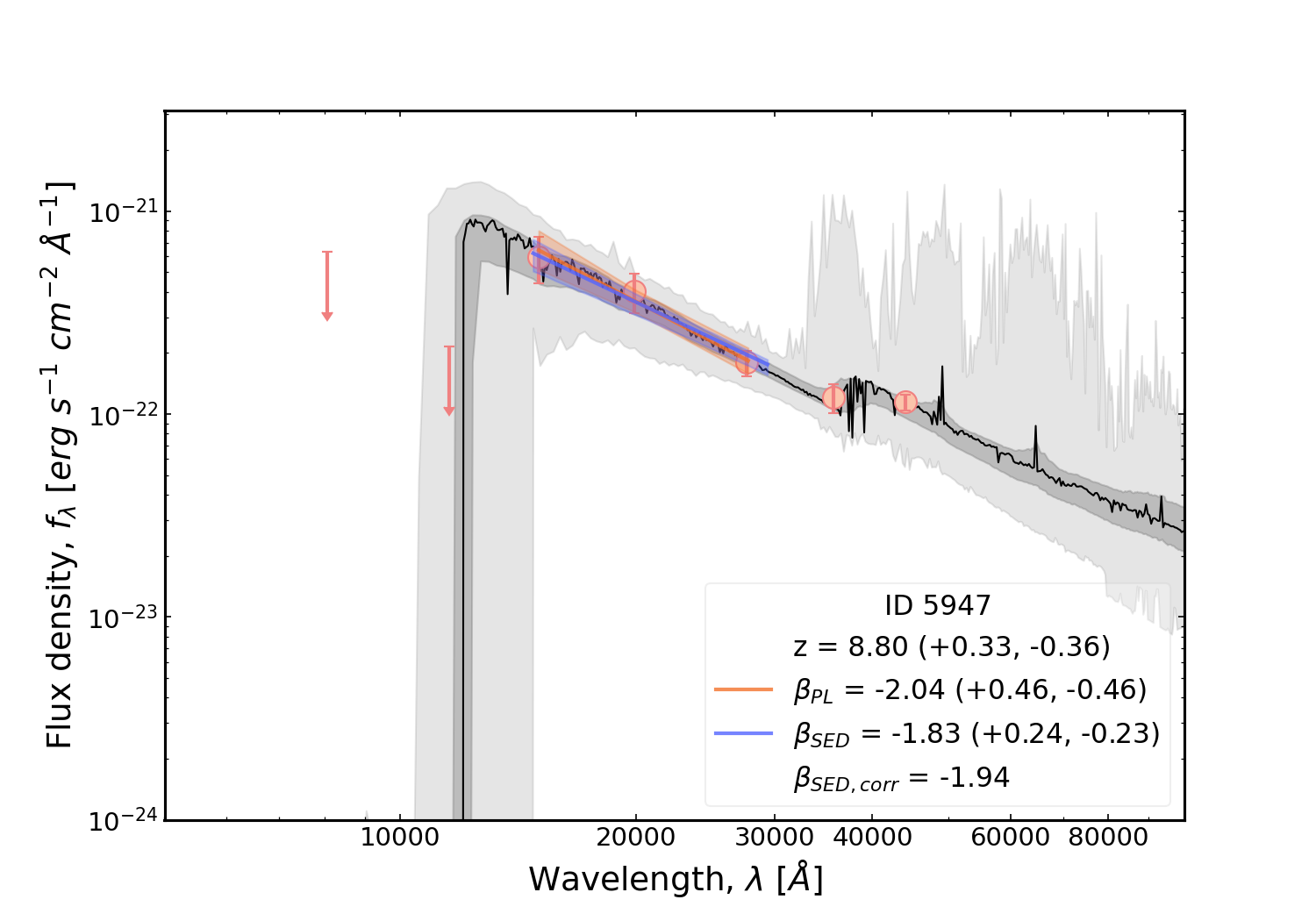}%
  \hfill
  \includegraphics[width=.33\linewidth]{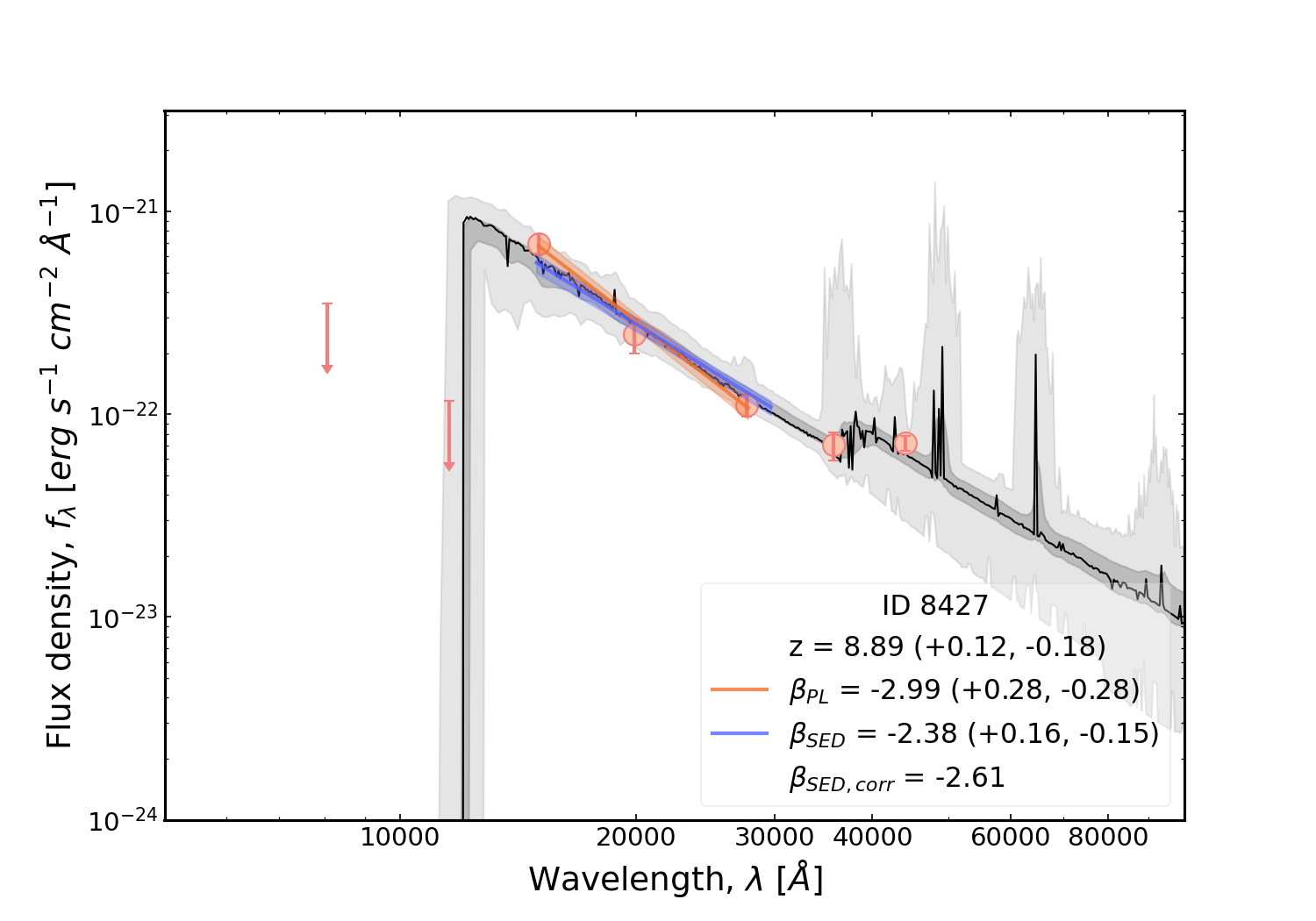}
\hfill
  \includegraphics[width=.33\linewidth]{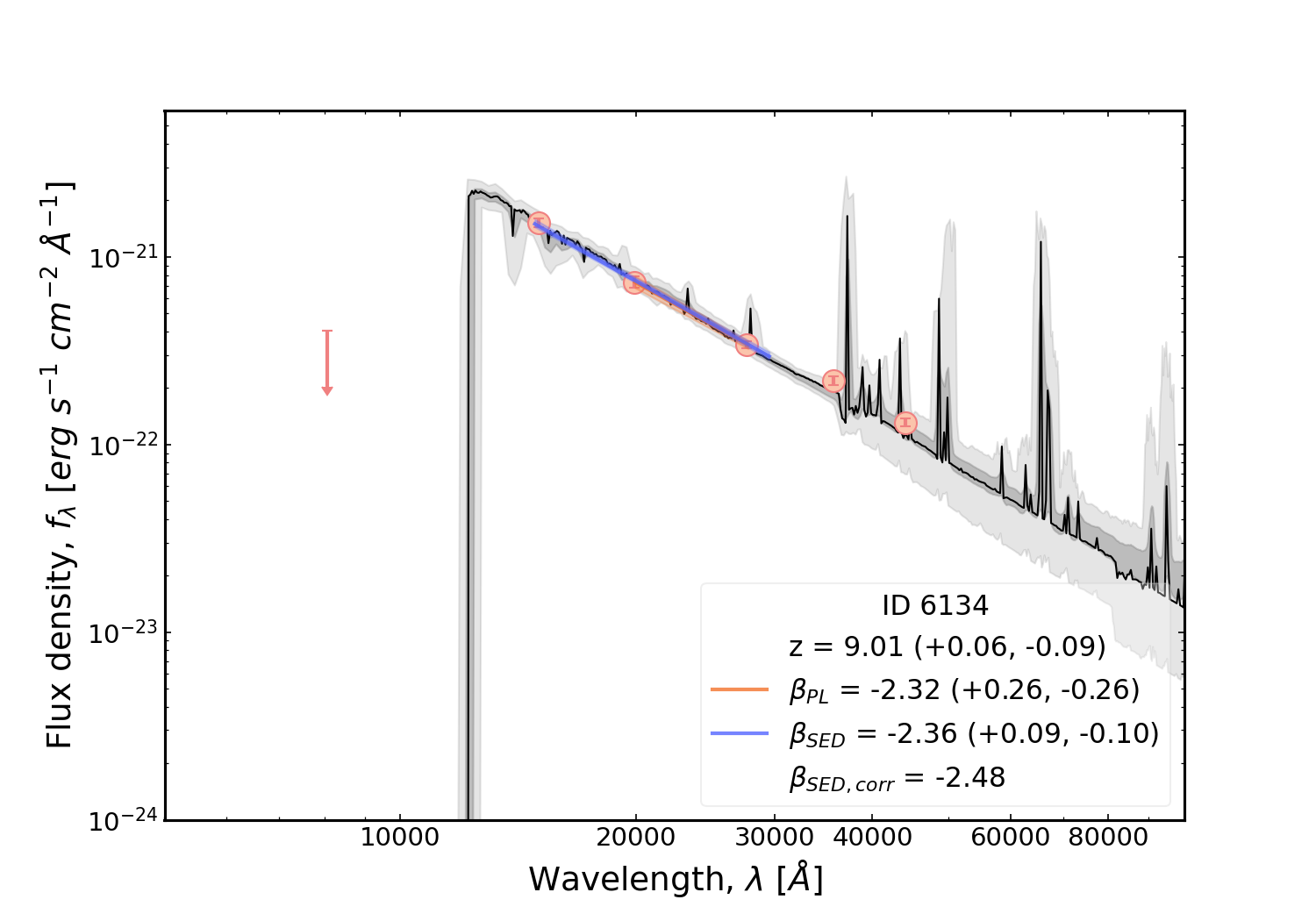}%
  \includegraphics[width=.33\linewidth]{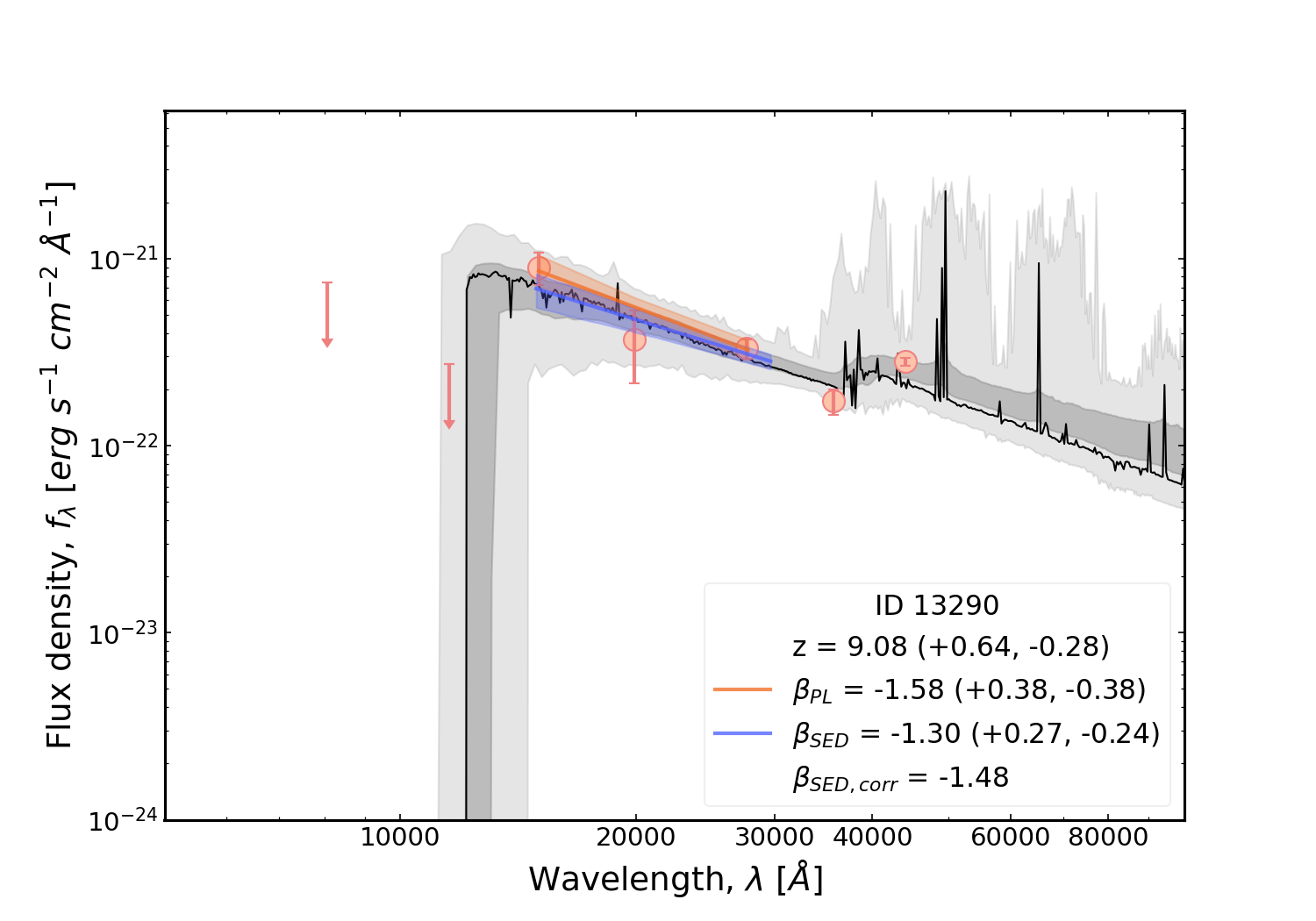}
\includegraphics[width=.33\linewidth]{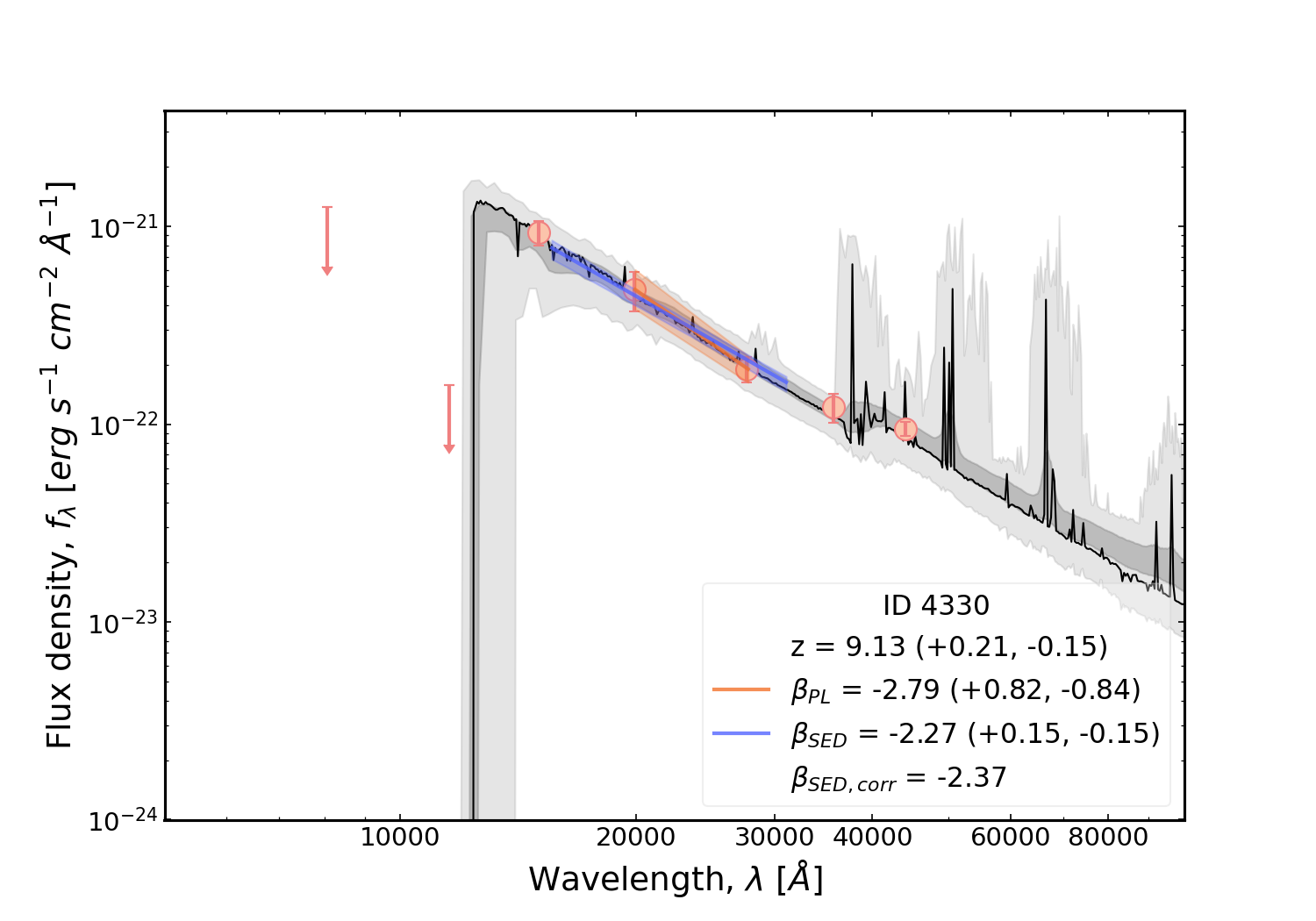} 
  \includegraphics[width=.33\linewidth]{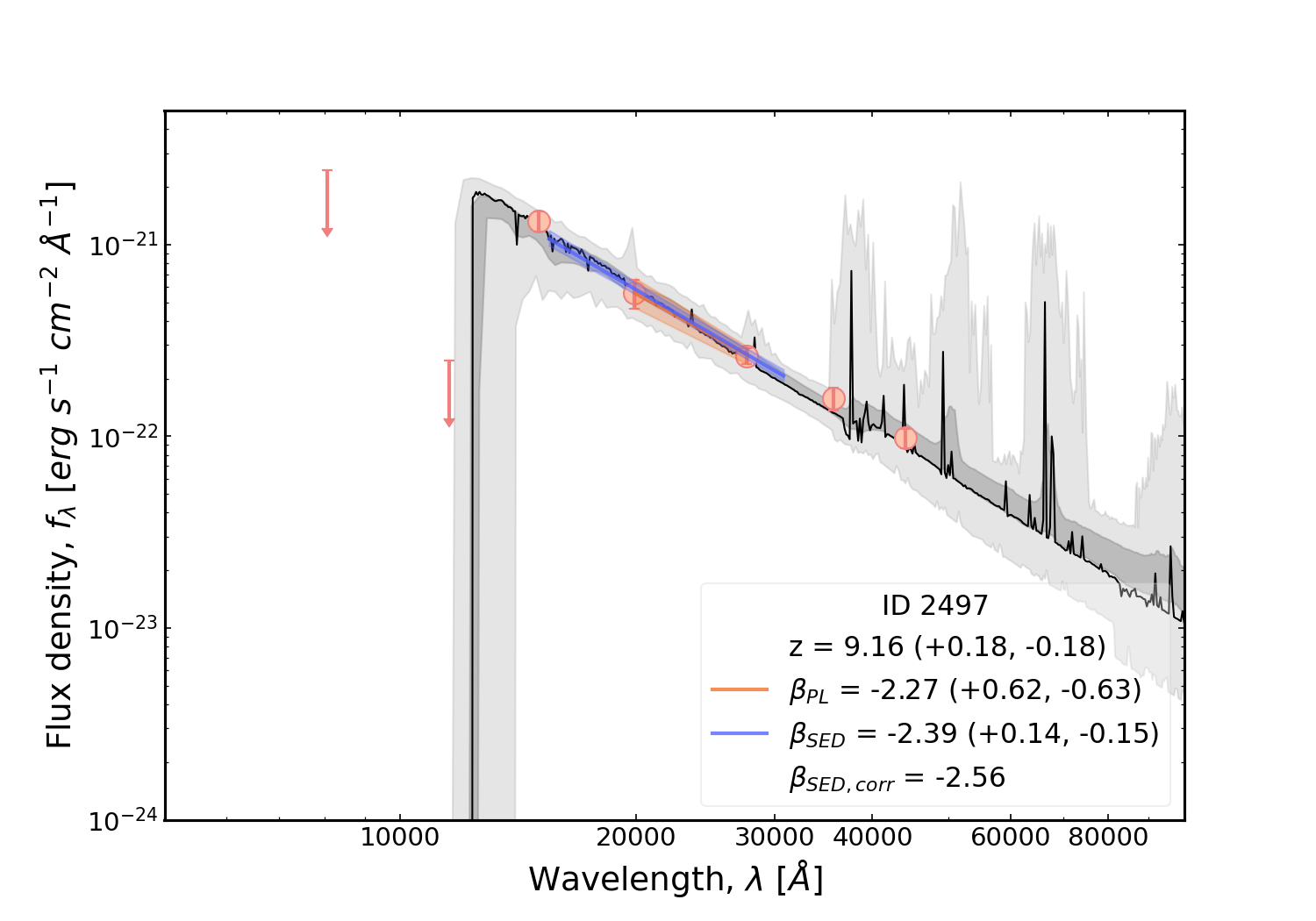}
\includegraphics[width=.33\linewidth]{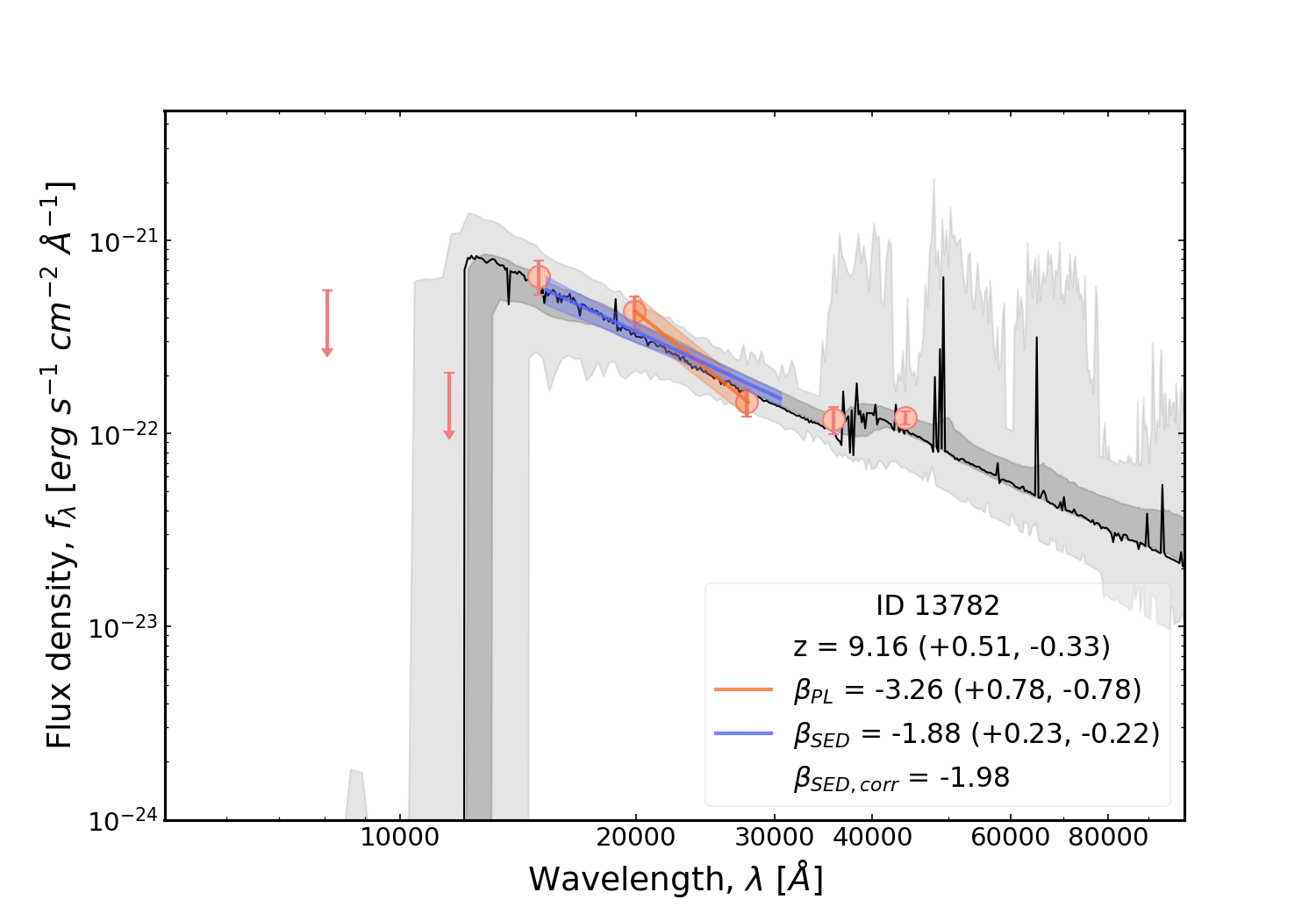}%
\hfill  
  \end{figure}
\pagebreak
  
  \begin{figure}[htb]

  \includegraphics[width=.33\linewidth]{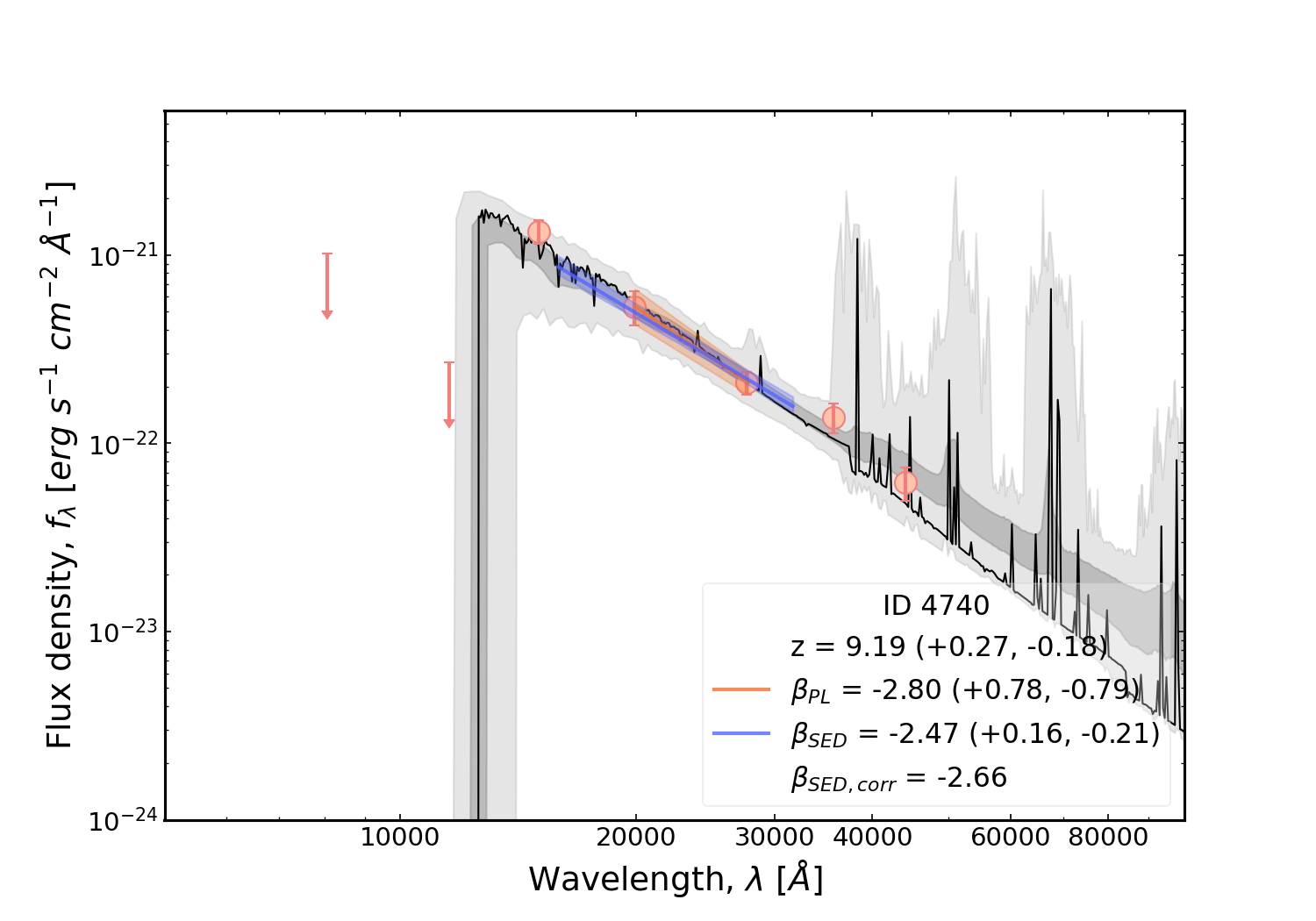}
    \hfill 
  \includegraphics[width=.33\linewidth]{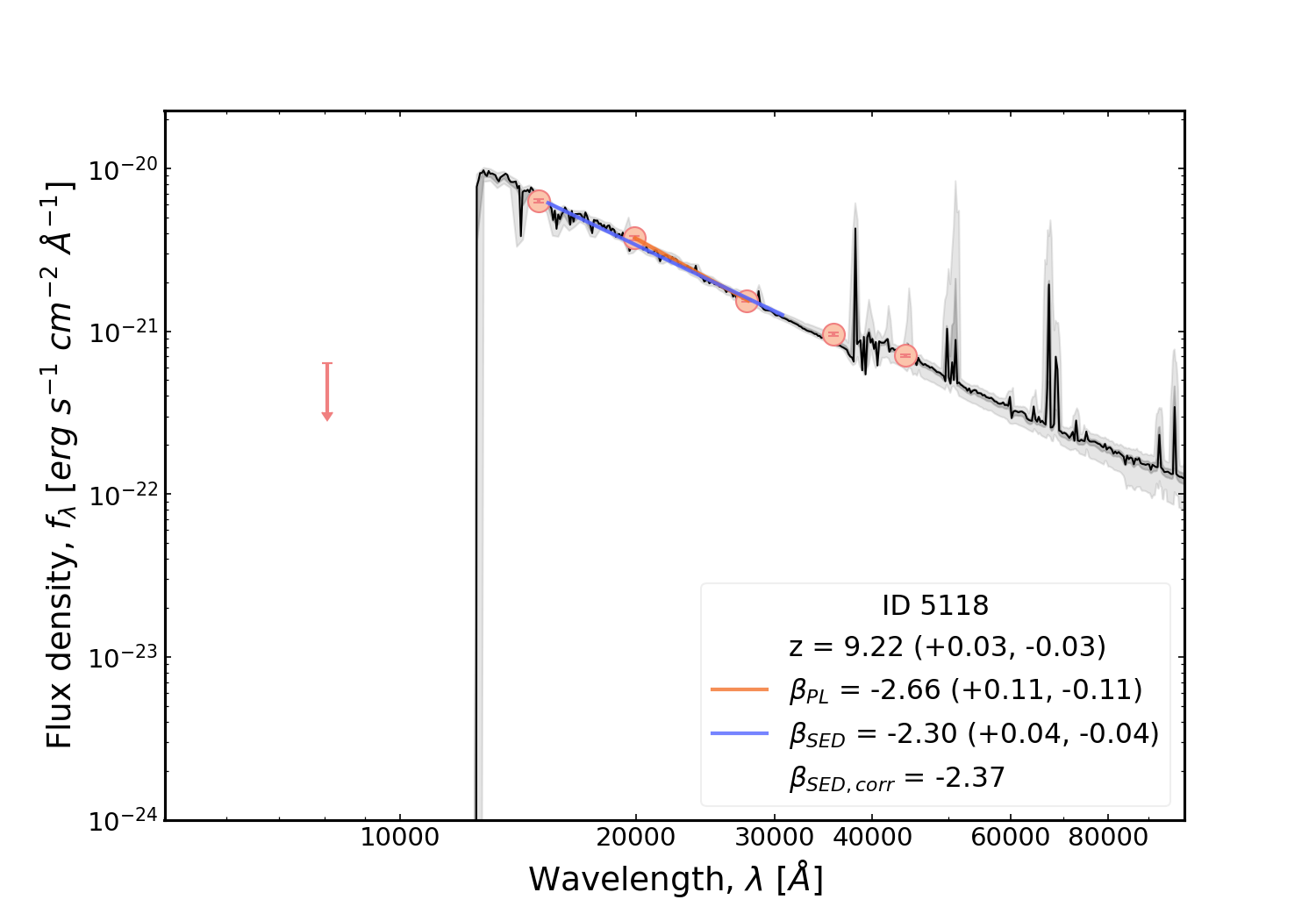}%
  \includegraphics[width=.33\linewidth]{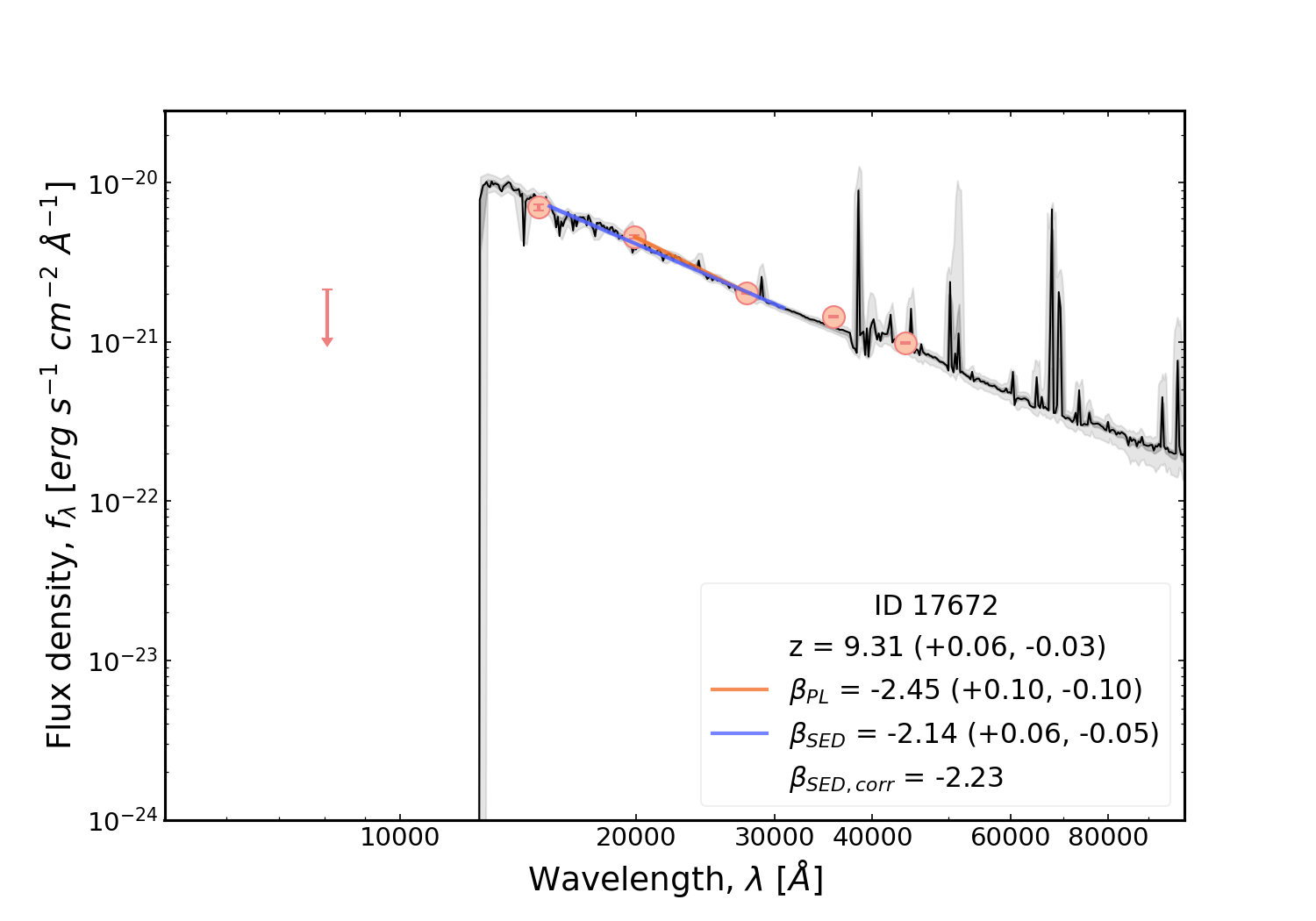}
    \hfill 
  \includegraphics[width=.33\linewidth]{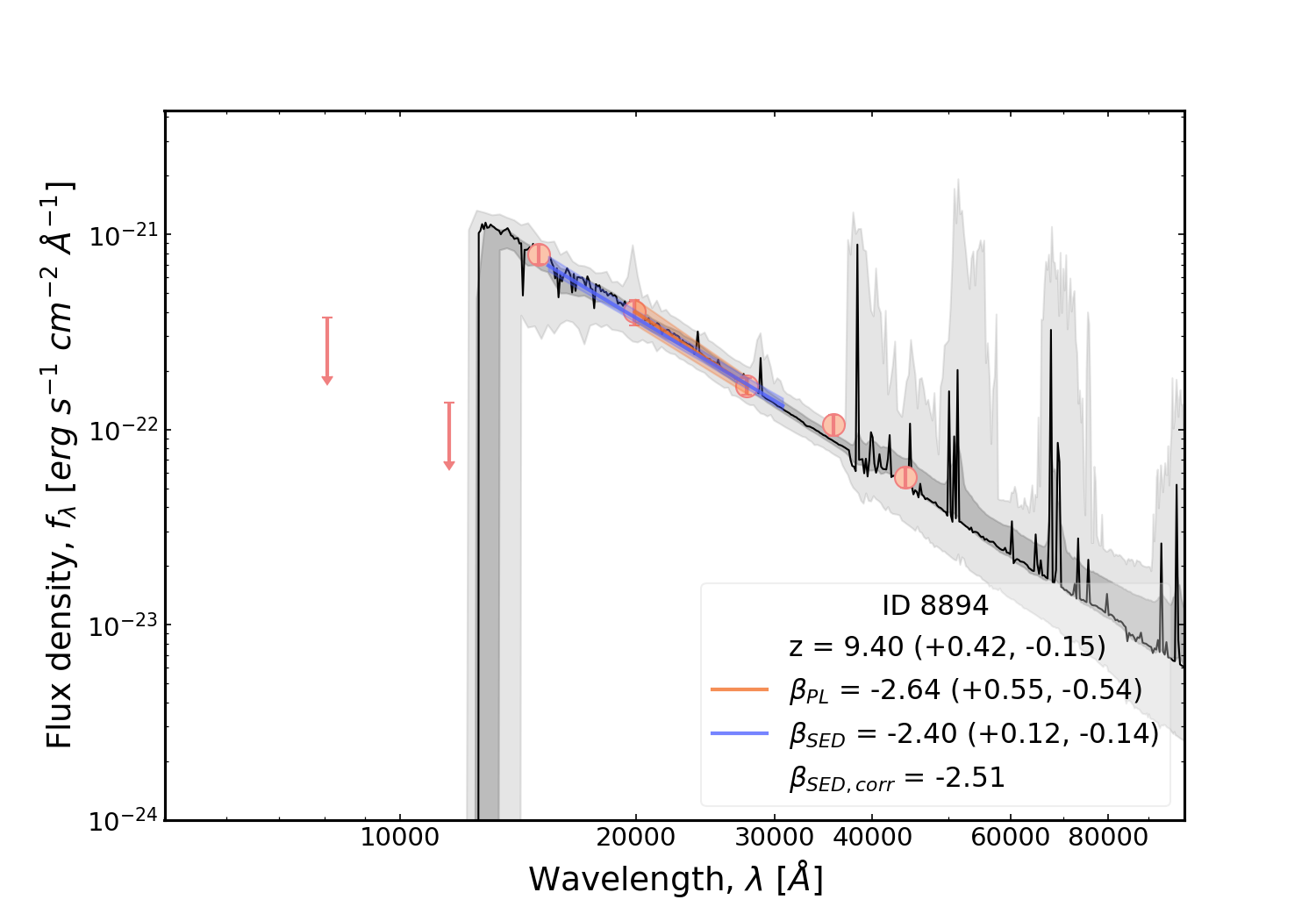}
  \includegraphics[width=.33\linewidth]{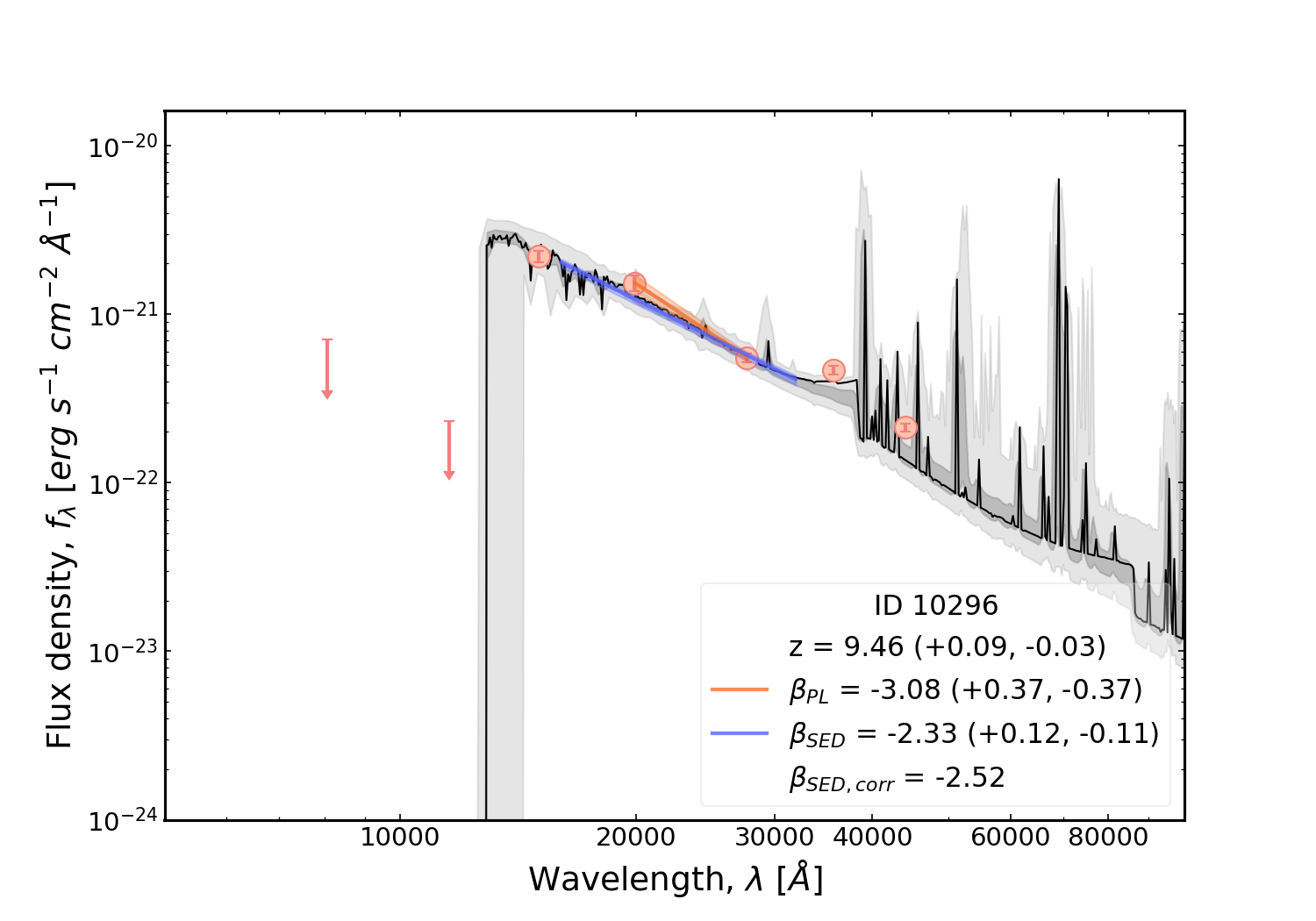}
  \hfill
  \includegraphics[width=.33\linewidth]{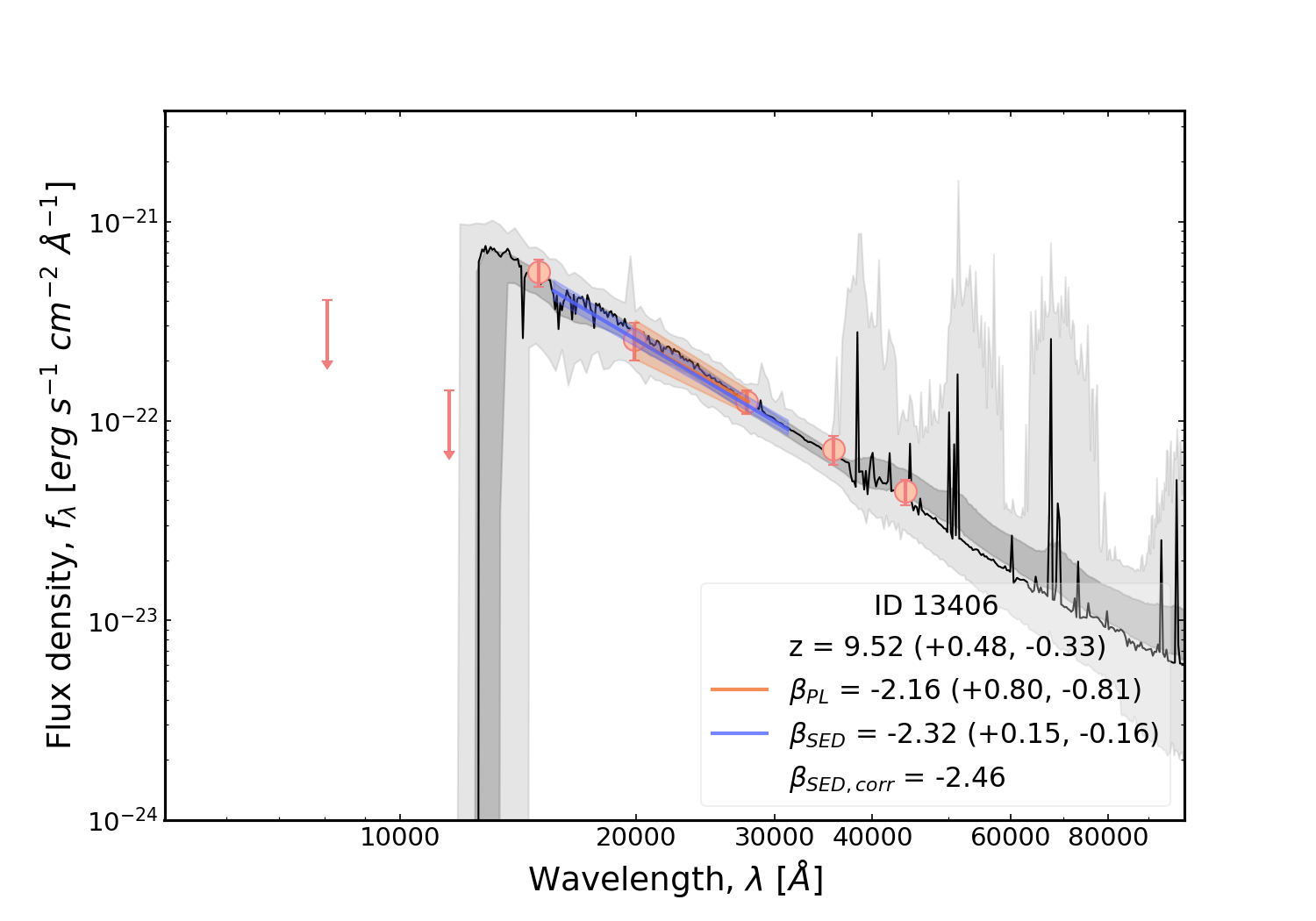}%
  \hfill
  \includegraphics[width=.33\linewidth]{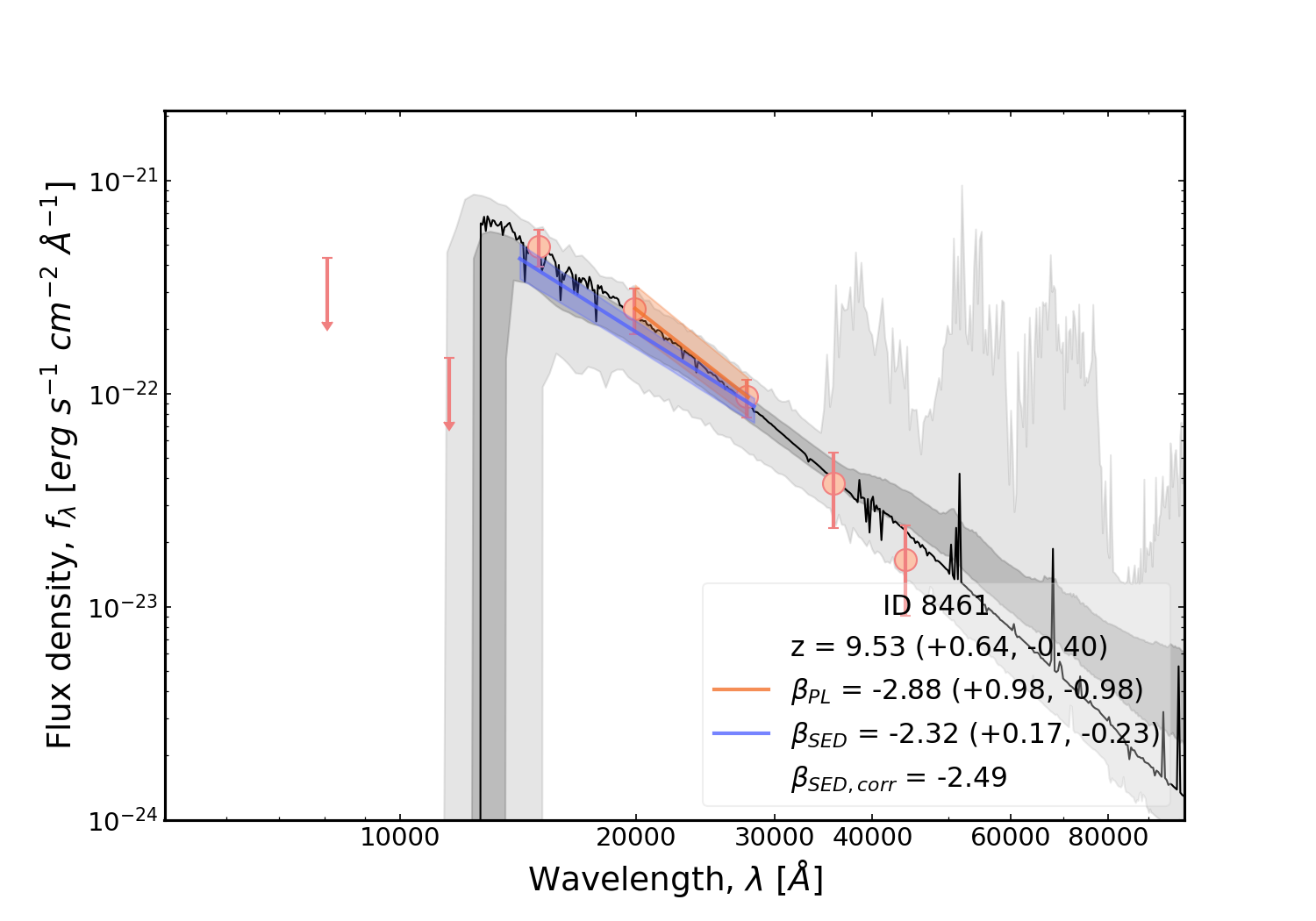}
  \hfill
  \includegraphics[width=.33\linewidth]{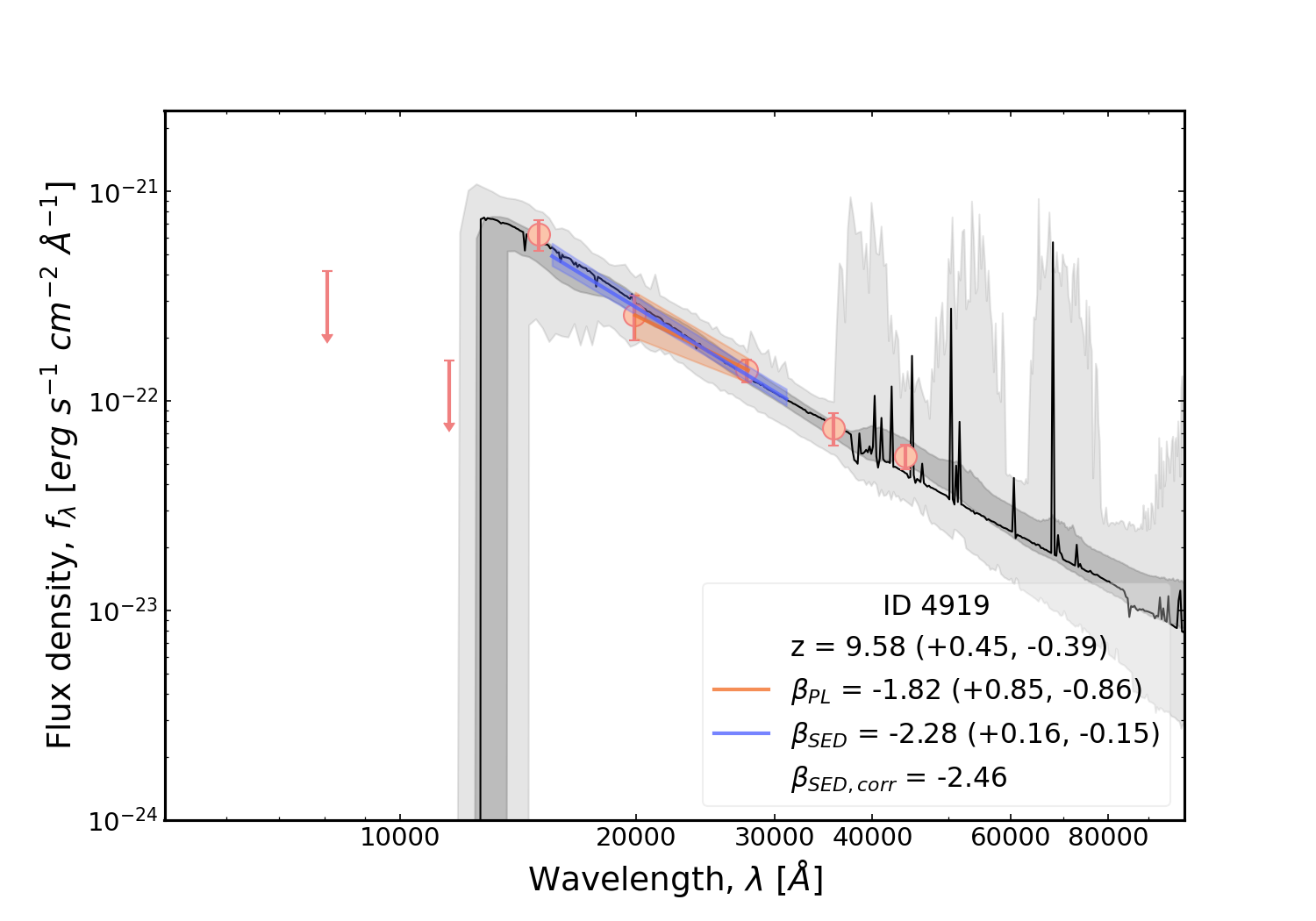}%
  \includegraphics[width=.33\linewidth]{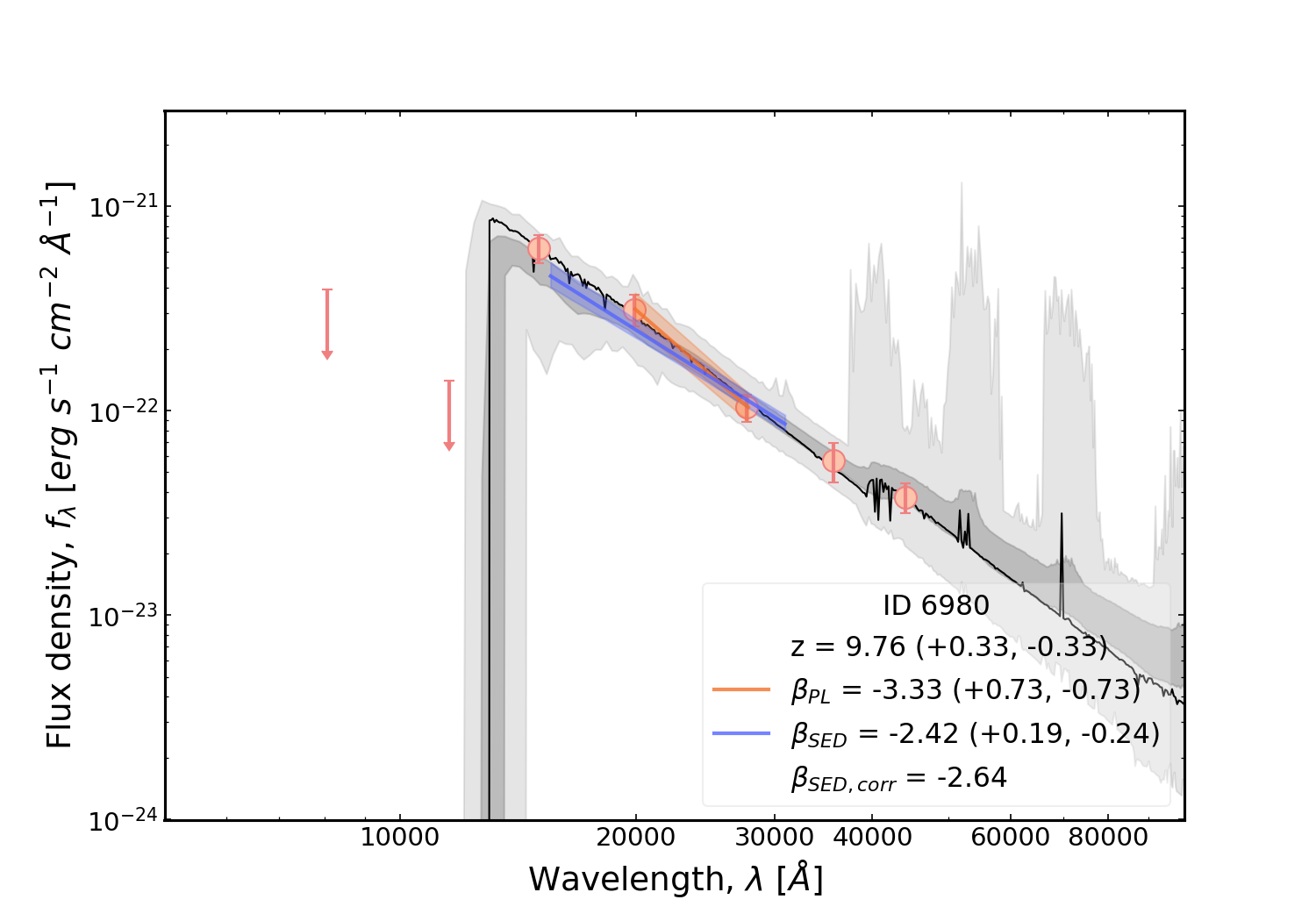}
  \hfill
  \includegraphics[width=.33\linewidth]{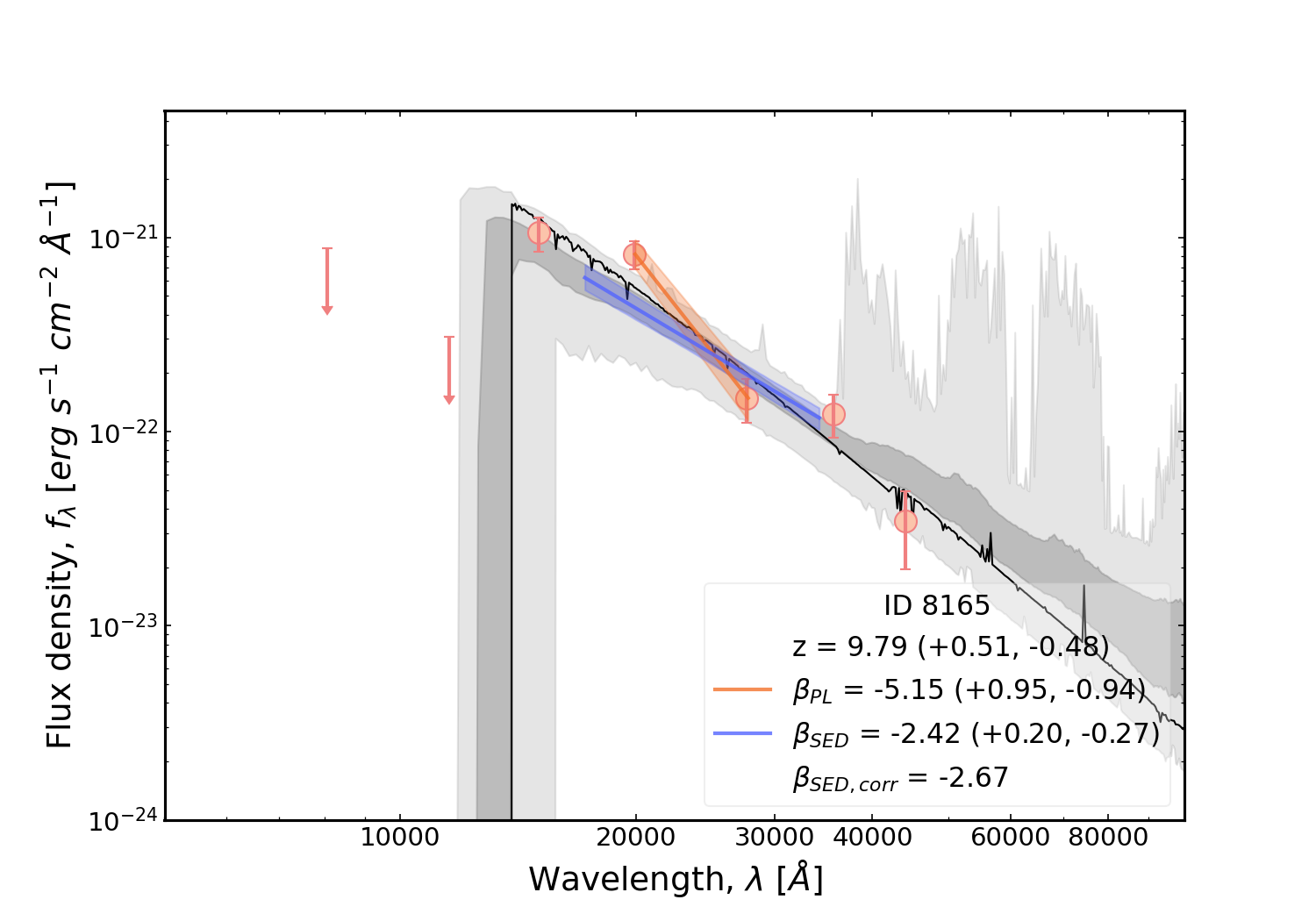}%
  \hfill
  \includegraphics[width=.33\linewidth]{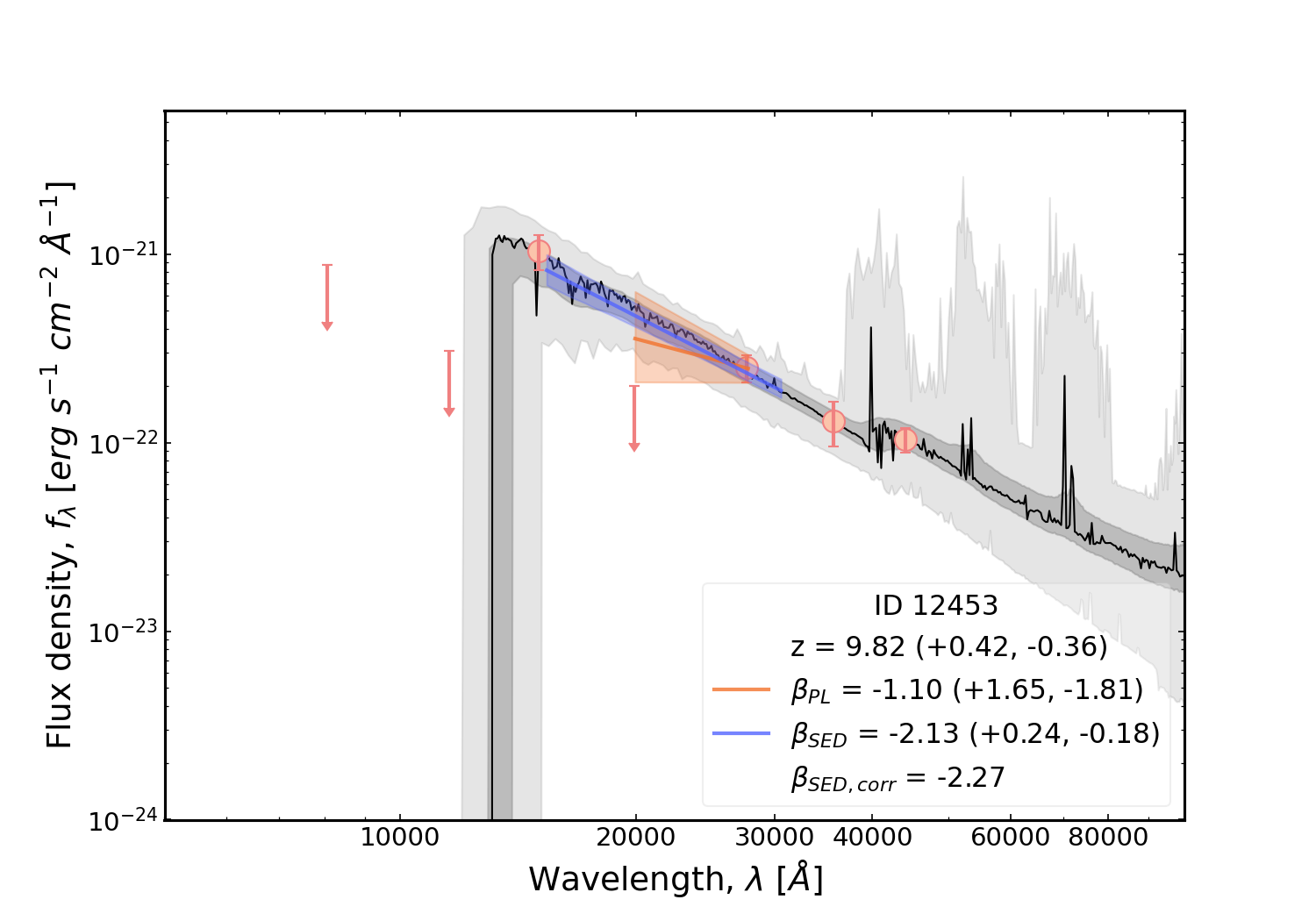}
  \hfill
  \includegraphics[width=.33\linewidth]{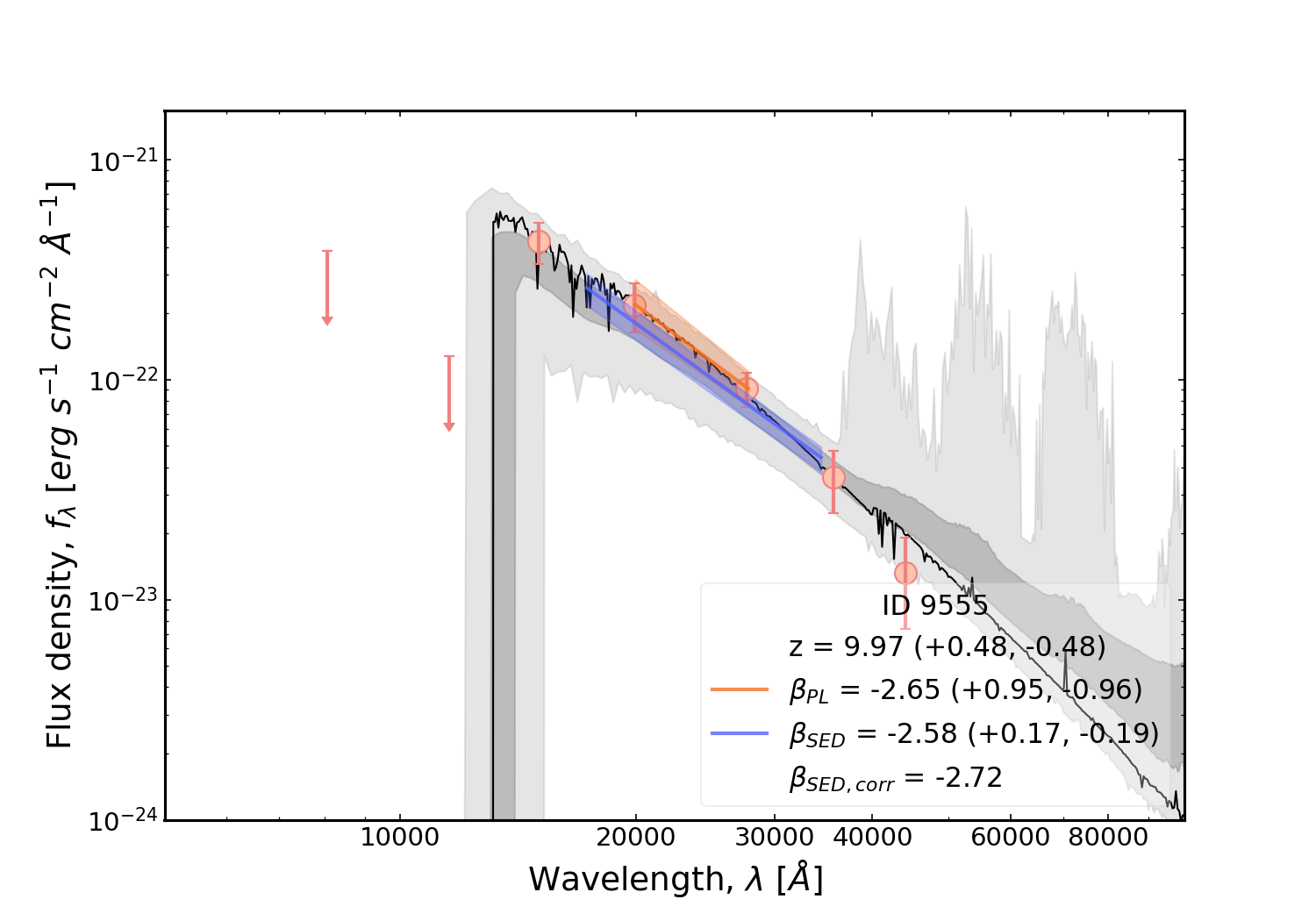}
  \includegraphics[width=.33\linewidth]{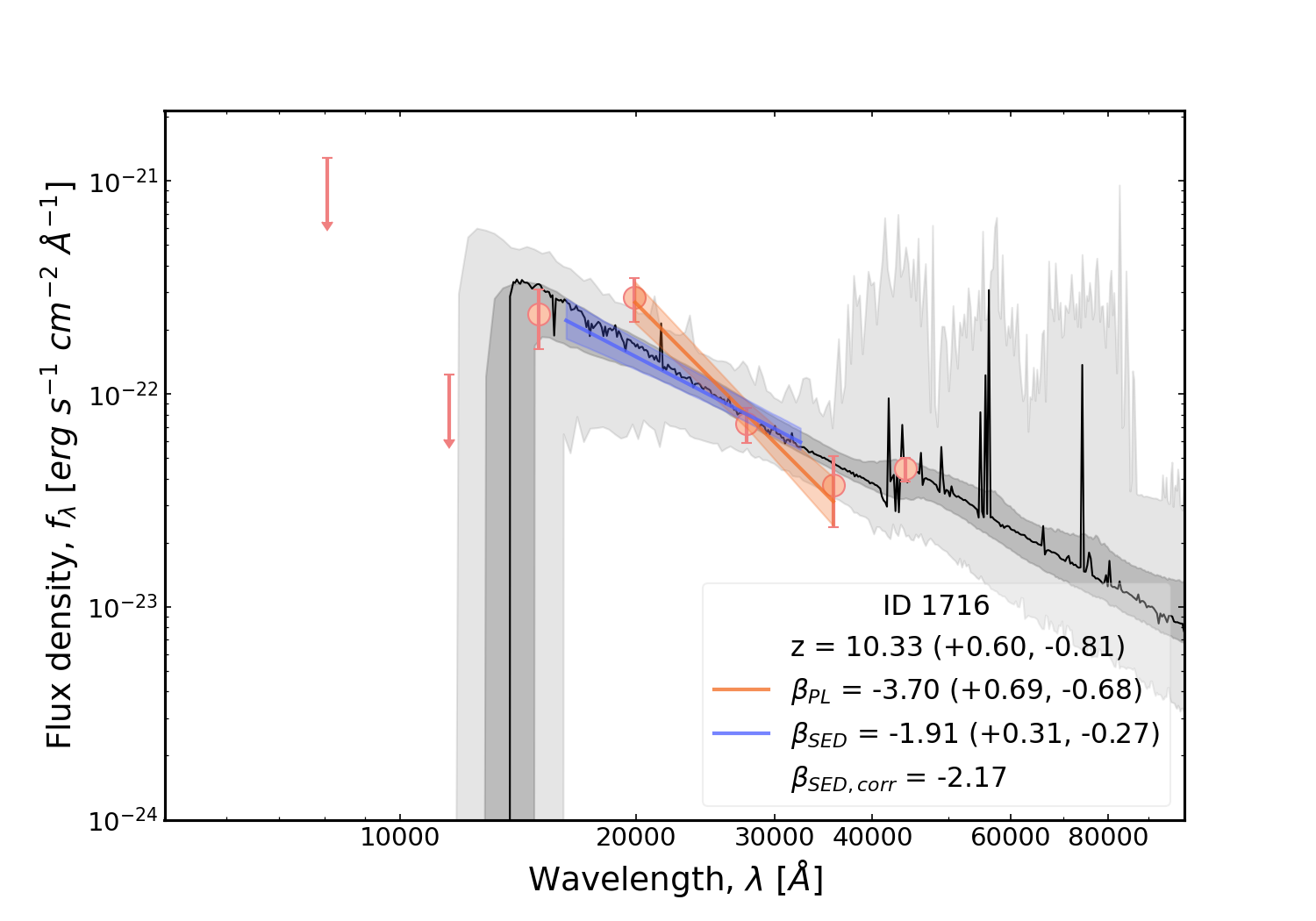}
  \includegraphics[width=.33\linewidth]{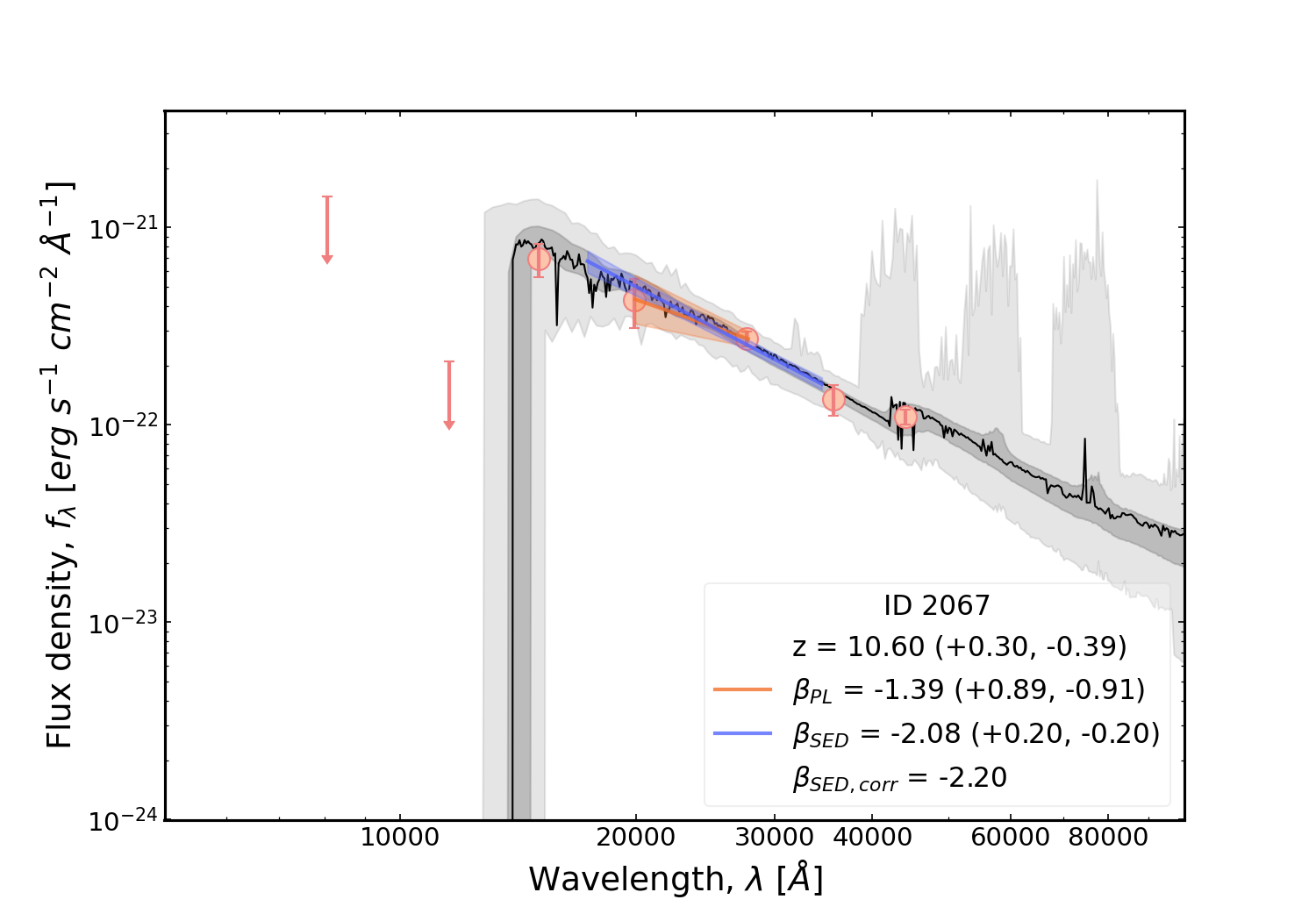}
  \includegraphics[width=.33\linewidth]{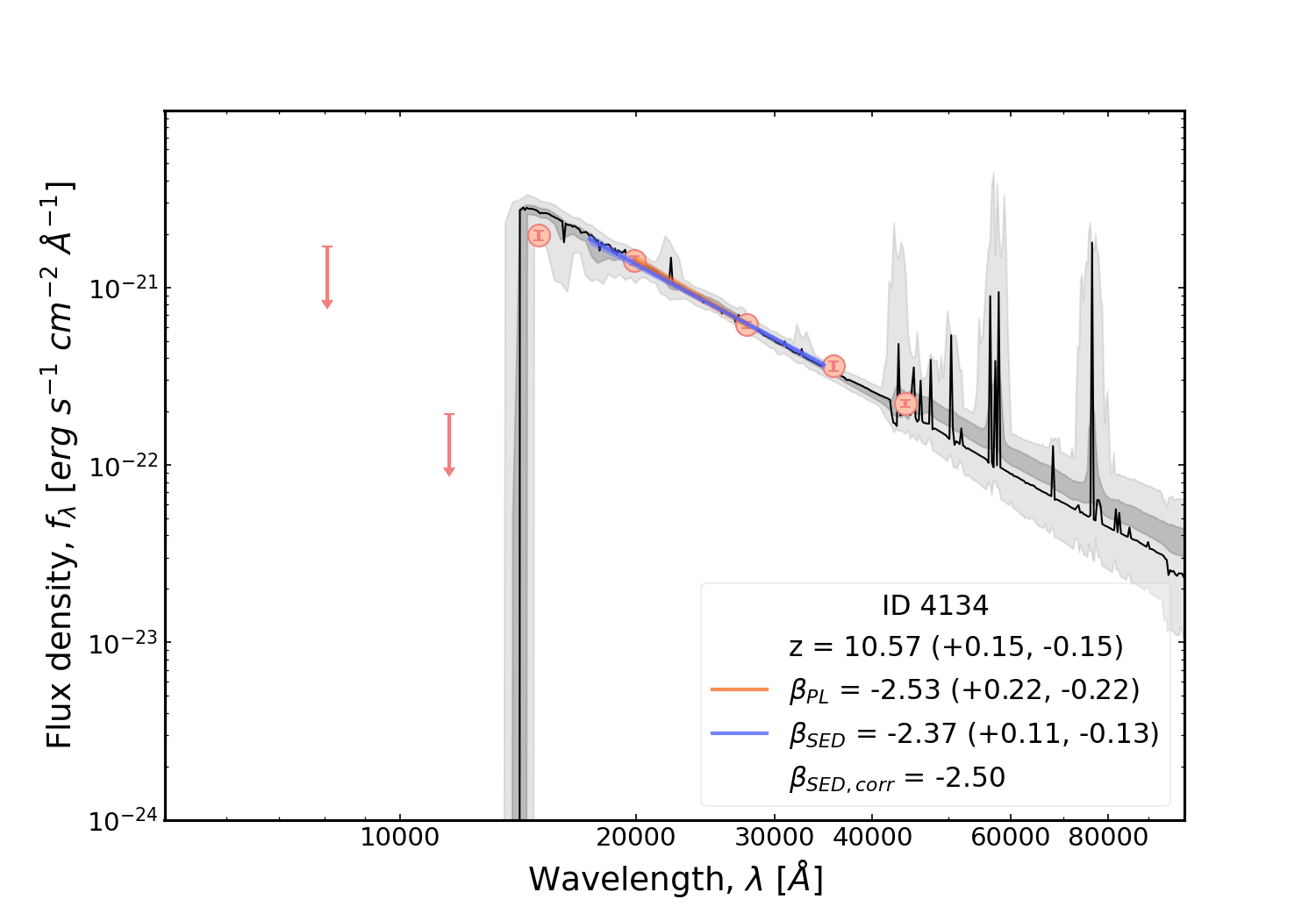} 

  \hfill 
\end{figure}
\pagebreak

\begin{figure}[htb]
    
    \includegraphics[width=.33\linewidth]{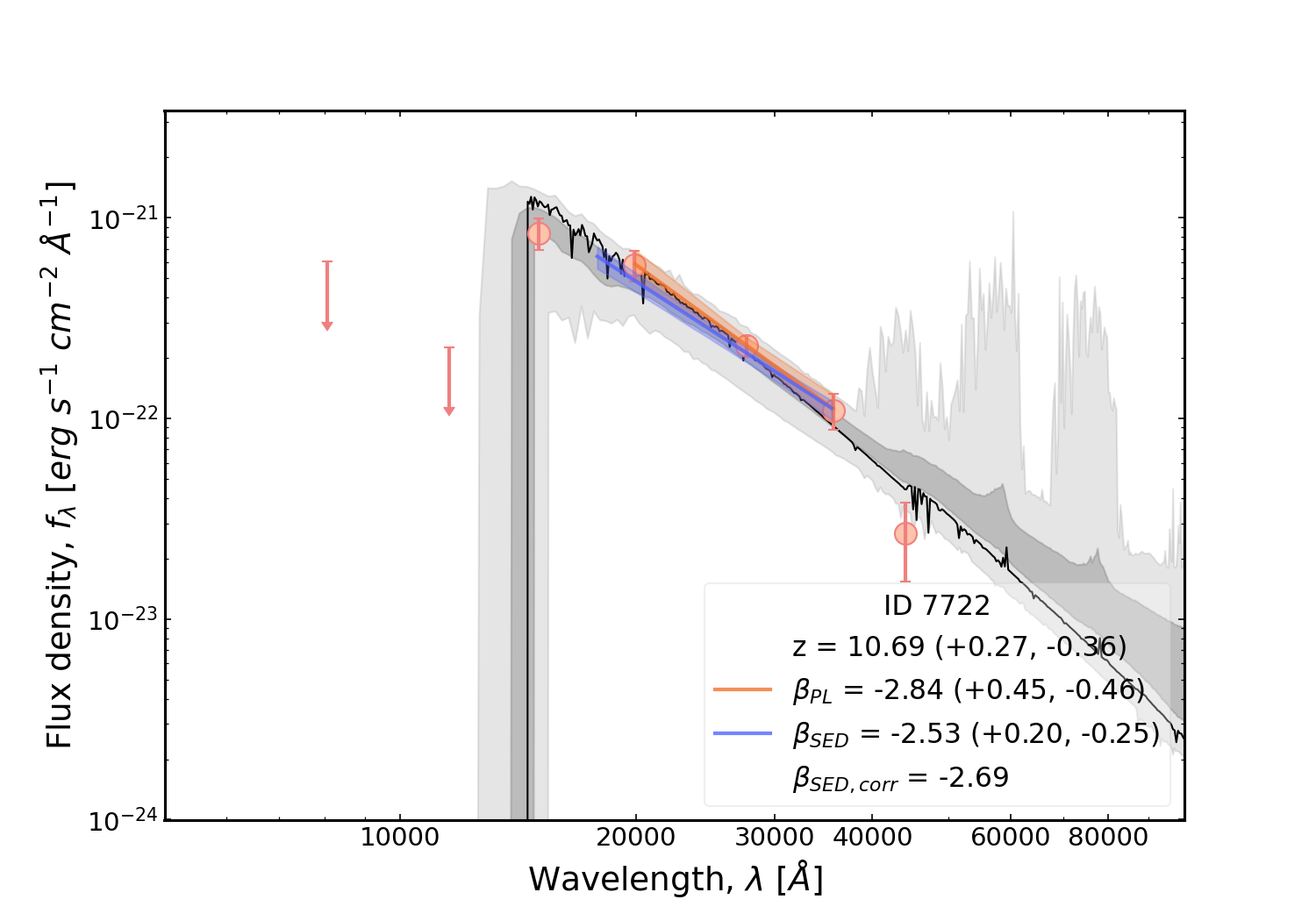}%
    % \hfill
    \includegraphics[width=.33\linewidth]{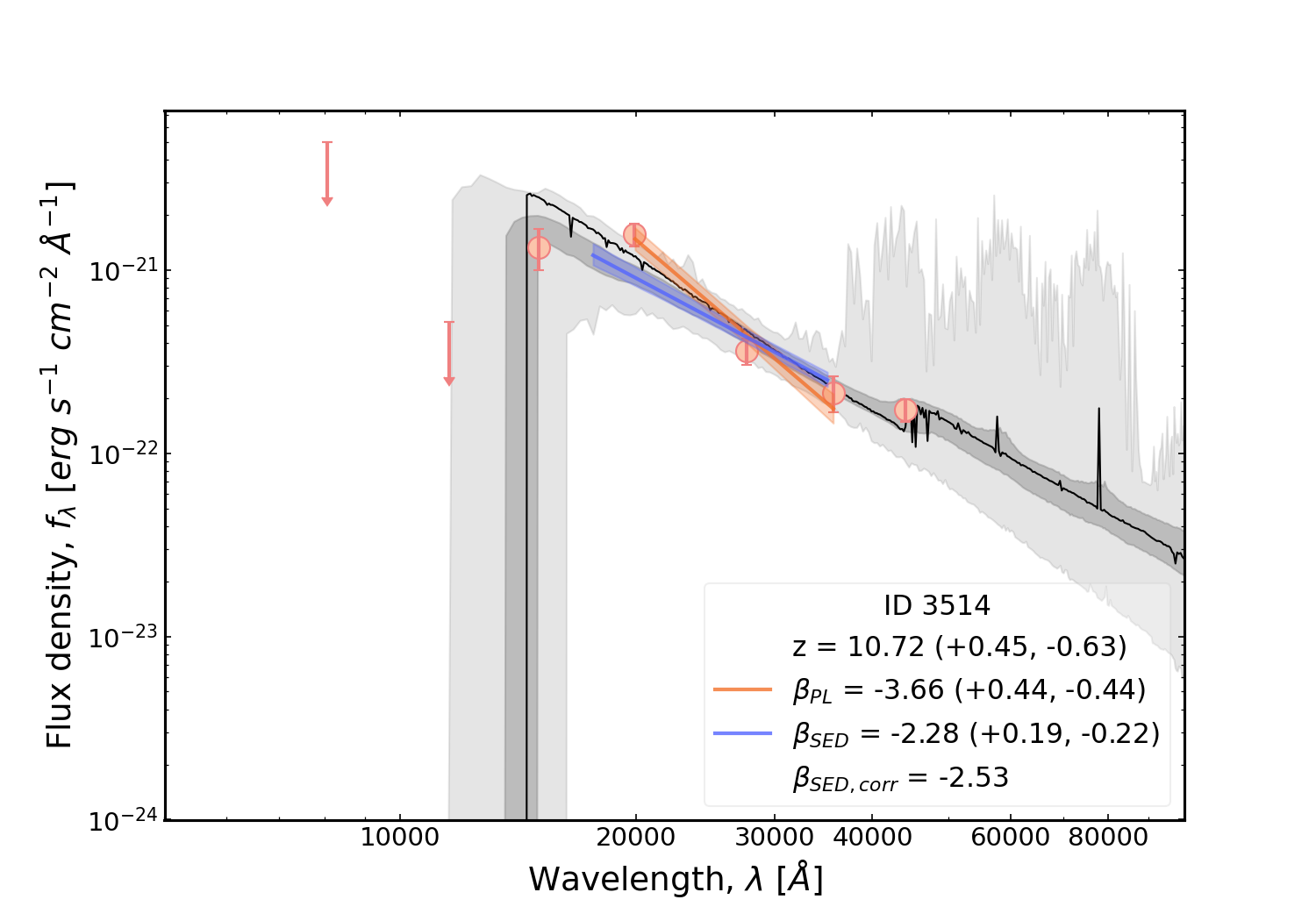}
    \hfill
    \includegraphics[width=.33\linewidth]{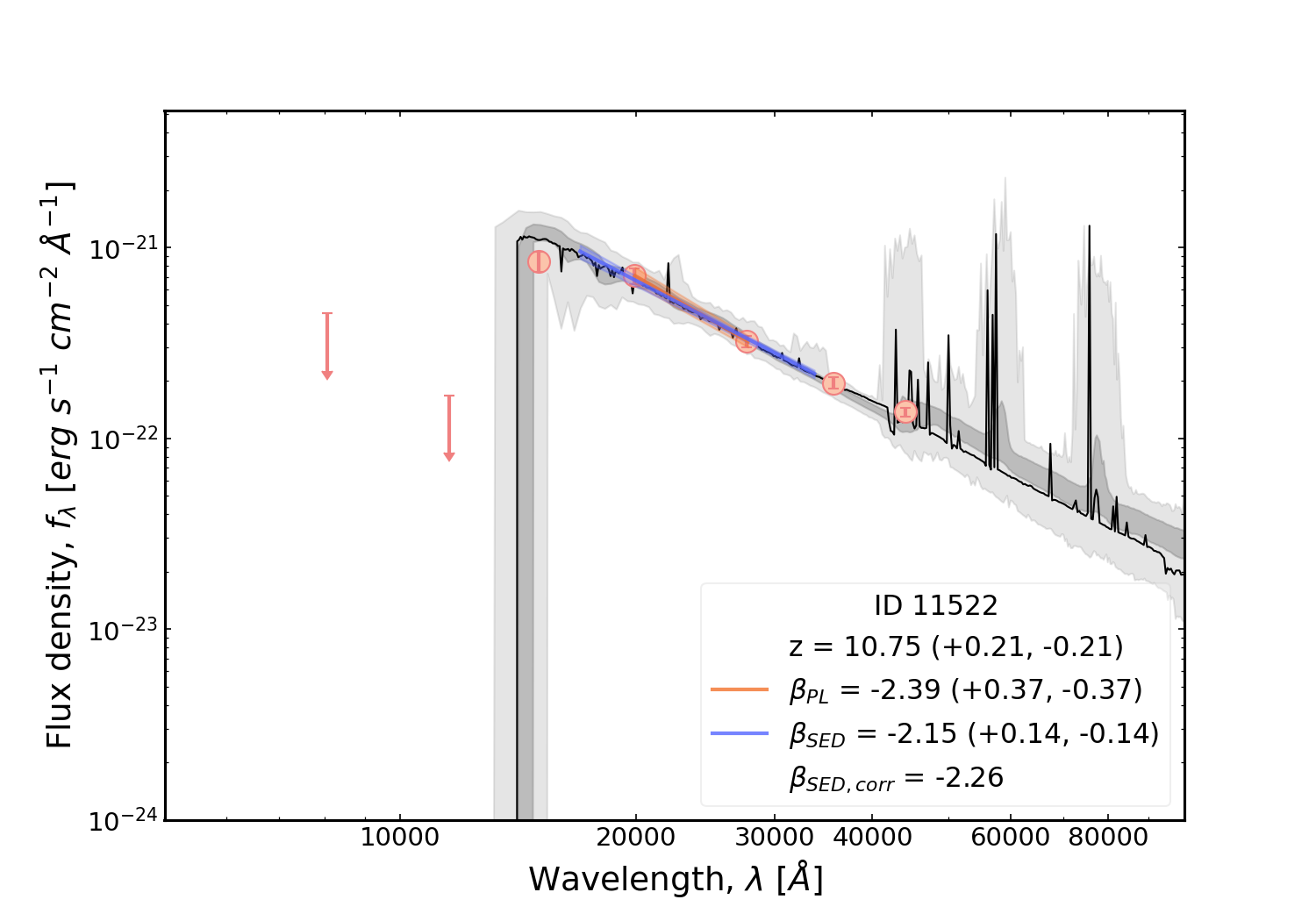}%
    \hfill
    \includegraphics[width=.33\linewidth]{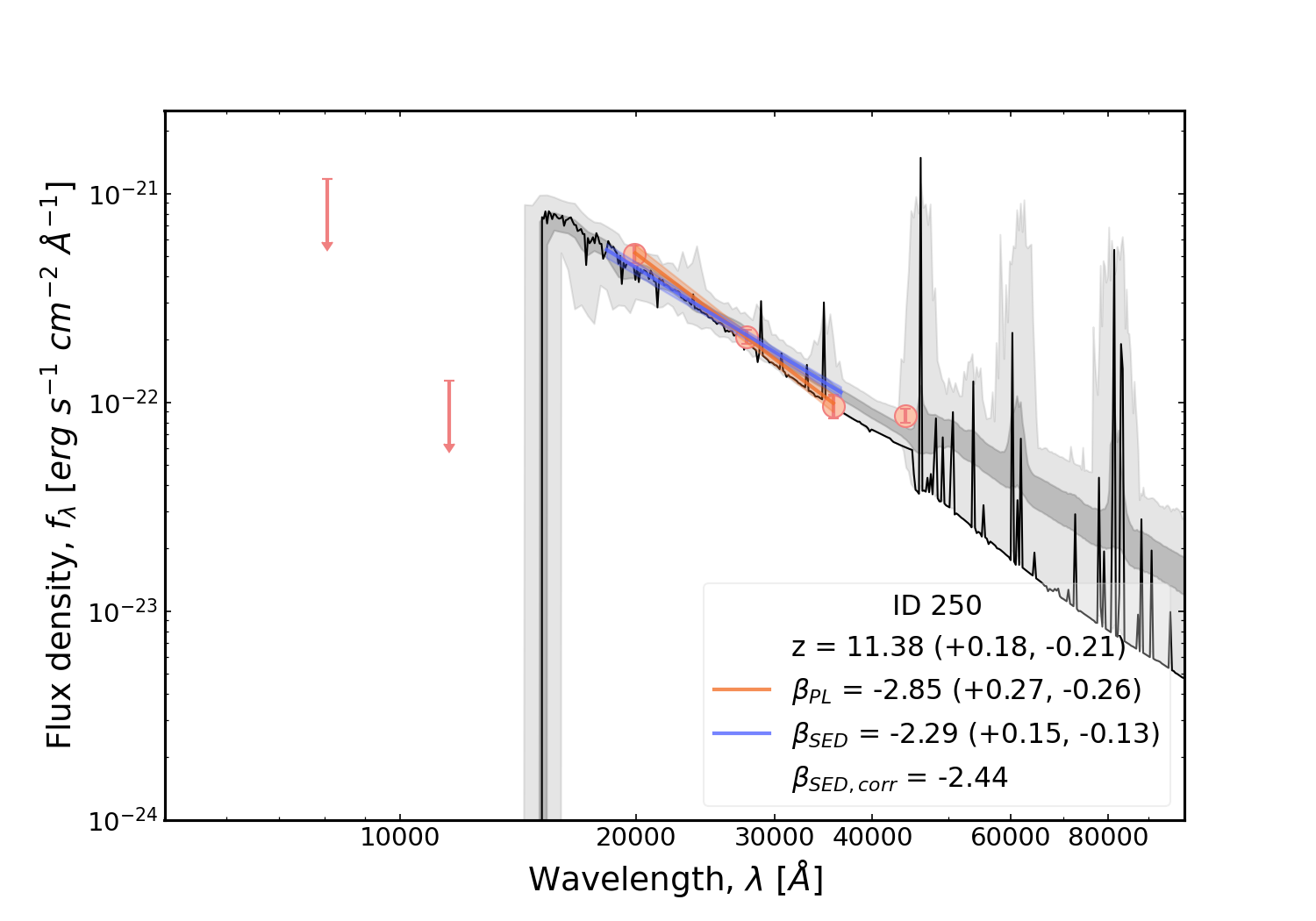}
    \includegraphics[width=.33\linewidth]{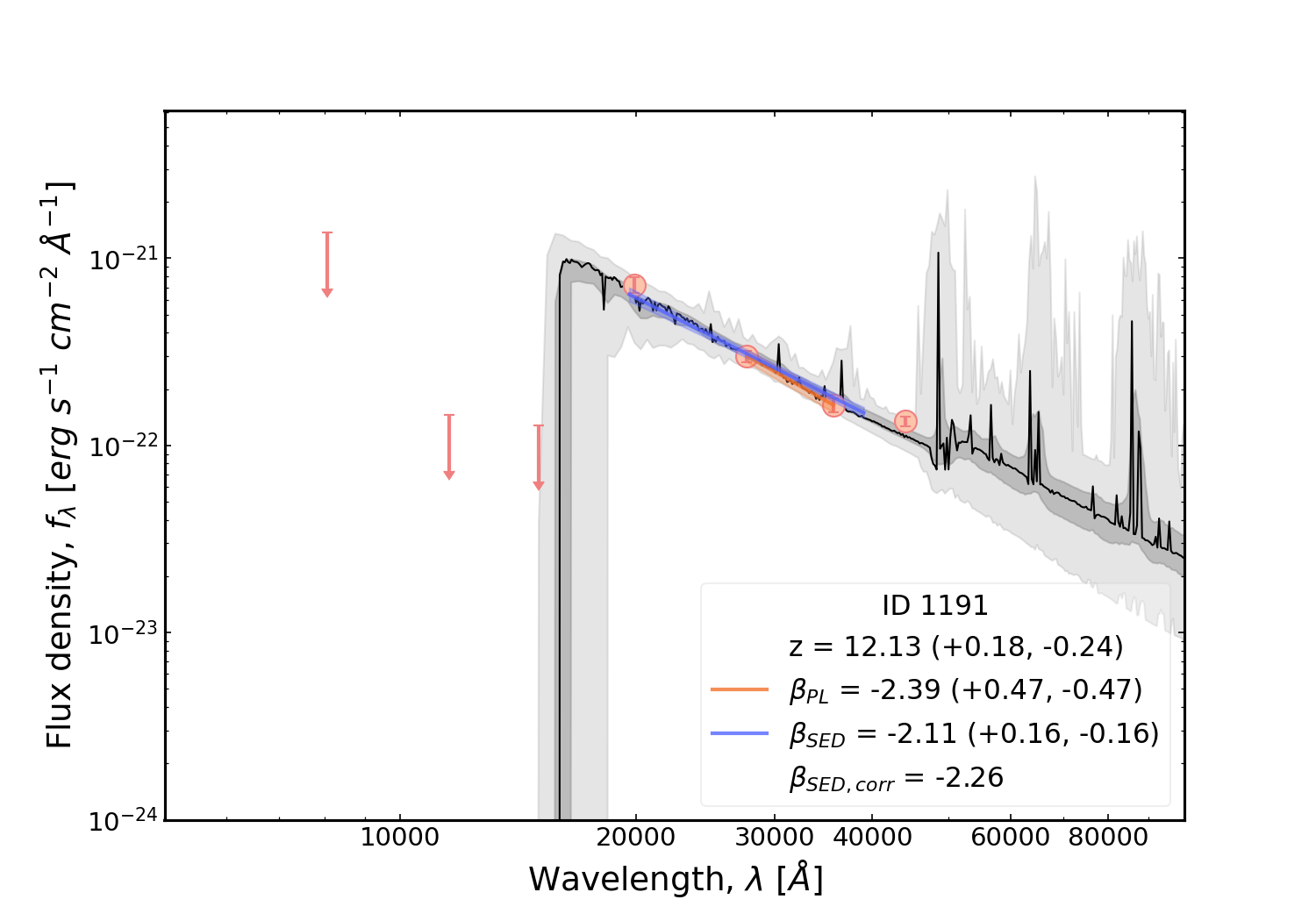}%
    \includegraphics[width=.33\linewidth]{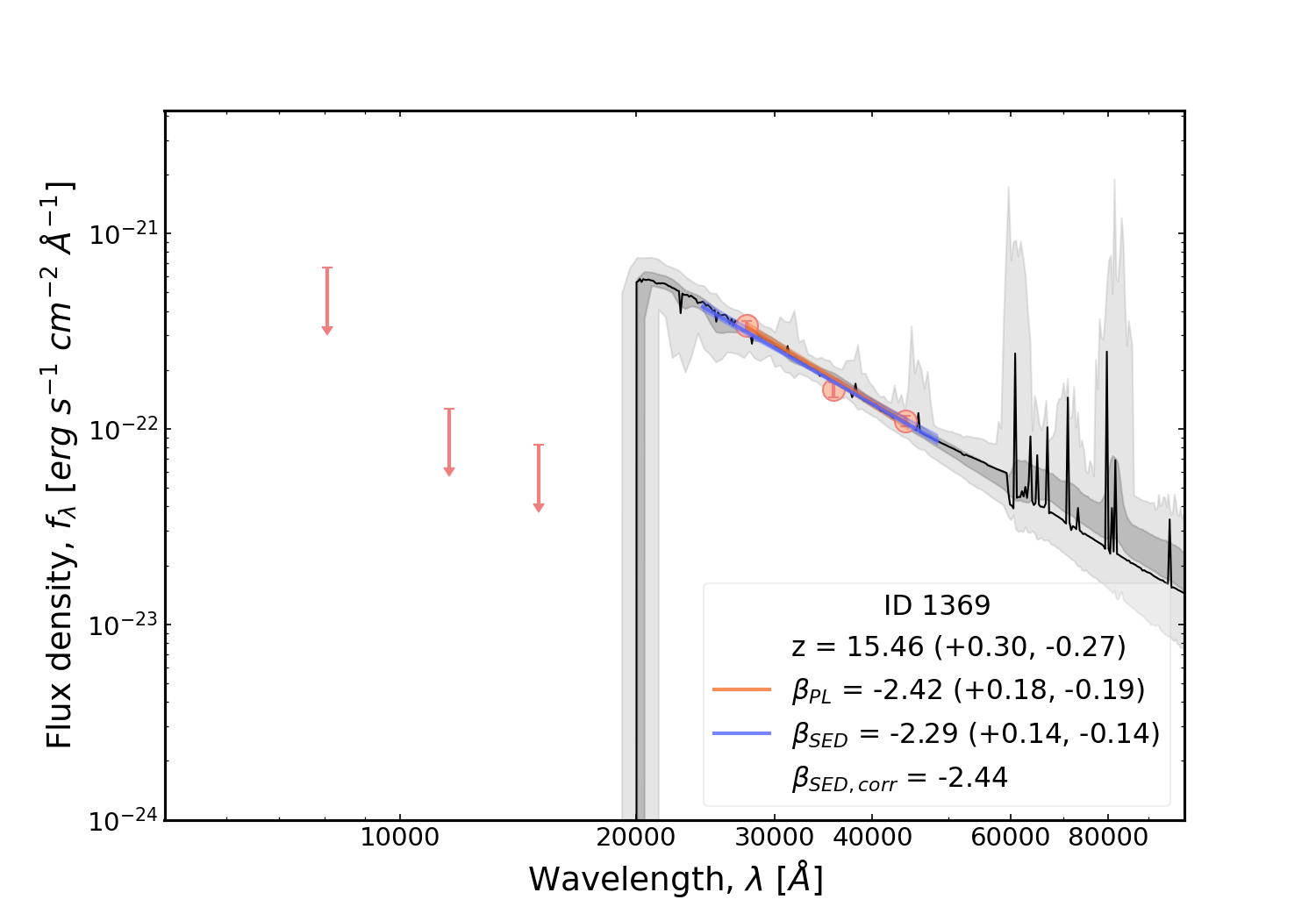} 
    \hfill
    
    \caption{\textsc{Bagpipes} SEDs to observed photometry for all 36 galaxies in our sample. The solid black line is the least $\chi^2$ SED model, the dark grey-shaded regions indicate the $1\sigma$ bounds of all 1000 SEDs fit to the photometry, while the light grey shaded regions are the full (100th percentile) range of the SED models. The orange points are the observed photometry taken with \textit{JWST}. The orange (and purple) lines and shaded regions are the median and $1\sigma$ confidence interval for the measured \bpl \ (and \bsed), as noted in the legends.}
    \label{fig:app_seds}
\end{figure}

\end{document}